\documentclass[conference]{IEEEtran}

\ifCLASSINFOpdf
\else
\fi
\DeclareMathAlphabet{\mathcal}{OMS}{cmsy}{m}{n}
\usepackage{cite}
\usepackage{float}
\usepackage{amsmath,amssymb,amsfonts}
\usepackage[hyphens]{url}
\usepackage[hidelinks]{hyperref}
\hypersetup{breaklinks=true}
\usepackage{stmaryrd}
\usepackage{xcolor}
\usepackage{graphicx}
\usepackage{wasysym}
\usepackage{pifont}% http://ctan.org/pkg/pifont
\newcommand{\cmark}{\ding{51}}%
\usepackage{subcaption}
\usepackage{mathtools}

\usepackage{empheq}
\usepackage{cases}
\usepackage{algorithm}
\usepackage[noend]{algpseudocode}
\usepackage{verbatim}
\usepackage{multirow}
\usepackage{booktabs}
\usepackage{tikz}
\usepackage{xspace}
\usepackage{tikz}
\usepackage{placeins}
\usepackage{hyperref}
\usepackage{sidecap}
\usepackage{makecell}
\sidecaptionvpos{figure}{c}
\newcommand\crule[3][black]{\textcolor{#1}{\rule{#2}{#3}}}

\definecolor{boxcolor}{HTML}{108f64}

\usepackage{threeparttable}
\usepackage{tabularx}
\usepackage{comment}
\usepackage{siunitx}
\usepackage{tikz}

\newcommand{\midsepremove}{\aboverulesep = 0mm \belowrulesep = 0mm}
\midsepremove
\newcommand{\midsepdefault}{\aboverulesep = 0.605mm \belowrulesep = 0.984mm}
\midsepdefault
\algrenewcommand\algorithmicfor{\textbf{for}} % <=======================
\algrenewcommand\algorithmicend{\textbf{end}} % <=======================

\newcommand{\norm}[1]{\displaystyle \left\| #1 \right\|}

\usepackage{cleveref}

\crefformat{section}{\S#2#1#3} % see manual of cleveref, section 8.2.1
\crefformat{subsection}{\S#2#1#3}
\crefformat{subsubsection}{\S#2#1#3}

\begin{document}
%
% paper title
% can use linebreaks \\ within to get better formatting as desired
\title{\LARGE \bf Magmaw: Modality-Agnostic Adversarial Attacks on \\ Machine Learning-Based Wireless Communication Systems %}
\vspace{-0.3cm}}

\def\jungwoo{\textcolor{black}}
\def\seira{\textcolor{green}}
\newcommand{\sys}{Magmaw\xspace}

% \setlength\abovedisplayskip{5pt}
% \setlength\belowdisplayskip{5pt}

% author names and affiliations
% use a multiple column layout for up to three different
% affiliations
% \author{\IEEEauthorblockN{Michael Shell}
% \IEEEauthorblockA{Georgia Institute of Technology\\
% someemail@somedomain.com}
% \and
% \IEEEauthorblockN{Homer Simpson}
% \IEEEauthorblockA{Twentieth Century Fox\\
% homer@thesimpsons.com}
% \and
% \IEEEauthorblockN{James Kirk\\ and Montgomery Scott}
% \IEEEauthorblockA{Starfleet Academy\\
% someemail@somedomain.com}}

% conference papers do not typically use \thanks and this command
% is locked out in conference mode. If really needed, such as for
% the acknowledgment of grants, issue a \IEEEoverridecommandlockouts
% after \documentclass

% for over three affiliations, or if they all won't fit within the width
% of the page, use this alternative format:
% 

\author{
\parbox{\linewidth}{\centering
Jung-Woo Chang$^*$,
Ke Sun$^*$,
Nasimeh Heydaribeni$^*$,
Seira Hidano$^\dagger$,
Xinyu Zhang$^*$,
Farinaz Koushanfar$^*$ \\
$^*$University of California San Diego \quad $^\dagger$KDDI Research, Inc. \\
}
%\vspace{-0.6cm}
}

% use for special paper notices
%\IEEEspecialpapernotice{(Invited Paper)}

\IEEEoverridecommandlockouts
\makeatletter\def\@IEEEpubidpullup{6.5\baselineskip}\makeatother
\IEEEpubid{\parbox{\columnwidth}{
    Network and Distributed System Security (NDSS) Symposium 2025\\
    24-28 February 2025, San Diego, CA, USA\\
    ISBN 979-8-9894372-8-3\\
    https://dx.doi.org/10.14722/ndss.2025.230336\\
    www.ndss-symposium.org}
    \hspace{\columnsep}\makebox[\columnwidth]{}
}

% make the title area
\maketitle

\begin{abstract}
Machine Learning (ML) has been instrumental in enabling joint transceiver optimization by merging all physical layer blocks of the end-to-end wireless communication systems. Although there have been a number of adversarial attacks on ML-based wireless systems, 
the existing methods do not provide a comprehensive view including multi-modality of the source data, common physical layer protocols, and wireless domain constraints. This paper proposes \sys, \jungwoo{a novel wireless attack methodology} capable of generating universal adversarial perturbations for any multimodal signal transmitted over a wireless channel. %that is robust the physical layer variations.
%We define the problem of generating adversarial perturbation signals by considering the design constraints for a practically feasible attack together with radio characteristics at the physical layer.
%We address several design challenges to generate practically feasible attacks on wireless systems.
%This proposed attack method allows for the creation of universal adversarial perturbations that are agnostic to the modality (e.g., image, video, text, and speech) of the transmitted signal and are robust to various transformations in the conditions of the physical layer.
%even though the adversarial device is not synchronized with a victim device.
We further introduce new objectives for adversarial attacks on downstream applications. We adopt the widely-used defenses to verify the resilience of \sys.
For proof-of-concept evaluation, we build a real-time wireless attack platform using a software-defined radio system. 
%, and then inject crafted adversarial signals into the wireless channel.
%Experimental results demonstrate that \sys causes significant performance degradation even in the presence of encrypted content, and is also effective in non-ML wireless communications.
Experimental results demonstrate that \sys causes significant performance degradation even in the presence of strong defense mechanisms. \jungwoo{Furthermore, we validate the performance of \sys in two case studies: encrypted communication channel and channel modality-based ML model. Our code is available at \textcolor{magenta}{\href{https://github.com/juc023/Magmaw} {https://github.com/juc023/Magmaw}}.} %and conventional communications.
% Surprisingly, \sys is also effective against encrypted communication channels.
%We further implement \sys real-time wireless attacks using software defined radio (SDR) and verify its superiority and practicality.
\end{abstract}

\vspace{-0.2cm}
\section{Introduction}
\label{sec:intro}
\vspace{-0.1cm}
%Structure of the introduction:
%What is the problem? Motivation? Why is it important?
% What has been done?
% Limitations
%what is it that we are doing?
%contributions
%paper organization

%What is the problem? Motivation? Why is it important?
Next-generation (NextG) networks promise to support ultra-reliable and low-latency communication for rapidly evolving wireless devices~\cite{chowdhury20206g}. Emerging networks are thus challenged to establish new features (e.g., adaptive coding and enhanced modulation) to overcome rapidly changing channel conditions and to achieve more efficient use of spectrum~\cite{zhang2022near,  qin2021semantic}. 
Machine Learning (ML) overcomes this barrier by revolutionizing the entire wireless network protocol stack~\cite{saad2019vision}. 

Recent research~\cite{bourtsoulatze2019deep} introduces joint source-channel coding (JSCC), an end-to-end wireless communication system leveraging deep neural networks (DNNs) for both transmitter and receiver. This ML approach jointly optimizes source and channel coding in a cross-layer framework to handle diverse and challenging channel conditions. 
To effectively cope with the multipath fading effects, the JSCC-encoded data can be further modulated into continuous signal waveforms through orthogonal frequency division multiplexing (OFDM)~\cite{yang2022ofdm}.
The DNN models for JSCC are tailored to specific modalities (e.g., texts, images, etc.), so as to convey semantic information more accurately than traditional communication systems (see \cref{sec:modality-specific}).  
%ML-based JSCC model can convey the semantic information of each modality more accurately than traditional communication systems.
The advantages of such ML-based communication systems are increasingly recognized by standardization bodies such as the Third Generation Partnership Project (3GPP)~\cite{wang2022transformer}. \jungwoo{Industry leaders, such as Apple~\cite{sambhwani2022transitioning}, Huawei~\cite{Huawei}, Nokia Bell Labs~\cite{Nokia}, Qualcomm~\cite{Qualcommwhite}, and ZTE~\cite{liu2023sst} are also investigating AI-native 6G communications. NVIDIA has established an ML-based, GPU-accelerated communication signal processing framework \cite{hoydis2022sionna} for 6G applications. These developments underscore the growing consensus that ML-based wireless communications will play a crucial role in shaping the future of 6G technology.}

%These efforts will enable most radio hardware to utilize ML-based communications.}
%

Unfortunately, ML is vulnerable to adversarial attacks~\cite{szegedy2013intriguing, biggio2018wild}, where small, imperceptible changes to input can yield substantial changes in the model's output. 
The susceptibility of the models to adversarial examples raises serious concerns for the safety of ML adoption in NextG. 

%As ML integration into wireless communication systems grows, we aim to investigate the real-world vulnerabilities of ML-based wireless systems through wireless adversarial perturbations.

%The semantic features extracted from JSCC can be trained to be robust to the channel model between transmitter and receiver. 

 %ML-based wireless systems have shown tremendous potential in efficiently transmitting different types of modalities, e.g., image~\cite{yang2022ofdm}, video~\cite{wang2022wireless}, speech~\cite{weng2021semantic}, and text~\cite{xie2021deep}.
 
%Due to its enhanced performance over conventional schemes, many standardization organizations, including the third generation partnership project (3GPP) and industry, engage in various activities to adopt ML in 5G and beyond~\cite{wang2020artificial, wang2022transformer, hoydis2022sionna, sambhwani2022transitioning}.

\begin{figure}[t]
    \centering
    \begin{tabular}{@{}c@{}}
        \includegraphics[width=0.9\columnwidth]{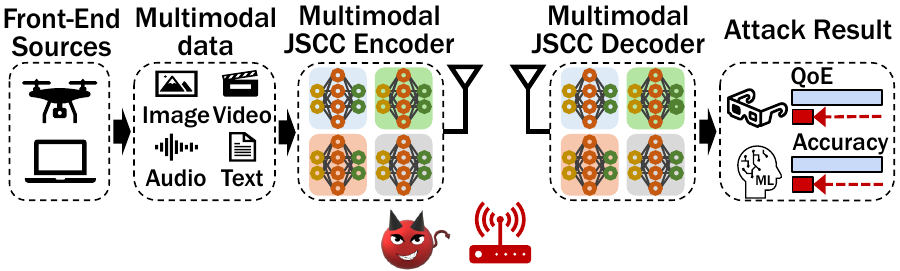} \\
    \end{tabular}
    \vspace{-0.2cm}
    \caption{High-level view of \sys.} 
    %By injecting small amounts of adversarial signals over the air into the channel, \sys can severely degrade the performance.}
    \label{fig:intro_0}
    \vspace{-0.4cm}
\end{figure}

 %~\cite{yang2022ofdm, wang2022wireless, weng2021semantic, xie2021deep}.

% What has been done?
%Unfortunately, DNN models have been shown to be vulnerable to adversarial attacks~\cite{szegedy2013intriguing, biggio2018wild}, where small, imperceptible changes to input can yield substantial changes in the model's output. 
%Such attacks were first identified in the image domain~\cite{goodfellow2014explaining, papernot2017practical} and were later extended to other modalities~\cite{ji2021poltergeist, pierazzi2020intriguing, sato2021dirty, carlini2018audio, jin2023pla, cao2023stylefool}. 
%The susceptibility of the models to adversarial examples raises serious concerns for the safety of ML adoption in NextG~\cite{bahramali2021robust, kim2021channel, liu2023exploring}. 

%
%\noindent\textbf{Existing Wireless Attacks and Limitations.} 
Traditional jamming or overshadowing attacks~\cite{xu2005feasibility,yang2019hiding, strasser2008jamming, erni2022adaptover} have been dedicated to developing a malicious RF device to disrupt legitimate wireless communications. However, these approaches typically rely on high-power transmissions to cause large-scale disruptions in the spectrum, leading spectrum owners to respond swiftly. 
%Even reactive jamming~\cite{wilhelm2011short}, which emits jamming signals only when the channel is in a non-ideal state, necessitates high signal strength to disrupt the channel.
Highly effective attacks that use low signal strengths are missing in this literature.

There have been recent works on small signal manipulations designed to target ML-based wireless systems~\cite{bahramali2021robust,  nan2023physical, li2023boosting, hu2023robust, liu2023exploring}.
However, they make unrealistic assumptions about the attacker's capabilities. For example, even though JSCC has a modality-specific structure, they assume that only a single modality (e.g., one-hot vector message or image) is wirelessly communicated. They also assume that the adversary knows which modality is sent by the transmitter. 
In practice, the above assumptions are not valid for the following reasons: 1) the transmitter typically incorporates data from all modalities into the data blocks and then sends them to the receiver; 2) if the adversary wants to recognize the modality of the signal, it needs to have access to the target ML model that carries out JSCC, and this is not always feasible, and 3) even if the adversary can detect the modality, high latency occurs until perturbations are generated and added to the victim signal. 
We propose \sys, a new hardware-driven wireless attack framework that creates universal adversarial perturbations (UAPs) to subvert ML-based wireless systems. We show for the first time that modulated multimodal data can be perturbed by adversaries, resulting in failure to restore the original data as well as subversion of downstream services. We consider examples of downstream services such as video classification (VC), which analyzes human activity from video, and audio-visual event recognition (AVE) which predicts the event label based on representations over multiple input modalities. 
%
%We assume OFDM is used as the underlying modulation scheme, which is common in modern wireless standards (4G LTE~\cite{abu2013uplink}, IEEE 802.11 family~\cite{hwang2008ofdm}, 5G NR~\cite{lin2022overview}). 
%OFDM divides a data stream into multiple sub-streams, which are transmitted in parallel. 
%We present a comprehensive framework that models the attack objective as an optimization problem and integrates the features of heterogeneous wireless networks as domain constraints. 
%xyz: Unclear what's reason why it works without synchronization
%jw: I added more explanation below.
%We incorporate a transformation mechanism in the objective function so that our UAP is trained to be robust to various transformations occurring in the physical layer.
%\sys can operate over a hardware-implemented platform without requiring prior knowledge of the sender and the receiver, and directly inject adversarial signals into the receiver's antenna. 
%
\sys can cause significant disruptions or threaten user safety in quality-sensitive applications, e.g., remote surgery~\cite{acemoglu20205g} and autonomous driving~\cite{he20206g}, as illustrated in Figure~\ref{fig:intro_0}. Emerging applications (e.g., XR~\cite{lin2021wireless}) would suffer even more from the corruption of multiple input modalities.

%\noindent\textbf{Design Challenges and Solutions.}
%Many recent works~\cite{bahramali2021robust, sadeghi2019physical, nan2023physical, li2023boosting, hu2023robust, liu2023exploring} aim at crafting adversarial signals on end-to-end wireless communication systems, but they make unrealistic assumptions about the attacker's capability. We summarize four new challenges that prevent existing attacks from being launched in practice. The overview of our challenges is depicted in Figure~\ref{fig:intro_workflow}. 
%Table~\ref{tab:summary} shows the comparison of \sys with the existing adversarial attacks on wireless networks. \sys includes the key elements required to create a practically feasible attack. In the following, we describe these challenging elements and our solutions in detail. 
%
\jungwoo{\sys must address four main design challenges.} Firstly, we assume that the adversary lacks prior knowledge about the data's modality and the exact channel model. %The diversity of JSCC models for different modalities makes crafting UAPs challenging. For instance, video JSCC models use spatiotemporal prediction to optimize transmission efficiency through a temporal chain of dependency among adjacent data symbols.
Additionally, the attacker's ability to adjust its transmit signal pattern effectively depends on knowing the channel matrix between the sender and receiver ($\textbf{H}_{\textbf{t}}$).
However, since $\textbf{H}_{\textbf{t}}$ varies due to factors like link distance, mobility, and environment, not having this information makes crafting an effective attack challenging.
%they assume that only a single modality (e.g., one-hot vector or image) is wirelessly communicated, and thus an adversary is aware of which modality is sent by the transmitter. One recent work~\cite{liu2023exploring} is devoted to attacking wireless sensing applications, but their approach aims at perturbing publicly-known inputs (preambles~\cite{biswas2004channel}). In practice, the above assumptions are not valid for the following reasons: 1) the data blocks in radio signals carry multimodal data; 2) if the adversary wants to recognize the modality of the signal, it needs to have access to the target ML model that carries out JSCC, and this is not always feasible, and 3) even if the adversary can detect the modality, high latency occurs until adversarial perturbations are generated and added to the victim signal. 
We solve the above challenges by designing a perturbation generator model (PGM) trained to create input- and channel-agnostic perturbations on surrogate wireless models. We adopt an ensemble learning approach that utilizes surrogate multimodal JSCC models to learn UAPs.
%perturbations that can be well generalized to all modalities.
%learn modality-agnostic perturbations.
%As a result, the realistic attacks must be modality-invariant.

\begin{comment}
\begin{figure}[t]
    \centering
    \begin{tabular}{@{}c@{}}
        \includegraphics[width=0.88\columnwidth]{Fig/fig_intro_workflow.png} \\
    \end{tabular}
    %\vspace{-0.2cm}
    \caption{Overview of the new challenges we have addressed.}
    \label{fig:intro_workflow}
    %\vspace{-0.5cm}
\end{figure}
\end{comment}

Secondly, previous attacks~\cite{bahramali2021robust,  nan2023physical, li2023boosting, hu2023robust, liu2023exploring} do not consider that the input signal can be adjusted by physical layer protocols (see \cref{sec:related_work_protocol}). They only focus on the scenarios where the adversary has prior knowledge of the protocol's full setup. In a practical scenario, the attacker does not know the constellation mapping or how the OFDM system assigns the complex symbols to multiple subcarriers. \jungwoo{It is possible to design an attacker that recognizes the protocol from the transmitted signals \cite{shi2018spectrum}. However, since wireless protocols change rapidly depending on the channel state, the analyzed output quickly becomes obsolete. A protocol-agnostic attack is required.}
%but the attack's real-time constraints prohibit online analysis. 
We address this challenge by incorporating multiple controllable parameters inside the ensemble learning to optimize perturbations generalizable across all modulated data.

%designing a Universal Adversarial Perturbation (UAP) trained to create protocol-agnostic perturbations on surrogate models.

%by designing a perturbation generator model (PGM) trained to create input-agnostic perturbations on surrogate multimodal JSCC models. 

%To ensure the transferability to attackers’ unknown synthesis models, an ensemble learning approach is adopted to improve the generalizability of the optimization process.

%Specifically, we build an ensemble of surrogate models with different modalities, modulation, and coding rates for transfer-based attack.

Thirdly, an adversarial wireless device may not be precisely synchronized with a legitimate transmitter or receiver in the time or frequency domain, reducing the effectiveness of perturbations.
%
%Adversarial signals are less effective due to time and phase offsets, which the adversary cannot predict.
%
We address de-synchronization issues between the adversarial device and the legitimate transmitter/receiver using our offline training procedure. Specifically, we train the PGM using time shift and phase rotation functions, ensuring that UAPs remain effective even with varying offsets.

Finally, previous studies~\cite{bahramali2021robust, liu2023exploring} are vulnerable to adaptive defenses. For instance, a perturbation detector~\cite{xu2021detecting} can exploit traces of perturbations to predict whether the input is perturbed. This is because their perturbations are overly rigid and lacking in variability due to overfitting~\cite{gulrajani2020towards}. To craft robust and diverse perturbations, we introduce a discriminator and diversity loss to regularize the learning process explicitly.
After integrating the above solutions, we implement \sys on the software-defined radio platform and validate its attack feasibility, as shown in Appendix~\ref{sec:appendix_real_world}. Our experiments show that \sys degrades the Peak Signal-to-Noise Ratio (PSNR) by up to 8.04dB and 8.29dB for image and video transmission, respectively, where PSNR is a representative image quality score. For speech transmission, \sys prevents receivers from recognizing the speech content, increasing the mean square error (MSE) by up to 3.91$\times$ compared to baseline attacks. Furthermore, \sys reduces the bilingual evaluation understudy (BLEU) score to 0.338 points for text transmission, indicating that the received text exhibits significant semantic errors and grammatical inaccuracies. Notably, we achieve up to 91.2$\%$ attack success rate on the downstream tasks. 
%More importantly, we evaluate the robustness of \sys against adaptive defenses, among which the detector accuracy is less than 7$\%$.
In our case study, we establish an encryption-based secure image transmission and prove that \sys leads to a reduction of up to 5.88dB in PSNR. \jungwoo{We also evaluate \sys with channel modality-based ML models. \sys introduces up to $2.2\times$ more error in the ML results than the baseline.}
%As a result, we confirm its transferability to case studies.
%induces a significant semantic error, resulting in up to 5.88dB reduction in PSNR, thus confirming its transferability.

%\noindent\textbf{Contributions.} 
%As ML integration into wireless communication systems grows, \sys plays a crucial role in elucidating the threats of adversarial perturbations on these systems.
%
In summary, we make the following contributions:

%\vspace{-0.1cm}
\begin{itemize} %[leftmargin=*]
\setlength\itemsep{0.3em}
    \item We introduce \sys, \jungwoo{a novel wireless attack framework} implemented over software-defined radio against ML-based multimodal communication systems and underlying downstream applications. 
    %that include the physical layers of modern Wi-Fi and cellular systems.
    \item We adopt an ensemble learning on a set of surrogate JSCCs to craft our UAP input- and protocol-agnostic, i.e., oblivious to the modality, constellation, coding rate, OFDM specifications, and channel conditions.
    %\item We optimize the adversarial perturbation to be robust to arbitrary radio signal with various number of data symbols.
    %\item Our attack is shown to fool the downstream tasks such as video classification and audio-visual event recognition. 
    \item \jungwoo{We evaluate \sys against various defense techniques, including adaptive ones. Extensive results from case studies further show Magmaw’s efficacy.}
    %\item The attack's effectiveness for encrypted communication channels is also evaluated.
    %\item Our attack is input- and protocol-agnostic. Specifically, our attack is agnostic to the input modality, channel conditions, constellation, coding rate, and OFDM specifications.  
    %\item 
    %\item We show the resiliency of \sys against the well-known defenses, adversarial training and perturbation subtraction.
    %\item We show that our attack works much better than baseline for encrypted wireless communication also work for non-ML-based wireless communication system. Our UAP is not only effective for ML-based communications, but can also achieve high bit error rates for secure communications and conventional communications.
\end{itemize}

\vspace{-0.1cm}
\section{Background}
\label{sec:system}
%\begin{comment}
\begin{figure*}[t]
\centering
\begin{tabular}{@{}cccc@{}}
\includegraphics[height=0.088\textheight]{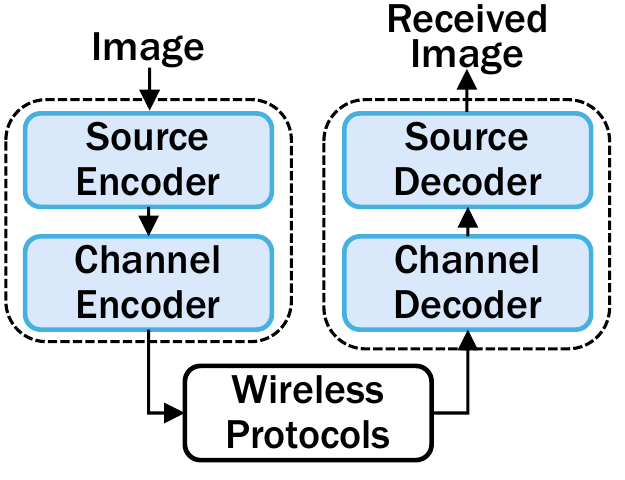} &
\includegraphics[height=0.103\textheight]{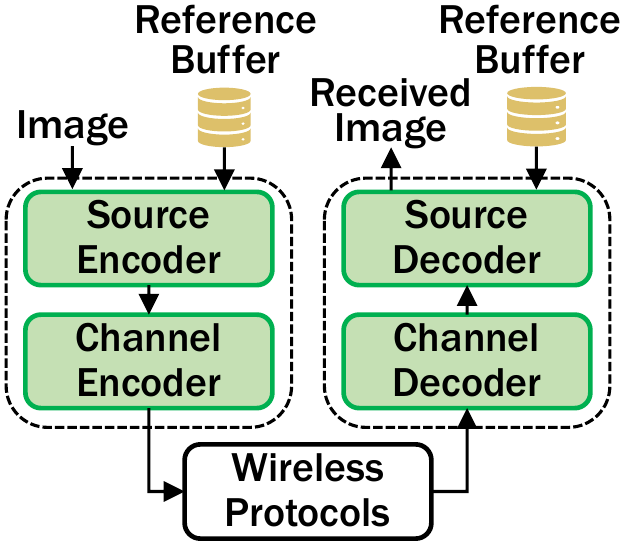} &
\includegraphics[height=0.103\textheight]{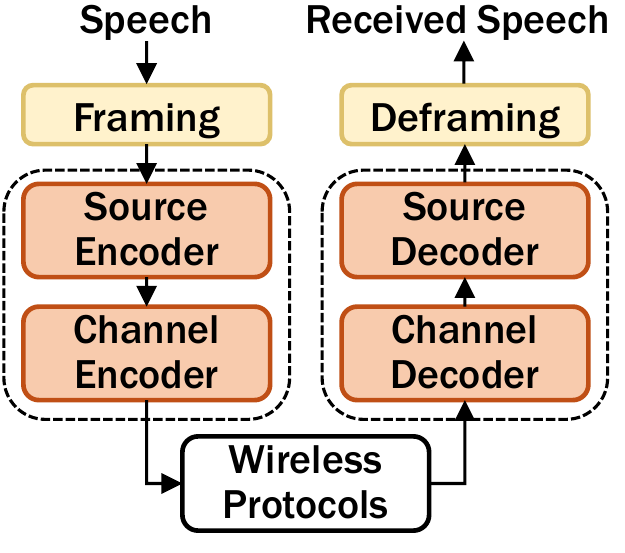} &
\includegraphics[height=0.103\textheight]{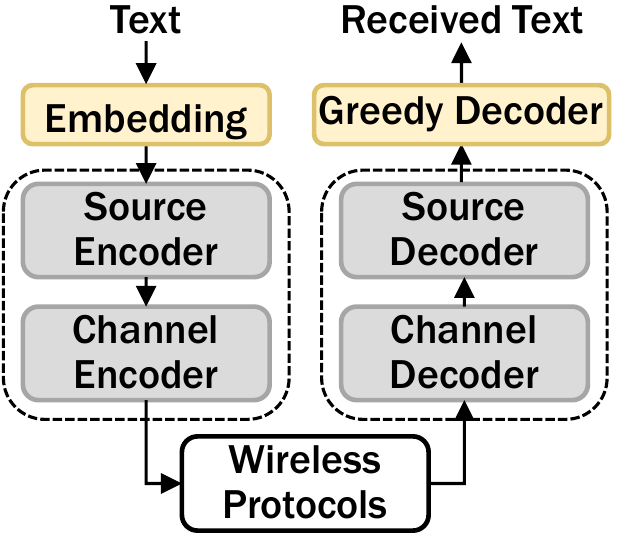} \\
\footnotesize{(a) Image JSCC model~\cite{yang2022ofdm}} & 
\footnotesize{(b) Video JSCC model~\cite{wang2022wireless}} &
\footnotesize{(c) Speech JSCC model~\cite{weng2021semantic}} &
\footnotesize{(d) Text JSCC model~\cite{xie2021deep}}  \\
\end{tabular}
\vspace{-0.1cm}
\caption{The modality-specific JSCC model for end-to-end wireless communication system.}
\label{fig:back}
\vspace{-0.4cm}
\end{figure*}
%\end{comment}

%JSCC is a classical topic in information theory and coding theory for wireless communication systems. However, the conventional JSCC schemes are based on explicit probabilistic models and manual designs, and the optimization complexity for complex sources is intractable~\cite{chen2018joint, guyader2001joint}. Recently, some pioneering works have demonstrated that ML-based JSCC systems can be realized end-to-end, optimizing all channel transformations in a data-driven manner with a non-trainable (deterministic) differential channel model~\cite{yang2022ofdm, wang2022wireless, weng2021semantic, xie2021deep}.
%We provide the background on ML-based wireless communications and state-of-the-art JSCC models.

\subsection{Wireless Communication Systems}
%We consider OFDM-based JSCC over a multipath fading channel with $L_{t}$ paths. We transmit multimodal source using $N_{s}$ OFDM symbols with $L_{fft}$ OFDM subcarriers. Note that $N_{s}$ depends on modality and coding rate. For channel estimation, the sender transmits a preamble, which is by standard and is public information, on the subcarriers. We denote the source data as $x^{Q}_{t}$ with modality $Q \in \{\mathcal{I}, \mathcal{V}, \mathcal{S}, \mathcal{T}\}$ at time step $t$, where $\mathcal{I}, \mathcal{V}, \mathcal{S}, \mathcal{T}$ denote the image, video, speech, and text, respectively. ML-based wireless models return reconstructed data $\hat{x}^{Q}_{t}$ by passing $x^{Q}_{t}$ through entire physical layer components. We describe the process by which $\hat{x}^{Q}_{t}$ is created in Section~\ref{sec:back_2}.
%\noindent\textbf{Conventional Method.}
Current communication standards (e.g., 4G LTE~\cite{abu2013uplink}, IEEE 802.11 family~\cite{hwang2008ofdm}, 5G NR~\cite{lin2022overview}) follow separate source and channel coding designs and require independent optimization of each component. The source encoder transforms the source data into the embedded source bits. The channel encoder adds redundancy to the transmitted signal, allowing the receiver to correct errors caused by noise. However, these conventional systems suffer from dramatic performance degradation due to the cliff effect where the receiver's error correction algorithm cannot recover the transmitted data if channel conditions are worse than a certain threshold~\cite{bourtsoulatze2019deep}.
%This is because error correction algorithms fail to correctly interpret the binary data.
%, resulting in a dramatic drop in output quality. \nas{the last sentence seems repetitive. }
%xyz: ML-based JSCC cannot solve these problems either
%jw: Thank you for letting me know.
%Moreover, existing wireless technologies  such as ultra-reliable and low-latency communication in unprecedentedly complex scenarios and low power requirements in ultra-dense networks. To tackle these challenges, 

%\noindent\textbf{ML-based Method.}
%JSCC is considered as a key enabler for NextG wireless systems to overcome such limitations~\cite{wang2020artificial, wang2022transformer}. %JSCC jointly designs the transmitter and receiver by optimizing various physical layer components in a cross-layer framework. 
ML-driven wireless systems aim to train a robust JSCC encoder and decoder on wireless channels infused with channel conditions similar to the physical world. The JSCC encoder directly maps the source to complex-valued symbols, and the JSCC decoder recovers its estimate directly from the noisy channel output. To adopt the widely used wireless standards, the JSCC models can be concatenated with OFDM to increase the spectral efficiency and reduce the multipath channel effects~\cite{yang2022ofdm}. Since multipath fading channels and OFDM blocks can be represented as differentiable layers, ML-based wireless systems are trained end-to-end. %Thus, ML-based communication systems have shown great potential in efficiently transmitting various modalities of data. 
\jungwoo{As such, JSCC can be built without modifying standard radio hardware (e.g., field test 6G with JSCC on 4G LTE~\cite{6g_field}). Furthermore, ML-based wireless communication can significantly save channel bandwidth costs compared to conventional systems while achieving the same end-to-end wireless transmission performance~\cite{you2024next}.}

\vspace{-0.2cm}
\subsection{Modality-Specific JSCC Models}
\label{sec:modality-specific}
%ML-based JSCC aims to train robust JSCC encoder and decoder on radio channel models infused with channel conditions similar to the real physical world.
%JSCC encoder is composed of a source encoder and a channel encoder to extract the semantic information and guarantee successful transmission over the wireless channel. Meanwhile, JSCC decoder includes a channel decoder and a source decoder for recovering from channel effect and returning symbols of the complex-values back to the source data. 
%xyz: citations? 
%jw: Thank you for letting me know. I added the references.
Existing JSCC systems~\cite{yang2022ofdm,wang2022wireless,weng2021semantic,xie2021deep} adopt modality-specific structures, with each modality requiring a specialized approach for accurate symbol recovery at the receiver. 
%Each modality needs a specialized way of extracting semantic information for accurate symbol recovery at the receiver.
%Consequently, the ML models should be tailored to the characteristics of each modality.
We consider four state-of-the-art JSCC models for image~\cite{yang2022ofdm}, video~\cite{wang2022wireless}, speech~\cite{weng2021semantic}, and text transmission~\cite{xie2021deep}.
Figure~\ref{fig:back} depicts the commonly used structures for each modality. 
%The modality of the source data plays an important role in designing JSCC systems. Typical examples of data modality include image, video, speech, and text.
%In order to learn the unique characteristics of each modality as efficiently as possible, JSCC adopts a modality-specific structure, and the architecture and parameters of the ML models are not shared between different modalities~\cite{yang2022ofdm, wang2022wireless, weng2021semantic, xie2021deep}. 
The image JSCC is trained to minimize distortion on a frame-by-frame basis. The video JSCC leverages spatiotemporal similarities between successive frames to remove the redundancy. To achieve this, the video JSCC adopts the temporal coding structure $\sigma$, which clusters each consecutive sequence of pictures into a group of pictures (GOP). Each frame within the GOP is entered into the video JSCC in coding order rather than display order. This means that the video JSCC encoder compresses frames in a specific order.
%xyz: What is coding order? 
%jw: Depending on the type of GOP, the display order and coding order may be different. Each frame within the GOP is entered into video JSCC in coding order, not display order.
For a total of $P$ frames included in the GOP, the coding order of each frame is determined by the mapping function $m_{\sigma}(t)$, where $1\leq m_{\sigma}(t)\leq P$.
%Specifically, for a given $\sigma$, the $n$-th coding order in each GOP is mapped to the $t$-th display order by a deterministic function $m_{\sigma}(n)$ = $t$. Here, $1 \leq n \leq P$ where $P$ is the number of frames in the GOP.
On the other hand, speech signals contain speaker characteristics such as speech rate and tone. The attention mechanism~\cite{weng2021semantic} is utilized for speech JSCC to identify the essential features to help accurately recover speech signals at the receiver. The text JSCC is designed to precisely encode context information and cope with semantic distortion based on Transformers~\cite{vaswani2017attention}. The text features recovered by the receiver are decoded into the text sentence through a greedy decoder~\cite{welleck2019neural}. A cross-entropy loss~\cite{li2021mixed} is used to understand semantic meaning while maximizing system capacity. 

\vspace{-0.2cm}
\subsection{Physical Layer Protocols}
\label{sec:related_work_protocol}

\noindent\textbf{Modulation.} \jungwoo{Wireless standards commonly adopt QPSK, 16-QAM, and 64-QAM to map bits to complex symbols~\cite{ieee2021wireless}. Therefore, the JSCC-encoded data are mapped to elements in a two-dimensional finite constellation diagram.} An adaptive modulation scheme can change the modulation type to balance reliability and spectral efficiency. For example, \cite{abdulla2014toward}  changes the modulation 
% applies to switch criteria 
based on the threshold of the channel state to meet the bit error rate (BER) requirement.

\noindent\textbf{OFDM.} To achieve high spectral efficiency, the OFDM transmitter may assign modulated symbols arbitrarily to the subcarriers rather than in a fixed order. Therefore, each subcarrier carries symbol vectors with a different distribution. %OFDM allocates each of the separate symbols onto different subcarriers in the frequency domain. 
% Then, OFDM signals are made up from a sum of time-domain signals, with each corresponding to a subcarrier.
%At the receiver side, the OFDM data are equalized using channel estimates.

\noindent\textbf{Coding Rate.} Adaptive encoding is essential to guarantee the reliability of wireless communications~\cite{xie2015pistream}. The JSCC encoder estimates the available bandwidth based on the channel state and employs adaptive algorithms to choose an optimal coding rate for efficient real-time streaming.

\vspace{-0.1cm}
\section{Related Work}
\label{sec:related_work}
\begin{table*}[t]
\caption{\jungwoo{A comparison of existing adversarial attacks against ML-based wireless communication and sensing.}} 
\centering
\scriptsize
\label{tab:summary}
%\makebox[\textwidth][c]{
\resizebox{0.94\textwidth}{!}{
\midsepremove
\begin{tabular}[t]{c|c|c|c|c|cc|ccc|cc|cccc}
\toprule
\multirow{2}{*}{\textbf{Attacks}}
%& \multicolumn{2}{c|}{\textbf{ML}}
& \multirow{2}{*}{\textbf{Type}}
& \multirow{2}{*}{\textbf{Channel}}
& \textbf{Non-WB}  %\multirow{2}{*}{\textbf{Black-box}}
& \textbf{HW} %\multirow{2}{*}{\textbf{HW demo}}
& \multicolumn{2}{c|}{\textbf{Input-Agnostic}}
& \multicolumn{3}{c|}{\textbf{Protocol-Agnostic}}
& \multicolumn{2}{c|}{\textbf{Sync-Free}}
& \multicolumn{4}{c}{\textbf{Defenses}}
 \\ \cline{6-16}
&     
& 
& \textbf{ML}
& \textbf{Demo}
& \textbf{Multimodal} %\multicolumn{1}{c}{}
& $\textbf{H}_{\textbf{t}}$
& \textbf{Constellation}
& \textbf{Coding Rate}
& \textbf{OFDM}

& \textbf{Time}  %\multicolumn{1}{c}{Time } 
& \textbf{Phase}  %\multicolumn{1}{c}{Frequency}  
& \textbf{RT}
& \textbf{PS} %\multicolumn{1}{c}{Image}
& \textbf{PD}
& \textbf{OD}
%& LC %\multicolumn{1}{c}{Video}
%& ST
\\ 
\midrule
\cite{sadeghi2019physical} & {\multirow{4}{*}{\begin{tabular}[c]{@{}c@{}} Offline \\ Attacks \end{tabular}}} & \multirow{2}{*}{AWGN} & \cmark &  &  & &  &  & & \cmark & & $\boxempty$ & $\boxempty$ & $\boxempty$ & $\boxempty$ \\ \cline{4-16} 
\cite{albaseer2020performance} &  & & \cmark & & & &  &  &  & & \cmark &  $\boxempty$& $\boxempty$ & $\boxempty$ & $\boxempty$\\ \cline{3-16} %\cline{1}
\cite{nan2023physical} & & {\multirow{3}{*}{\begin{tabular}[c]{@{}c@{}} Multipath\\ Fading \end{tabular}}} &  & &  &  & &  &  &  &  & $\boxempty$ & $\boxempty$ & $\boxempty$ & $\boxempty$\\ \cline{4-16} %\cline{1}
\cite{flowers2019evaluating} & & &  & &  & &  & & &  &  &  $\boxempty$ &  $\boxempty$ & $\boxempty$ & $\boxempty$ \\ \cline{4-16}
\cite{li2023boosting} & & &  & &  & &  & & &  &  &  $\boxempty$ &  $\boxempty$ & $\boxempty$ & $\boxempty$ \\ \hline
\cite{apruzzese2022wild} & {\multirow{6}{*}{\begin{tabular}[c]{@{}c@{}} Online\\ Attacks \end{tabular}}} & - & \cmark &  &  & &  &  & &  & & \crule{0.145cm}{0.145cm} & $\boxempty$ & $\boxempty$ & $\boxempty$\\ \cline{3-16} 
\cite{sadeghi2018adversarial} &  & {\multirow{2}{*}{\begin{tabular}[c]{@{}c@{}} AWGN \end{tabular}}} & \cmark
  & & & &  &  & & \cmark &  &  $\boxempty$ & $\boxempty$ & $\boxempty$ & $\boxempty$ \\ \cline{4-16}
\cite{bahramali2021robust} &  &  & \cmark
  & & & &  &  & &  & \cmark &  \crule{0.145cm}{0.145cm} & \crule{0.14cm}{0.14cm} & $\boxempty$ & \crule[lightgray]{0.145cm}{0.145cm} \\ \cline{3-16}
\cite{hu2023robust} &  & {\multirow{3}{*}{\begin{tabular}[c]{@{}c@{}} Multipath\\ Fading \end{tabular}}} & \cmark  & & & \cmark &  &  &  &  & & \crule[lightgray]{0.145cm}{0.145cm} & $\boxempty$ & $\boxempty$ & $\boxempty$\\ \cline{4-16}
\cite{kim2021channel} & & & \cmark & 
& & \cmark & &  & &  &  & \crule[lightgray]{0.145cm}{0.145cm} & $\boxempty$ & $\boxempty$ & $\boxempty$\\ \cline{4-16}
\cite{liu2023exploring} & & & \cmark & \cmark
& & \cmark & &  & & \cmark & \cmark & \crule[lightgray]{0.145cm}{0.145cm} & $\boxempty$ & $\boxempty$ & $\boxempty$\\ \cline{4-16}
Ours & & & \cmark & \cmark
& \cmark & \cmark & \cmark & \cmark & \cmark & \cmark & \cmark & \crule{0.145cm}{0.145cm} & \crule{0.145cm}{0.145cm} & \crule{0.145cm}{0.145cm} & \crule[lightgray]{0.145cm}{0.145cm} \\ \bottomrule
\end{tabular}}
     \begin{tablenotes}
     %\tiny
       \item [1] $\textbf{H}_{\textbf{t}}$: a channel matrix between the sender and the receiver; \cmark: the item is supported; \textbf{WB}: White Box. 
       \item [2]\textbf{RT}: Robust Training (Adversarial Training, Defensive Distillation, and Randomized Smoothing); \textbf{PS}: Perturbation Subtraction; \textbf{PD}: Perturbation Detection; \textbf{OD}: Oracle Defense.
       %\item [2] \cmark: the item is supported by the attack
       \item [3] \crule{0.145cm}{0.145cm}: the attack can compromise the defense; \crule[lightgray]{0.145cm}{0.145cm}: the defense was considered, but the attack was ineffective.; $\boxempty$: not mentioned in the paper.
     \end{tablenotes}
\vspace{-0.4cm}
\end{table*}

\begin{comment}
\noindent\textbf{Adversarial Attacks on Other Domains.} Adversarial attacks have been studied to validate and analyze the robustness of the ML model across multiple domains of computer vision, such as image classification~\cite{carlini2017towards, chen2020hopskipjumpattack}, speech recognition~\cite{carlini2018audio, o2022voiceblock}, human activity recognition~\cite{cao2023stylefool, xie2022universal}, etc. 
Most of these studies give an attacker the ability to perform a man-in-the-middle attack where she can intercept data in the middle and then injects a small amount of adversarial perturbation. Therefore, the perturbation is not physically feasible and only exposes theoretical vulnerabilities. As the demand for physically feasible adversarial research grows, recent studies~\cite{jin2023pla, abdullah2021hear, sato2021dirty} more pragmatically define attack methodologies so that adversarial attacks can be implemented in the real world. However, the adversary can more easily generate input and protocol invariant perturbations compared to wireless communication.
\end{comment}

\subsection{Conventional Wireless Attacks}  
%We analyze two representative attacks in wireless systems:
%in wireless communication: (1) jamming attack and (2) overshadowing attack. 

\noindent\textbf{Jamming Attacks.} RF jamming transmits radio signals indiscriminately across a range of frequencies, causing interference and disrupting communication. Jamming can be broadly categorized as active jamming and reactive jamming~\cite{xu2005feasibility,pirayesh2022jamming}. Active jamming continuously emits powerful interference signals, but its continuous operation leaves detectable traces, making it vulnerable to defensive techniques~\cite{xu2005feasibility}. Reactive jamming~\cite{liu2022physical} adjusts its jamming behavior according to observed signals in the environment. It remains silent when the channel is idle but initiates high-power signal transmission upon detecting activity on the channel. 
%\jungwoo{Shi~\textit{et al.}~\cite{shi2018spectrum} predicts when the victim occupies the channel and then leverages this to efficiently attack the target channel.}
%
%Thus, the jammer may induce the channel to register as busy, resulting in delays for legitimate transmissions. Even if entire packets are transmitted, the receiver might struggle to decode them accurately due to interference. 
%For the jammer to be effective, it must transmit high-power radio signals. A jammer operating at a lower power level is less effective in interfering with network operations. Adversarial perturbation aims to manipulate the behavior of machine learning models, whereas RF jamming directly disrupts wireless communication channels.
The drawback of these approaches is that spectrum owners may promptly detect the presence of an attack and respond accordingly.

\noindent\textbf{Overshadowing Attacks.} Cellular networks are vulnerable to overshadowing attacks~\cite{wen20245g}. Recent works~\cite{yang2019hiding, erni2022adaptover} can force the victims to receive the attacker's symbols/subframes by sending high-powered signals to a base station. However, the adversary must have the capability to receive and decode the messages transmitted by the victim. These attacks typically require signal strengths ranging from -3.4dB to +3dB over the benign signal~\cite{erni2022adaptover}. 
%Thus, the attacker must deploy amplification and high-gain antennas. 
Since the adversary's signal strength is comparable to or even stronger than the legitimate signals, it becomes easier for the legitimate nodes to identify the attack.

%We aim to raise awareness of a new vulnerability called adversarial attacks~\cite{cao2019adversarial} for the safety of ML adoption in NextG. We demonstrate that \sys can successfully disrupt the ML-based JSCC via much weaker adversarial signals [-20,-10]dB than legitimate signals. Our adversary does not coordinate its transmissions with a benign sender or access to the signal sent by a benign receiver.

%The susceptibility of the models to adversarial perturbations raises serious concerns   
%We are interested in small changes of the signal, but ML-based models have shown to be susceptible to a small magnitude of adversarial perturbations~\cite{cao2019adversarial}
%Therefore, an attacker must deploy amplification and high-gain antennas to achieve a successful attack.
%against all users within a cell
%or specifically target a victim based on its phone number.

%These conventional overshadowing and jamming attacks cause 

%However, we aim to inject smaller signals[-20,-10]dB compared to legitimate signals by exploiting the vulnerabilities of ML.

%\noindent\textbf{Adversarial Attacks on Other Domains.}
\vspace{-0.2cm}
\subsection{Adversarial Attacks on Computer Vision Domains} Adversarial ML has been studied to analyze the robustness of the model across multiple areas, such as image classification~\cite{ carlini2017towards, quiring2020adversarial}, speech recognition~\cite{alzantot2018did}, activity recognition~\cite{cao2023stylefool, chang2023netflick}, video compression~\cite{chang2022rovisq, chang2023videoflip}, etc. 
%The scope of adversarial attack research is expanding as artificial intelligence (AI) services, such as autonomous driving~\cite{qian2021robust, li2022modality} and health care diagnosis~\cite{xu2021intelligent}, are increasingly used in real life.
Most studies provide an attacker with the capabilities to perform a man-in-the-middle attack where he/she intercepts data in the middle and then injects small perturbations. These are not physically feasible and only expose theoretical vulnerabilities. As the demand for physically feasible attacks grows, recent studies~\cite{jin2023pla, lovisotto2021slap, sato2021dirty} define practical approaches so that attacks can be realized in the real world. SLAP~\cite{lovisotto2021slap} applies a projector to superimpose light onto an object, causing the model to misclassify the object. %Some works create real-world adversarial examples, leveraging the knowledge that the camera's image sensor first captures the image before the ML models in the vision domain are run.
Compared to wireless domains, physical attacks in vision domains are less susceptible to signal distortion and have relatively fewer domain constraints. %\nas{what do you mean by fewer protocols to change?}.

%\noindent\textbf{Adversarial Attacks on Wireless Systems.}
\vspace{-0.2cm}
\subsection{Adversarial Attacks on Wireless Domains}
\jungwoo{
%With the increasing ML integration into wireless systems, it becomes important to investigate vulnerabilities in ML-based wireless systems.
There are two types of target wireless systems: (1) wireless networking, which concentrates on efficient dataflow management between networked devices,
and (2) wireless communication and sensing for restoring and analyzing radio signals at the physical layer. 
In this paper, we focus on the second point.}
%~\cite{bahramali2021robust, sadeghi2019physical, nan2023physical, li2023boosting, hu2023robust, liu2023exploring, sadeghi2018adversarial, flowers2019evaluating, hameed2020best}.

\jungwoo{Attackers targeting wireless networking seek to deceive the ML-based network devices into making wrong decisions (e.g., for resource allocation). Certain attacks~\cite{manoj2021adversarial, kim2021adversarial} operated in a white-box setting with complete knowledge of the target ML model. In contrast, Apruzzese~\textit{et al.}~\cite{apruzzese2022wild} devised a realistic threat model by assuming a constrained attacker and demonstrated their performance across various ML systems.}

\jungwoo{When attacking wireless communication and sensing, it is crucial to design physically realizable perturbations. 
Table~\ref{tab:summary} summarizes existing attacks in two categories: offline attacks and online attacks. %The literature falls into two categories: offline and online attacks.
%These works can be classified into two branches: offline and online attacks.
%
Offline attacks are impractical as they allow attackers unlimited access to inputs and models. Online attacks address this by adding UAPs to victim signals.
%
%Appendix Table~\ref{tab:summary} shows a comparison of several online adversarial attacks against wireless communication and sensing.
%
Several works~\cite{sadeghi2018adversarial, kim2021channel} studied methods for crafting UAPs against radio signal classifiers but aimed to identify theoretical vulnerabilities rather than design physically feasible attacks.
Flowers \textit{et al.}~\cite{flowers2019evaluating} identified victim's transmissions by sniffing the signal strength of the target channel, but sniffing does not provide an accurate time offset due to latency and cannot reveal the modality and wireless protocol.}
%We assume realistic and challenging multipath fading channels. We also consider a black-box adversary who does not have access to victim ML models.
%
%
%Shi~\textit{et al.}~\cite{shi2018spectrum} conducted a spectrum data poisoning attack to manipulate the channel occupancy status. 
%
%The most recent studies~\cite{bahramali2021robust, liu2023exploring} have tried to make the attack viable by limiting the adversary's abilities. 
%
\jungwoo{Bahramali~\textit{et al.}~\cite{bahramali2021robust} adopted a generative model to produce diverse UAPs, but they made the unrealistic assumption that the target system sends only one-hot vector messages~\cite{o2017introduction}. Their attacks are evaluated individually on each physical layer component rather than on an end-to-end system.
%
%do not take into account the physical layer components, which are traditionally used in end-to-end wireless communication, thus their attacks are not practical. 
%Hu~\textit{et al.}~\cite{hu2023robust} studied the adversarial vulnerability of image JSCC, but did not take into account the practicality constraints (i.e., multi-modality, protocols, synchronization, etc.) required to attack the image JSCC model in the real world.
%
%More importantly, most studies do not demonstrate their attacks in hardware-implemented platforms.
%
RAFA~\cite{liu2023exploring} designed a practically feasible UAP in a limited-knowledge setting. They solely target the publicly-known preambles, so their attacks are not applicable to JSCC which transmits unknown data symbols. In addition, the JSCC-encoded data are modulated by various protocols (e.g., modulation, coding rate, and OFDM). Furthermore, due to the lack of diversity in its perturbations, RAFA can be directly mitigated by the adaptive defense with high accuracy (see \cref{sec:csi}).}
\jungwoo{Additionally, a recent study~\cite{li2024practical} attacked wireless sensing systems by assuming that an adversary could install malicious firmware on the victim transmitter and change pilot packets. However, we are interested in a more realistic scenario where an adversarial signal is injected into the target channel.}

%To solve new challenges in JSCC, we propose a new adversarial attack which crafts our UAP input- and protocol-agnostic to JSCC, i.e., oblivious to the input modality, modulation schemes, channel conditions, coding rate, and OFDM specifications.

%their perturbations are not suitable for attacking our target system, end-to-end wireless transmission. Furthermore, their attack method does not take into account multi-modality of the source data and different settings of wireless protocols (constellation, coding rate, and OFDM specifications).

%However, since their target is pilot symbols that are much less constrained than data symbols, their perturbation signals are not effective at perturbing data symbols of OFDM signals. %Our adversarial perturbation prevents the receiver from equalizing the perturbed channel output and causes the ML-based JSCC to fail to recover the multimodal source data as well as subvert the downstream service.
%In this paper, we present, \sys, a black-box attack framework  to generate universal multi-modal perturbations that target both OFDM pilots and data symbols. We show, for the first time, that modulated multimodal data can be exploited by adversaries, resulting in subverting downstream services without restoring the original data.

\vspace{-0.1cm}
\section{Threat Model}
\label{sec:threatmodel}
\subsection{Attack Scenario}
%To explore the adversarial robustness of NextG wireless communications, we focus on ML-based wireless systems. 
%The ML-based wireless systems adopt the widely used wireless standards, such as constellation mapping~\cite{tung2022deepjscc} and OFDM~\cite{felix2018ofdm, yang2022ofdm, wu2022channel}. 
\sys is targeted towards radio signals created by front-end sources that are used to transmit the multimodal source to back-end user(s). The attacker deploys commercial off-the-shelf (COTS) hardware (e.g., software-defined radios) to send the attack signals. We focus on vulnerabilities unique to ML in wireless environments, leading to the failure of the receiver's JSCC decoder to correctly decode the received packet. Note that we exclude the jamming effect~\cite{pirayesh2022jamming}, a brute-force solution that disrupts all communication within the medium.

\jungwoo{Multiple transmitters and receivers can share the spectrum. As described in Appendix~\ref{sec:appendix_real_world}, the standard Wi-Fi protocol ensures only one device uses the wireless channel at a time within a cell to avoid collision. \sys can thus inject adversarial signals to target different transmitter-receiver pairs sequentially.
\sys can also be positioned in a selective attack~\cite{aras2017selective} that only targets a specific wireless device, leaving other devices unaffected. Specifically, \sys can identify the victim by sniffing the MAC address in the packet, and launch the attack whenever the victim device transmits the packet. Please note that all the above cases are equivalent to applying Magmaw's adversarial perturbation to a single transmitter-receiver pair (as shown in Appendix Figure~\ref{fig:real_world}).}

\vspace{-0.2cm}
\subsection{Adversary's Goal}
\sys aims to transmit well-crafted perturbations over the target wireless channel to prevent legitimate receivers from recovering the source data and performing target downstream tasks. 
To ensure stealthiness, \sys sends adversarial signals with a small magnitude. As a result, the victim cannot differentiate between adversarial perturbations and natural noise from wireless channels. %Figure~\ref{fig:threat_psd} depicts the difference in power spectrum density of each OFDM signal to analyze whether the attack is unnoticeable. Note that the multipath fading channel induces the distortion in the transmitted signal.
Following the previous studies~\cite{bahramali2021robust, sadeghi2019physical}, we utilize a perturbation-to-signal ratio (PSR) metric to compare the power of the perturbation at the receiver with the received legitimate signal power. The PSR is set to be [-20,-10] dB~\cite{bahramali2021robust, sadeghi2019physical} so that the perturbation is not distinguishable from the expected natural noise in the channel.
%As shown in Figure~\ref{fig:threat_psd}, we show that the receiver's perturbed signal has little change compared to the benign signal transmitted from the sender. Thus, the receiver will perform a restoration process on the signal that appears seemingly undamaged but actually perturbed.

\vspace{-0.2cm}
\subsection{Adversary's Capability and Knowledge}
\label{sec:capability}
\jungwoo{We envision a constrained attacker~\cite{apruzzese2023real} with limited knowledge of ML-based wireless systems as described below.}

\noindent\textbf{Wireless System.} \jungwoo{We assume that the adversary has no prior knowledge about the ML model architecture/parameters, but knows the category of target models (e.g., autoencoder which is the de facto model for JSCC) 
%how to process each modality, 
and the physical layer techniques being used (e.g., OFDM modulation which is specified in the communication standard).} This is a realistic assumption for the following reasons: 1) standard documentation usually describes the core technology and is open to the public, and 2) \jungwoo{specialized operations (see \cref{sec:modality-specific})} for each modality have already been widely known in the ML community. The adversary trains surrogate ML-based JSCC models using a large amount of publicly available data. \jungwoo{Note that the attacker cannot access the target JSCC model or observe the output.}

\noindent\textbf{Knowledge about Input and Protocols.} 
%The adversary has no knowledge of the incoming wireless signals. Moreover, 
We assume that the adversary does not know the modality and the constellation mapping method due to the following reasons: 1) all the application-layer source data, regardless of modalities, need to multiplex the transmitter radio and wireless channel, 2) the transmitter can adapt several types of modulation techniques according to channel conditions. %To understand the importance, we study the transferability of adversarial perturbations between modalities and modulation schemes. As shown in Figure~\ref{fig:comp}, the lack of learning generalized adversarial features limits both the cross-modal and cross-domain transferability.
Additionally, the JSCC model can dynamically adjust the coding rate in real time based on the current channel conditions, so the adversary has no prior knowledge about the number of OFDM symbols encoded by the JSCC model in the transmitted signal.
%Furthermore, the adversary has no prior knowledge of the number of OFDM symbols encoded by the JSCC model because the JSCC model varies the coding rate according to the state of the channel. %For example, at a high SNR, a model with a low coding rate is used because source data can be transmitted with high throughput. 
However, we assume that the adversary can refer to the possible coding rates specified in the standards documents. %This allows the attacker to know the lowest coding rate.
Lastly, we do not assume that the adversary knows how the transmitter maps the OFDM symbol to the subcarriers. %At the OFDM transmitter stage, the benign transmitter may assign modulated symbols arbitrarily to the subcarriers rather than in a fixed order.

\noindent\textbf{Target Wireless Channel.} 
We consider a real-world attack scenario where the attacker cannot have access to the channel matrix between the transmitter and the receiver, i.e., $\textbf{H}_{\textbf{t}}$. In addition, we do not assume that the adversary is synchronized with either the transmitter or the receiver, leading
to random time and frequency offsets. %herefore, the adversary does not know when the transmitter sends the wireless signal, which leads to a random time offset. When the perturbation signal arrives at the receiver, there is a phase difference with the victim wireless signal. 
Furthermore, we assume that the attacker can determine the carrier frequency used by the targeted channel. The attacker can overhear the victim's signals by arbitrarily adjusting its waveform bandwidth and carrier frequency using a software-defined radio~\cite{liu2023time, zhang2015wireless}.
%This allows an attacker to distinguish the target channel from other channels~\cite{liu2023time}.

\noindent\textbf{Attacker's Wireless Channel.} The attacker employs a single antenna to send the adversarial signal. %The perturbation signal undergoes a multipath fading channel. For simplicity, we assume the adversary employs a single antenna. 
%which prevents the adversary from making attacks that is robust to the channel effects. 
%xyz: Clarify: Is this the channel matrix between the adversary and the victim receiver? 
%jw: Yes, it is
We denote the channel matrix for the attacker as $\textbf{H}_{\textbf{a}}$. 
%xyz: Are you assuming the vicim link is using 802.11 standard? Clarify. In fact, 802.11 is not using ML-based comm. It's better if the attacker method is general across different standards.  
%jw: I change the "802.11 standard" to "Wi-Fi protocol".
According to the Wi-Fi protocol, the receiver periodically sends beacons to wireless devices within the range~\cite{banerji2013ieee}. An attacker can overhear this transmission and estimate the channel matrix from the receiver to itself. Due to the principle of reciprocity, this channel is the same as $\textbf{H}_{\textbf{a}}$. In contrast to recent work~\cite{liu2023exploring}, we relax the assumption that the adversary knows the exact channel matrix between the attacker and the receiver. We make a weaker assumption that the adversary has limited information, 
%xyz: Do you assume it knows the "distribution" or the exact channel matrix? 
%jw: I would collect the channel matrix without changing the adversary's location. But the training-time channel information may be different with the test-time channel information.
i.e., the distribution of the channel between the attacker and the receiver. 
%To realize the UAP, we generate $N_{a}$ random samples $\{\textbf{H}^{1}_{\textbf{a}}, \dots, \textbf{H}^{N_{a}}_{\textbf{a}} \}$ from the distribution. 
%xyz: It sounds like you are using self-generated data for testing. Aren't you using the real data from software radios?
%jw: Thank you for finding the error. I change the term test to validation. I would measure the attack under the sdr. 

%Then, we divide the samples into training and validation samples. We train the UAP with channel information from the training set and determine the best model from the validation set. After training is complete, we evaluate the UAP in real-world scenarios.

%The adversary cannot simultaneously transmit the perturbation signal when the sender transmits a benign signal, and cannot manipulate the adversarial signal to have the same phase offset as when the receiver receives the benign signal. Furthermore, the attacker does not know the length of the information symbols contained in the sacrificial OFDM packet.

\begin{figure*}[t]
\centering
\includegraphics[width=0.80\textwidth]{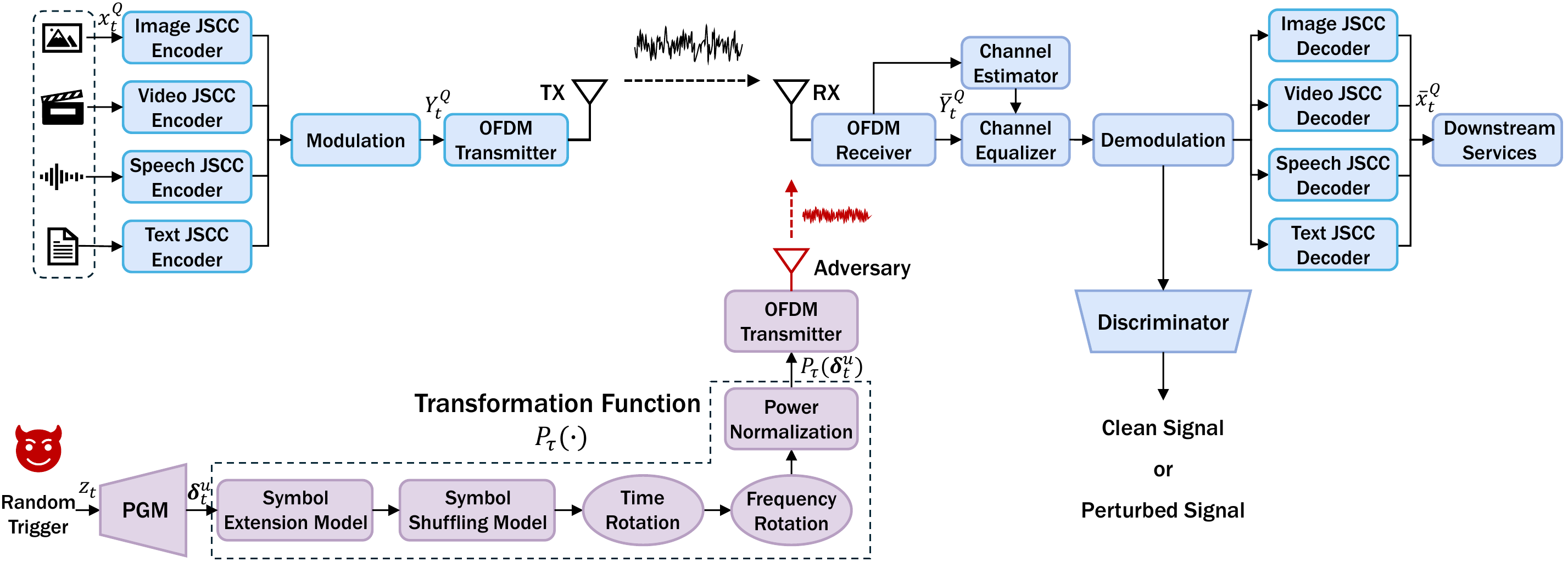}
\vspace{-0.1cm}
\caption{Overview of \sys. During the PGM training, the attacker employs the surrogate JSCC model (blue modules). 
%Note that the discriminator is solely utilized in the training process.
}
\label{fig:main}
\vspace{-0.4cm}
\end{figure*}

\vspace{-0.1cm}
\section{System Model}
\label{sec:sys_model}
Figure~\ref{fig:main} illustrates the core processing blocks in the victim communication link along with the \sys attacker. 

\noindent\textbf{ML-based Transmitter.}
We consider OFDM-based JSCC over a multipath fading channel with $L_{t}$ paths. The multimodal data are transmitted using $N_{s}$ OFDM symbols with $L_{fft}$ OFDM subcarriers. Note that $N_{s}$ has different values depending on the modality and the coding rate. For channel estimation, the sender transmits a preamble (according to the wireless communication standards) on the subcarriers. We denote the source data as $x^{Q}_{t}$ with modality $Q \in \{\mathcal{I}, \mathcal{V}, \mathcal{S}, \mathcal{T}\}$ at time step $t$, where $\mathcal{I}, \mathcal{V}, \mathcal{S}, \mathcal{T}$ denote the image, video, speech, and text, respectively. We describe a JSCC encoder for processing a modality $Q$ with a given coding rate $\lambda$ and modulation scheme $C$ as a function $E_{Q,C,\lambda}(x^{Q}_{t},\mathcal{B}^{Q}_{t})$, where $\mathcal{B}^{Q}_{t}$ is the transmitter's 
%xyz: What's "reference buffer"? 
%jw: video codec requires two inputs (current frame and previously decoded frame) to improve the coding efficiency.
reference frame buffer used for the video JSCC, as depicted in Figure~\ref{fig:back} (b). We define $\mathcal{B}^{Q}_{t}$ containing the previously decoded frame $\tilde{x}^{\mathcal{V}}(\cdot)$ as:
\begin{equation} \label{eq:buffer}
\mathcal{B}^{Q}_{t} = 
\begin{cases}
\{\tilde{x}^{\mathcal{V}}_{m_{\sigma}(1)}, \cdots, \tilde{x}^{\mathcal{V}}_{m_{\sigma}(t-1)}\}, & \text{if } Q = \mathcal{V}, \\
\emptyset, & \text{if } Q \neq \mathcal{V}.
\end{cases}
\end{equation}
%\nas{doesn't the buffer have a limited capacity? We can ignore it.}
Recall that $m_{\sigma}(t)$ is a function that finds the coding order of the $t$-th image in the given GOP structure $\sigma$.
%Recall that $m_{\sigma}(\cdot)$ is a function that maps the $n$-th coding order image from the given GOP structure $\sigma$ to the $t$-th image in temporal order.
$\mathcal{B}^{\mathcal{V}}_{t}$=$\emptyset$ when $t$=1. This is because the first frame is coded by the image JSCC. Note that $\tilde{x}^{\mathcal{V}}(\cdot)$ is reconstructed as the output of a video JSCC decoder that takes encoded video sequence $E_{\mathcal{V},C,\lambda}(x^{\mathcal{V}}_{t},\mathcal{B}^{\mathcal{V}}_{t})$ as input.
%is reconstructed through the transmitter's video JSCC decoder.
$\lambda$ is the coding rate to control the number of symbols.

Then, a constellation mapper $M_{C}(\cdot)$ moves symbols to the nearest points in a finite constellation diagram $C$. The modulated symbol, $Y^{Q}_{t} \in \mathbb{C}^{N_{s} \times N_{fft}}$, can then be obtained as: 
\begin{equation} \label{eq:trans}
Y^{Q}_{t} = 
M_{C}(E_{Q, C,\lambda}(x^{Q}_{t},\mathcal{B}^{Q}_{t})).
%\mathcal{T}_{C}(E_{Q}(s^{Q}_{t},\mathcal{B}^{Q}_{t},\lambda)).
\end{equation}
%where $Y^{\mathcal{V}}_{t}$ is fed into the video JSCC decoder on the transmitter side and restored to $\tilde{x}^{\mathcal{V}}_{t}$ which will be stored in $\mathcal{B}^{Q}_{t}$ for encoding of the next frame.

Without loss of generality, we assume the target transmitter/receiver uses a single antenna following the 802.11a/g/n Wi-Fi standard~\cite{ieee2021wireless}. We split $Y^{Q}_{t}$ into a number of signal vectors with dimension of $N_{fft}$. Afterwards, an OFDM transmitter allocates divided signals on each subcarrier. Each OFDM symbol passes through an inverse discrete Fourier transform (IFFT), then a cyclic prefix (CP) is added and transmitted to the receiver over a multipath fading channel. 

\noindent\textbf{ML-based Receiver.} The receiver obtains the complex-valued symbols from the channel output by removing the CP and applying FFT with an OFDM receiver. The received signal of the $k$-th subcarrier in the $i$-th OFDM symbol is given by:
\begin{equation} \label{eq:channel}
\hat{Y}^{Q}_{t}[i,k] = 
\textbf{H}_{\textbf{t}}[k]Y^{Q}_{t}[i,k] + W[i,k],
\end{equation}
where $\textbf{H}_{\textbf{t}}\in \mathbb{C}^{N_{fft} \times N_{fft}}$ is the frequency-domain channel matrix, which is a diagonal matrix, and $W \in \mathbb{C}^{N_{s} \times N_{fft}}$ is the frequency-domain AWGN matrix. %During JSCC training, $\textbf{H}_{\textbf{t}}$ is randomly sampled from the multipath fading distribution.

Given the FFT output of the pilot signals, the channel estimation and equalization are performed to compensate the channel-induced transformation. We adopt a least squares (LS) algorithm to predict the channel state information. After equalizing all of the divided signals with the channel equalizer $R(\cdot)$, we quantize the phase and amplitude of the signal on each subcarrier with $M_{C}(\cdot)$. Finally, we employ the decoder $D_{Q, C,\lambda}(\cdot)$ to reconstruct an estimate $\hat{x}^{Q}_{t}$ of the original signal. We express the entire process after OFDM receiver as follows:
\begin{equation} \label{eq:receiv}
\begin{aligned}
\hat{x}^{Q}_{t} = &
D_{Q, C,\lambda}(M_{C}(\mathcal{R}(\hat{Y}^{Q}_{t})),\hat{\mathcal{B}}^{Q}_{t}) \\
%D_{Q}(M_{C}(\hat{x}^{Q}_{t}),\hat{\mathcal{B}}^{Q}_{t},\lambda)
= & \mathcal{F}_{Q,C, \lambda}(\hat{Y}^{Q}_{t}, \hat{\mathcal{B}}^{Q}_{t}),
\end{aligned}
\end{equation}
where $\hat{\mathcal{B}}^{Q}_{t}$ is the receiver's decoded frame buffer for the video JSCC. $\hat{\mathcal{B}}^{\mathcal{V}}_{t}$=$\{\hat{x}^{\mathcal{V}}_{m_{\sigma}(1)}, \cdots, \hat{x}^{\mathcal{V}}_{m_{\sigma}(t-1)} \}$, where $\hat{\mathcal{B}}^{\mathcal{V}}_{t}$=$\emptyset$ when $t$=1. $\hat{\mathcal{B}}^{Q}_{t}$=$\emptyset$ for other modalities. For simplicity, we denote all processes after the OFDM receiver as $\mathcal{F}_{Q,C, \lambda}(\cdot)$.

%performed by the receiver 

\vspace{-0.1cm}
\section{Attack Construction}
\label{sec:attack_construction}
The framework of \sys is illustrated in Figure~\ref{fig:main}. 
Our attack methodology follows a hardware/algorithm co-design to ensure \sys is robust against various signal distortions.

\vspace{-0.2cm}
\subsection{Our Attack Formulation}
\noindent\textbf{General Attack Formulation.}
%We aim to find adversarial perturbations with $Y^{Q}_{t}$ generated from a set of surrogate models and transfer the learned perturbations to the target ML system, resulting in high reconstruction error. Instead of computing signal-wise perturbations, the adversary performs an offline process to train a single UAP that can target all benign signals. Moreover, the UAP must be agnostic to the input signals generated by all modalities and constellation mapping techniques. 
Our adversary aims to find an input-agnostic perturbation $\boldsymbol{\delta}^{s} \in \mathbb{C}^{N_{s} \times N_{fft}}$, with a magnitude bounded by the attacker's power budget $\epsilon \in \mathbb{R}$.  When $\boldsymbol{\delta}^{s}$ is injected into the victim wireless channel, the receiver obtains the frequency-domain channel output $\bar{Y}^{Q}_{t}$ as:
\begin{equation} \label{eq:general1}
\begin{gathered}
%\bar{y}^{Q}_{t} = \mathcal{F}_{Q,C, \lambda}(\bar{x}^{Q}_{t}, \bar{\mathcal{B}}^{Q}_{t}),  \\
\bar{Y}^{Q}_{t}[i, k] = \textbf{H}_{\textbf{t}}[k]Y^{Q}_{t}[i,k] + \textbf{H}_{\textbf{a}}[k]\boldsymbol{\delta}^{s}[i, k] + W[i,k],
\end{gathered}
\end{equation}
where 
%$\textbf{H}_{\textbf{t}}$ is randomly generated based on multipath fading distribution to make $\boldsymbol{\delta}^{s}$ channel-agnostic. $Y^{Q}_{t}$ is obtained from the surrogate JSCC model. The magnitude of perturbation $\boldsymbol{\delta}^{s}$ is bounded by the attacker's power budget $\epsilon \in \mathbb{R}$. 
$\bar{Y}^{Q}_{t}[i, k]$ and $\boldsymbol{\delta}^{s}[i,k]$ represent the frequency-domain perturbed response and the adversarial perturbation at the $k$-th subcarrier of the $i$-th OFDM symbol, respectively. $\textbf{H}_{\textbf{a}}$ is the channel matrix between the attacker and the receiver. The attacker can obtain $\textbf{H}_{\textbf{a}}$ by leveraging channel reciprocity. However, the attacker does not have access to the target wireless system and therefore does not know modality $Q$, modulation scheme $C$, $Y^{Q}_{t}$, $\textbf{H}_{\textbf{t}}$, and JSCC models. \jungwoo{One method to address such a lack of knowledge is to utilize a set of surrogate models with different configurations (i.e., $Q, C, \lambda)$ and diverse channel matrix $\textbf{H}_{\textbf{t}}$.} Specifically, we seek to generate $Y^{Q}_{t}$ from a set of surrogate JSCC models, train UAPs using ensemble learning, and transfer the learned UAPs to the target system.
%find UAPs with $Y^{Q}_{t}$ generated from a set of surrogate models and transfer the learned UAPs to the target ML-driven wireless system.
During this offline UAP training, we randomly sample $\textbf{H}_{\textbf{t}}$ from multipath fading model to make $\boldsymbol{\delta}^{s}$ channel-agnostic. %Furthermore, the UAP must be robust to the input signals generated by all modalities and modulation schemes. \nas{it looks good now thanks. Just this last sentence seems disconnected from the rest. We don't say how it should be robust to the modality and modulation.}

Using Equation~\eqref{eq:receiv}, the receiver in the surrogate model then feeds this perturbed signal $\bar{Y}^{Q}_{t}$ to the remaining physical layer elements to reconstruct the source with modality $Q$ as:
\begin{equation} \label{eq:general2}
\begin{gathered}
\bar{x}^{Q}_{t} = \mathcal{F}_{Q,C, \lambda}(\bar{Y}^{Q}_{t}, \bar{\mathcal{B}}^{Q}_{t}),  \\
%\mathrm{where} \ \bar{x}^{Q}_{t}[i, k] = \hat{x}^{Q}_{t}[i, k] + \textbf{H}^{l}_{\textbf{a}}[k]\boldsymbol{\delta}[i, k]
\end{gathered}
\end{equation}
where $\bar{\mathcal{B}}^{Q}_{t}$ is the perturbed decoded frame buffer to be used in the video JSCC model. $\bar{\mathcal{B}}^{\mathcal{V}}_{t}$=$\{\bar{x}^{\mathcal{V}}_{m_{\sigma}(1)}, \cdots, \bar{x}^{\mathcal{V}}_{m_{\sigma}(t-1)}\}$, where $\bar{\mathcal{B}}^{\mathcal{V}}_{t}$=$\emptyset$ when $t$=1. $\bar{\mathcal{B}}^{Q}_{t}$=$\emptyset$ for other modalities. 

As mentioned before, we aim to find the adversarial signals in a \jungwoo{limited-knowledge setting (\cref{sec:capability})}. A representative way to handle this is to exploit the fact that adversarial examples exhibit good transferability between different ML models~\cite{moosavi2017universal, liu2023exploring, bahramali2021robust}. By adopting the attack transferability, we first train a surrogate JSCC model for each modality using publicly available datasets that have different distributions from the target model's training data. Then we use an ensemble learning approach to find a modality-agnostic adversarial perturbation $\boldsymbol{\delta}^{s}$ by solving the following optimization problem:
\begin{equation} \label{eq:general3}
\begin{gathered}
\operatorname*{arg\,max}_{\boldsymbol{\delta}^{s}} %\underset{{z \sim p_{z}, (Q,C,\lambda)\in p_{c}}}{\mathbb{E}} 
[\sum_{w \in \Psi^{s}} \mathcal{L}(w)], 
\ \mathrm{s.t.} \ \norm{\boldsymbol{\delta}^{s}}_{2} < \epsilon,\\
%\text{where} \ \bar{x}^{Q}_{t}[i, k] = \hat{x}^{Q}_{t}[i, k] + \textbf{H}_{\textbf{a}}(G(z)[i, k]
\end{gathered}
\end{equation}
where $\Psi^{s}$ is a set of all wireless signals that can be created by physical layer elements. $\mathcal{L}(w)$ is the loss function of ML-based JSCC model when $w$ is sampled from $\Psi^{s}$.

However, this attack formulation is not suitable for making the UAPs physically realizable for the following reasons. First, having a single $\boldsymbol{\delta}^{s}$ as the UAP allows the receiver to estimate the perturbation signal using OFDM pilot signals, resulting in low robustness and persistence of adversarial attacks. Second, the adversary has no prior knowledge of the number of OFDM symbols in the target signal and thus is unable to define $\boldsymbol{\delta}^{s}$ as a matrix of the same size as the transmitted signal. %We find that the effectiveness of an adversarial attack can be minimized if there is a mismatch in the number of symbols between the target and perturbation signals.
Third, the video JSCC model has a network structure that forms a temporal chain between all video frames within the same GOP, so the model encodes current source data based on previous encoding results. This constructs the inter-frame dependency within a video sequence and it should be considered in crafting the UAPs. Fourth, the adversary does not know the distribution of the channel inputs carried by each OFDM subcarrier. Finally, when the perturbation signal overlaps with the benign signal, time or phase offsets may occur.

%\subsection{Practical Attack Formulation}
\noindent\textbf{Practical Attack Formulation.}
To address the problem of Equation~\eqref{eq:general3}, we construct a Perturbation Generator Model (PGM) $G(z_{t})=\boldsymbol{\delta}^{u}_{t}$ that generates a UAP signal by receiving a 
%xyz: What does the "trigger" mean? Where does it come from? 
%jw: Similar to generative adversarial networks, perturbation generator model (PGM) would be defined as G(z_{t}) that receives the random trigger z_{t}. Random trigger can be also called random noise.
random trigger $z_{t}$ at time step $t$. We adopt a ResNet-based generator~\cite{choi2020stargan}. The adversary changes $z_{t}$ and injects a new perturbation signal into the target channel each time. Compared with using a single $\boldsymbol{\delta}^{s}$ as the UAP, the adversary creates an extremely large set of perturbations, which makes it difficult for the receiver to predict the perturbations. The following equation holds for frequency-domain complex-valued symbols at the receiver in the attacker's surrogate models:
\begin{equation} \label{eq:general4}
\small
\begin{gathered}
%\bar{y}^{Q}_{t} = \mathcal{F}_{Q,C, \lambda}(\bar{x}^{Q}_{t}, \bar{\mathcal{B}}^{Q}_{t}),  \\
\bar{Y}^{Q}_{t}[i, k] = \textbf{H}_{\textbf{t}}[k]Y^{Q}_{t}[i,k] + \textbf{H}_{\textbf{a}}[k]P_{\tau}(\boldsymbol{\delta}^{u}_{t})[i, k] + W[i,k],\\
\end{gathered}
\end{equation}
where $\boldsymbol{\delta}^{u}_{t} \in \mathbb{C}^{N_{g} \times N_{fft}}$ denotes a UAP which contains $N_{g}$ data symbols. Since the attacker does not know the number of target symbols, $N_{g}$ may not be equal to $N_{s}$. We define a novel transformation function $P_{\tau}$ which enables the PGM-generated wireless signals to model the distribution of real wireless data.
%xyz: Briefly explain each of this
%jw: I added the below explanation.
The transformation function consists of several steps: 1) symbol extension model, 2) symbol shuffling model, 3) time rotation, and 4) frequency rotation.
The symbol extension model concatenates multiple PGM-generated perturbations such that the symbol-extended perturbations can perturb all OFDM symbols of the target radio signal. The symbol shuffling model makes our attack robust against unknown target symbols by randomly shuffling symbols between the OFDM subcarriers of the adversarial signal. The time and phase rotation changes the offset of the adversarial signal during offline training so that the adversarial signals are agnostic to random time and phase shifts in the real world.
%We define a novel transformation function $P_{\tau}$ that encourages diverse outputs to cope with multiple wireless constraints, such as perturbation signal extension, arbitrary symbol assignment, and coordination-free perturbations.
We also incorporate the power normalization into the transformation to make \sys undetectable from natural noise. The wireless properties controlled by the transformation function are parameterized with $\tau$. Figure~\ref{fig:main} shows all the modules included in the transformation function. With the help of $P_{\tau}$, the PGM can be optimized to produce the perturbation signals that are resilient to real-world transformations. In \cref{sec:attack_design}, we explain the internal mechanisms of $P_{\tau}$. 
%how these solutions model the actual effects on wireless systems.

%By including the power normalization function inside $P_{\tau}$, we can define the unconstrained problem.
We define an optimization problem to train the PGM that generates a hardware-implementable perturbation signal as: 
%Then, we convert Equation~\ref{eq:general3} to the following optimization problem:
%formulate an optimization problem for simultaneously attacking a video JSCC model that leverage inter-frame dependency properties between time-adjacent source data and other models that encode source data individually.
\begin{equation} \label{eq:general5}
\resizebox{0.87\columnwidth}{!}{
$\begin{aligned}
\operatorname*{arg\,max}_G \quad & \underset{{z_{t} \sim p_{z}}}{\mathbb{E}}  [   
%\sum_{Q\in\mathcal{Q}}\sum_{C\in\mathcal{C}}\sum_{\lambda\in\Lambda}\sum_{j=1}^{N_{a}} \sum_{T_{\tau}\in\mathcal{T}_{\tau}} 
%\mathcal{L}_{Q, C, \lambda, \textbf{H}_{\textbf{a}}^{j}, T_{\tau}}] \\
\sum_{w \in \Psi^{u}} \mathcal{L}_{rx}(z_{t},w)], \\
%(X^{\mathcal{V}}_{g}, \hat{Y}^{\mathcal{V}}_{g}, x^{Q}_{t}, \hat{y}^{Q}_{t}, \boldsymbol{\delta}^{u}_{t}, \boldsymbol{\delta}^{u}_{g}, \textbf{H}_{\textbf{a}}^{j})] \\
%\mathrm{s.t.} \quad \norm{\boldsymbol{\delta}_{t}}_{2} \leq \epsilon \\
\mathcal{L}_{rx}(z_{t},w) = &
\begin{cases}
\mathcal{L}_{mse}(x^{\mathcal{I}}_{t}, \bar{x}^{\mathcal{I}}_{t}), & \text{if} \ Q = \mathcal{I}, \\
\sum_{t = m_{\sigma}(1)}^{m_{\sigma}(P)} \mathcal{L}_{mse}(x^{\mathcal{V}}_{t}, \bar{x}^{\mathcal{V}}_{t}), & \text{if} \ Q = \mathcal{V}, \\
\mathcal{L}_{mse}(H_{F}(x^{\mathcal{S}}_{t}), H_{F}(\bar{x}^{\mathcal{S}}_{t})), & \text{if} \ Q = \mathcal{S}, \\
\mathcal{L}_{CE}(H_{G}(x^{\mathcal{T}}_{t}), H_{G}(\bar{x}^{\mathcal{T}}_{t})), & \text{if} \ Q = \mathcal{T},
\end{cases}
\end{aligned}$}
\end{equation}
where $\Psi^{u}$ is a set containing all radio signals that can be generated by the surrogate ML models. The perturbed signals at the receiver are computed from Equation~\eqref{eq:general2}. We use mean-squared error (MSE) loss as the distortion function $\mathcal{L}_{mse}$. We train the PGM to maximize distortion on a frame-by-frame basis for the image JSCC model. For the video JSCC model, we consider the inter-frame dependency between adjacent frames as the sum of the distortions over all frames within the GOP. This allows the PGM to adapt to any GOP without the need to reconfigure the attack. As for speech, we transform the speech data into a one-dimensional vector via the deframing function $H_{F}$ before the loss is calculated. 
%The perturbed speech sample is obtained from $\bar{v}_{t} = H_{F}(\bar{x}^{\mathcal{S}}_{t}))$. 
%$\bar{w}^{\mathcal{T}}_{t}=S_{M}(\bar{y}^{\mathcal{T}}_{t})$ indicates the predicted probability of word candidates when the text JSCC model is attacked.  
Since the text JSCC model completes sentence restoration by sequentially finding the probabilities that words will appear with a greedy decoder $H_{G}$, we use a cross-entropy loss $\mathcal{L}_{CE}$ between the predicted sentence $H_{G}(\bar{x}^{\mathcal{T}}_{t})$ and the ground truth sentence $H_{G}(x^{\mathcal{T}}_{t})$.

\setlength{\textfloatsep}{2pt}
\begin{algorithm}[t]
\footnotesize
\caption{\sys}\label{alg:pseudocode}
\begin{algorithmic}
\State \textbf{Input:} Dataset $\mathbb{T}^{Q}$, Surrogate JSCC model, Power constraint $\epsilon$
\State \textbf{Output:} PGM $G(\cdot)$
% , $\mathbb{G} \gets \{\text{non-hierarchy}, \text{hierarchy-P}, \text{hierarchy-B}\}$
\For {\text{epoch} $l < \mathrm{MaxIter}$}
    %\State $C, \lambda, \textbf{H}_{\textbf{t}} \gets$ uniformly at random
    %\State $\textbf{H}^{l}_{\textbf{a}}, \text{are sampled uniformly from} \{\textbf{H}^{1}_{\textbf{a}}, \cdots, \textbf{H}^{0.8 \cdot N_{a}}_{\textbf{a}} \}$
    \For {\text{each modality} $Q \in \{\mathcal{I},\mathcal{V},\mathcal{S},\mathcal{T}\}$}
        \For {\text{each batch} $\textbf{B}^{Q} \in \mathbb{T}^{Q}$}
            \State $C, \lambda \gets$ is sampled uniformly from candidates
            \State $\textbf{H}_{\textbf{t}}$ is randomly sampled from channel model
            \State $\textbf{H}_{\textbf{a}} \text{ is sampled uniformly from training set}$
            \If {$ Q = \mathcal{V}$}
                \For {$\mathbf{x}^{\mathcal{V}}_{t} \in \textbf{B}^{\mathcal{V}}(=\{\mathbf{x}^{\mathcal{V}}_{1}, \cdots, \mathbf{x}^{\mathcal{V}}_{P}\})$}
                    %\{\mathbf{s}^{\mathcal{V}}_{t}\}^{m_{\sigma}(G)}_{t=m_{\sigma}(1)} \in
                    %\State // Transmitter
                    \State $Y^{\mathcal{V}}_{t} \gets$ Equation~\eqref{eq:trans}
                    \State $\mathcal{B}^{\mathcal{V}}_{t}$.append($\tilde{x}^{\mathcal{V}}_{t}$)
                    %\State // Adversary
                    \State $z_{t} \sim \text{Uniform}(0,1), z'_{t} \sim \text{Uniform}(0,1)$
                    %\State $\{\mu, \zeta, \phi, \Delta t\} \gets$ uniformly at random
                    %\State $\tau \gets \{\mu, \zeta, \epsilon, \phi, \Delta t\} $
                    \State $\tau \gets $ uniformly at random 
                    \State $\bar{Y}^{\mathcal{V}}_{t}[i,k] \gets$ Equation~\eqref{eq:general4}
                    %\State // Receiver
                    \State $\bar{x}^{\mathcal{V}}_{t} \gets$ Equation~\eqref{eq:general2}
                    \State $\bar{\mathcal{B}}^{\mathcal{V}}_{t}$.append($\bar{x}^{\mathcal{V}}_{t}$)
                \EndFor
            \Else
                \State $Y^{Q}_{t} \gets$ Equation~\eqref{eq:trans}
                \State $z_{t} \sim \text{Uniform}(0,1), z'_{t} \sim \text{Uniform}(0,1)$
                %\State $\{\mu, \zeta, \phi, \Delta t\} \gets$ uniformly at random
                %\State $\tau \gets \{\mu, \zeta, \epsilon, \phi, \Delta t\} $
                \State $\tau \gets $ uniformly at random 
                \State $\bar{Y}^{Q}_{t}[i,k] \gets$ Equation~\eqref{eq:general4}
                \State $\bar{x}^{Q}_{t} \gets$ Equation~\eqref{eq:general2}
            \EndIf
            %\State $\mathcal{L}_{total} \gets$ Equation~\ref{eq:general7}
            \State Update PGM $G$ and $\mathcal{D}$ by solving Equation~\eqref{eq:general7}
        \EndFor
    \EndFor
\EndFor
\State \textbf{Return:} PGM $G$
\end{algorithmic}
\end{algorithm}

%\subsection{Downstream Attack Formulation}
\noindent\textbf{Downstream Attack Formulation.}
Figure~\ref{fig:main} depicts downstream tasks appended to the wireless communication pipeline. We consider two ML models as examples: 1) VC and 2) AVE. Let $F^{N}$ denote a discriminant function for the receiver's downstream task $N \in \{\text{VC}, \text{AVE}\}$. After the receiver demodulates incoming perturbed signals into data, the discriminant function takes the data $\bar{X}_{N}$ and outputs a probability distribution over a set $K_{N}$ of class labels. Note that the VC takes a video clip
$\bar{X}_{\text{VC}}=\{\bar{x}^{\mathcal{V}}_{t}\}^{T}_{t=1}$ consisting of $T$ consecutive frames and the AVE receives $\bar{X}_{\text{AVE}}=\{\bar{x}^{\mathcal{I}}_{t}, \bar{x}^{\mathcal{S}}_{t}\}$ as two inputs. A classifier for task $N$, $\mathcal{C}^{N}$, points $\bar{X}_{N}$ to the class with the maximum probability: $\mathcal{C}^{N}(\bar{X}_{N})=\text{arg max}_{c \in K_{N}} F^{N}_{c}(\bar{X}_{N})$, where $F^{N}_{c}$ is the probability of the perturbed input belonging to a specific class $c$. 
%xyz: Does the attacker have to know the type of downstream task in order to decide on its loss function? This seems to conflict with the claim that it is modality agnostic
%jw: Our main goal is to attack the multimodal reconstruction. Because of the page limit, we show the experimental results of representative downstream tasks. The adversary can include losses from all of the downstream tasks.
We define a loss $\mathcal{L}^{N}_{cls}$ to subvert classifiers:
%\begin{equation} \label{eq:general6}
%\resizebox{0.90\columnwidth}{!}{$
%\mathcal{L}_{cls} =\begin{cases}
% F^{N}_{\mathcal{C}^{N}(\hat{X}_{N})}(\{\bar{x}^{\mathcal{V}}_{t}\}^{T}_{t=1}) - \max\limits_{c \neq \mathcal{C}^{N}(\hat{X}_{N})}F^{N}_{c}(\{\bar{x}^{\mathcal{V}}_{t}\}^{T}_{t=1}) & \text{(VC)}\\
% F^{N}_{\mathcal{C}^{N}(\hat{X}_{N})}(\{\bar{x}^{\mathcal{I}}_{t}, \bar{x}^{\mathcal{S}}_{t}\}) - \max\limits_{c \neq \mathcal{C}^{N}(\hat{X}_{N})}F^{N}_{c}(\{\bar{x}^{\mathcal{I}}_{t}, \bar{x}^{\mathcal{S}}_{t}\}) & \text{(AVE)} 
%\end{cases}
%$}
%\end{equation}
\begin{equation} \label{eq:general6}
\mathcal{L}^{N}_{cls} = \max\limits_{c \neq \mathcal{C}^{N}(\hat{X}_{N})}F^{N}_{c}(\bar{X}_{N}) - F^{N}_{\mathcal{C}^{N}(\hat{X}_{N})}(\bar{X}_{N}), \\
\end{equation}
where $\hat{X}_{N}$ denotes the reconstructed data when there is no attack. $\hat{X}_{\text{VC}}=\{\hat{x}^{\mathcal{V}}_{t}\}^{T}_{t=1}$ and $\hat{X}_{\text{AVE}}=\{\hat{x}^{\mathcal{I}}_{t}, \hat{x}^{\mathcal{S}}_{t}\}$. The attack succeeds when $\mathcal{L}^{N}_{cls}>0$. With the ensemble learning, we find UAPs that maximize $\mathcal{L}^{N}_{cls}$ for the surrogate model with different architectures from the target model. We then fool the downstream services by 
%xyz: What does this mean? 
%jw: We do not assume that adversary knows the ML architecture of downstream tasks. Therefore, we train the perturbation from surrogate models that are different with target ML models.
transferring the attacks calculated from the surrogate model to the target model.

\noindent\textbf{Stealthy Attack Formulation.} Existing works~\cite{bahramali2021robust, liu2023exploring} have a problem that an adaptive defender can devise an anomaly classifier~\cite{xu2021detecting} that identifies the attacks by analyzing the perturbation’s statistical behavior. To enforce the generator to produce undetectable perturbations, we explicitly regularize our PGM with the discriminative loss~\cite{goodfellow2014generative}:
\begin{equation} \label{eq:general_div}
\small
%\begin{aligned}
\mathcal{L}_{ds} = \log \mathcal{D}(M_{C}(\mathcal{R}(\hat{Y}^{Q}_{t}))) + \log (1 - \mathcal{D}(M_{C}(\mathcal{R}(\bar{Y}^{Q}_{t}))), \\
%\end{aligned}
\end{equation}
where $\mathcal{D}$ is a discriminator~\cite{he2016deep} that distinguishes clean signals from perturbed signals. We aim to minimize $\mathcal{L}_{ds}$ for forcing our PGM to explore the latent space and discover robust adversarial examples. To guarantee that the PGM properly produces the diversified perturbation, we utilize the diversity sensitive loss~\cite{choi2020stargan}:
\begin{equation} \label{eq:general_div2}
%\begin{aligned}
\mathcal{L}_{dv} = \underset{{z_{t}, z'_{t}}}{\mathbb{E}}[\norm{G(z_{t}) - G(z'_{t})}_{1}], \\
%\end{aligned}
\end{equation}
where $z_{t}$ and $z'_{t}$ are two different random latent codes.

%$\mathcal{L}_{dv}$ is the diversity sensitive loss obtained from differences between two different perturbations.

\begin{comment}
\begin{equation} \label{eq:general_div}
\small
\mathcal{L}_{ds} = \mathcal{L}_{adv} + \underset{{z_{t} \sim p_{z}}, {z'_{t} \sim p_{z}}}{\mathbb{E}} [\norm{G(z_{t}) - G(z'_{t})}_{1}], \\
\end{equation}
\end{comment}

%\subsection{Unified Attack Formulation}
\noindent\textbf{Unified Attack Formulation.}
Finally, we integrate all losses into the objective function so that UAPs generated by the PGM can perturb wireless communication and downstream services simultaneously. Specifically, our goal is to solve the following objective function:
\begin{equation} \label{eq:general7}
\resizebox{0.88\columnwidth}{!}{$
\begin{aligned}
\operatorname*{max}_G \operatorname*{min}_\mathcal{D} \ \underset{{z_{t}}}{\mathbb{E}} [\sum_{w \in \Psi^{u}} [\mathcal{L}_{rx} + \sum_{N \in \mathcal{N}}\beta^{N}_{cls}\mathcal{L}^{N}_{cls}  - \beta_{ds}\mathcal{L}_{ds}]]+ \beta_{dv}\mathcal{L}_{dv}, 
\end{aligned}$}
\end{equation}
where $\beta^{N}_{cls}$, $\beta_{ds}$, and $\beta_{dv}$ weigh the relative importance of each term and $\mathcal{N} = \{\text{VC}, \text{AVE}\}$. PGM generates a perturbation conditioned on the latent code and multiple controllable parameters of the wireless protocols, while $\mathcal{D}$ tries to distinguish between perturbed and clean signals.

In Algorithm~\ref{alg:pseudocode}, we outline the training process. \jungwoo{Please refer to \cref{sec:exp_setup} for the parameters selected in the experiment.} Our goal is to train the PGM $G$ that generates UAPs to subvert ML-based JSCC models. We ensemble outputs of multimodal JSCC models to find generalizable adversarial signals that can transfer between modalities and protocols. The ML model used for training is a surrogate model that is different from the target model. We utilize the transformation function $P_{\tau}$ to change the outputs of PGM to practically feasible adversarial signals. At each training iteration, the algorithm selects a batch from the training dataset $\mathbb{T}$ with a different distribution from the training dataset of the target model. 
\jungwoo{We ensure that the PGM learns effective UAPs leveraging ensemble learning, which integrates a set of JSCC models with different ${Q,C,\lambda}$. The loss values derived from each JSCC model are jointly backpropagated to optimize the PGM using the Adam optimizer~\cite{kingma2014adam}.}
%
%Specifically, we utilize ensemble learning incorporating a set of JSCC models with different parameters $(Q, C, \lambda)$ to optimize perturbations generalizable across various synthesizers.
%
%We apply the Adam optimizer~\cite{kingma2014adam} to find the UAPs. %This algorithm applies the fundamental constraints of a wireless system to produce a perturbation signal that can be implemented in the real world. 
%
\jungwoo{As a result, we solve four technical challenges described in \cref{sec:intro}: (1) multi-modality and unknown $\mathbf{H}_{\textbf{t}}$, (2) unknown protocols, (3) de-synchronization, and (4) susceptibility to adaptive defense.}

\vspace{-0.2cm}
\subsection{Design of Our Transformation Function}
\label{sec:attack_design}
To cope with challenging real-world scenarios, the adversary should craft input-agnostic UAP signals regardless of synchronization with the legitimate receiver. The transformation function $P_{\tau}$ helps PGM learn to produce perturbations with a distribution similar to that of adversarial signals that can be realized in the real environment.
%We define $P_{\tau}$ as a transform function that modifies the adversarial signal to solve the design challenges. Through the $P_{\tau}(\cdot)$ function, 
Therefore, our adversarial signals are agnostic to 1) inconsistency of the number of data symbols between the benign signal and the adversarial signal, 2) unknown symbol allocation across the OFDM subcarriers, 3) time misalignment, and 4) unknown phase rotation. We additionally include a power regularization for undetectability. The modules included in the transformation function are shown in Figure~\ref{fig:main} and detailed below. 

\begin{figure}[t]
    \centering
    \begin{tabular}{@{}c@{}}
        \includegraphics[width=0.71\columnwidth]{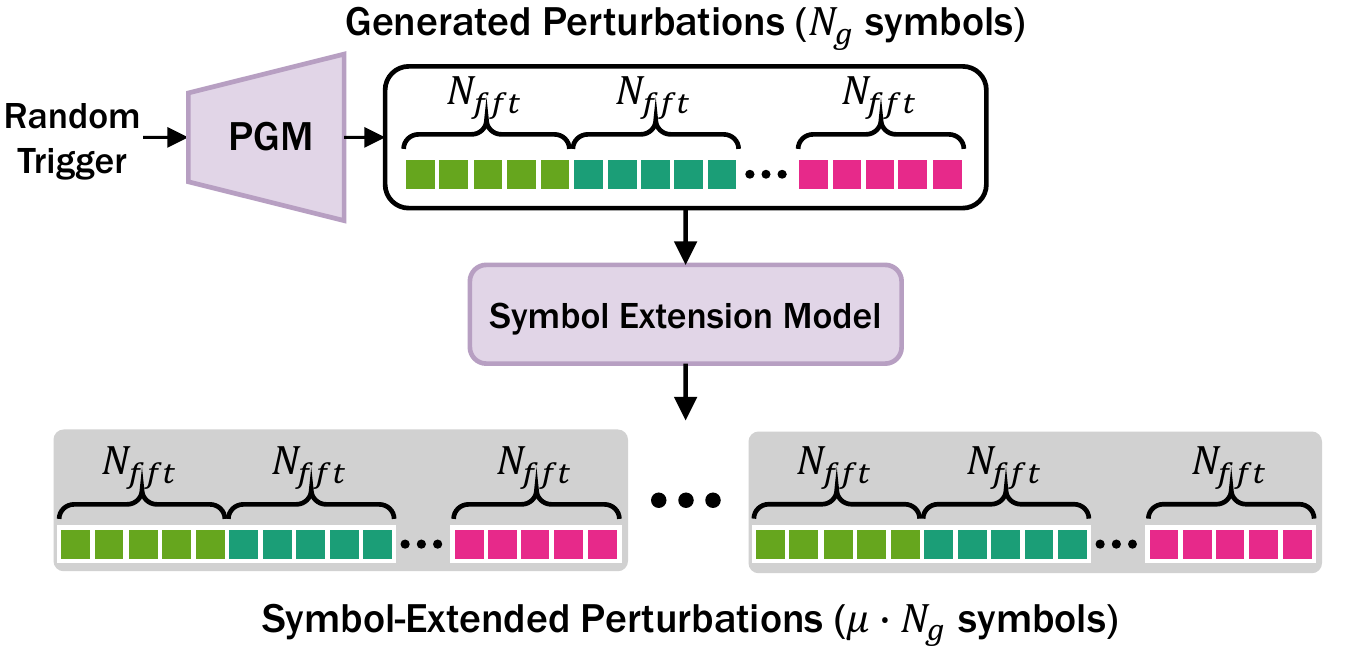} \\
    \end{tabular}
    \vspace{-0.2cm}
    \caption{\jungwoo{Symbol extension mechanism. The perturbation length is extended to match the maximum length of the target signal.}}
    \label{fig:operations}
    %\vspace{-0.5cm}
\end{figure}

\begin{figure}[t]
    \centering
    \begin{tabular}{@{}c@{}}
        \includegraphics[width=0.68\columnwidth]{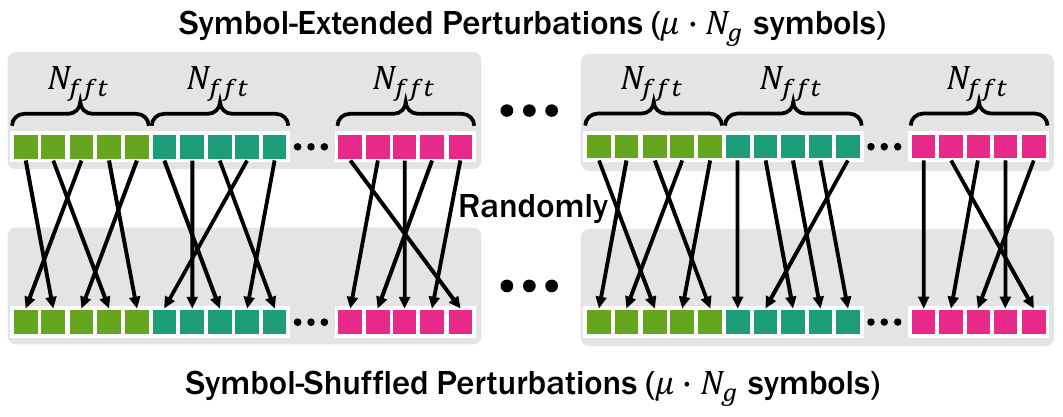} \\
    \end{tabular}
    \vspace{-0.2cm}
    \caption{\jungwoo{Symbol shuffling mechanism. The complex-valued symbols assigned to the subcarriers are randomly shuffled.}}
    \label{fig:operations3}
    %\vspace{-0.2cm}
\end{figure}

\noindent\textbf{Symbol Extension Model.} The number of OFDM symbols varies greatly depending on the modality and coding rate. %For instance, the number of symbols generated after encoding a video clip is proportional to the number of frames and the resolution.
Furthermore, the coding rate of the JSCC encoder determines the amount of data compressed. In an online attack, the modality and coding rate are unknown. This leads the adversary to make the attack signal invariant to the number of OFDM symbols contained in the target signal. 
%Therefore, the adversary's goal is to train the PGM to maintain attack performance even when the adversarial examples are repeated multiple times.
%We present two new methodologies to make our attack invariant to the benign sender.
%We define $K(\cdot)$ as a perturbation extension function, which increases the number of OFDM symbols by repeatedly appending the same perturbation to the original perturbation signal.
%The OFDM transmitter divides the output of the JSCC encoder into several OFDM symbols with length $L_{fft}$ and transmits them.
%xyz: How does the attacker know the number of symbols in the target signal? 
%jw: Thank you for finding this issue. I clarify the contents.
As the information about the coding rate is publicly available (see~\cref{sec:capability}), we can find the maximum value of $N_{s}$. \jungwoo{As shown in Figure~\ref{fig:operations}, we concatenate the PGM-generated signal multiple times through function $K(\cdot)$ such that $\mu \cdot N_{g}$ is equal to the maximum value of $N_{s}$, where $\mu$ is a parameter to adjust the number of symbols.}
%As a result, the perturbation signal has the maximum number of symbols and can all be added to the symbols of the unknown target signal.
Hence, our symbol-extended perturbations can perturb all symbols of the target signal without prior knowledge of the target signal's symbol count.
Then, we ensure that the concatenated perturbations achieve high generalizability for multiple coding rates. In Algorithm~\ref{alg:pseudocode}, we randomly select the coding rate $\lambda$ for each training epoch.

%Therefore, the concatenated perturbation signal can perturb all target OFDM symbols without knowing $N_{s}$. %Then, we train the augmented perturbation signal to be invariant to different $N_{s}$.
%To meet this requirement, we concatenate the perturbation signal multiple times through function $K(\cdot)$ to convert it into a perturbation signal with the same number of symbols as the number of symbols in the target radio signal. We adjust the length of adversarial signal through the parameter $\mu$. %Therefore $K(\boldsymbol{\delta}^{u}_{t}, \mu) \in \mathbb{C}^{N_{s} \times N_{fft}}$ has the same matrix size as $\hat{Y}^{Q}_{t}$.

\begin{figure}[t]
    \centering
    \begin{tabular}{@{}cc@{}}
        \includegraphics[width=0.44\linewidth]{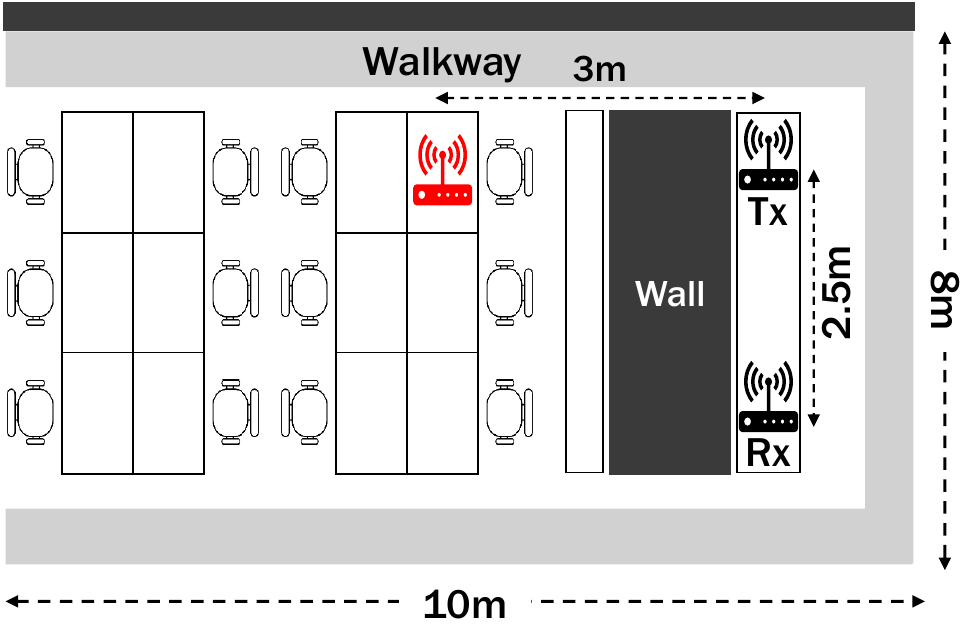} &
        \includegraphics[width=0.44\linewidth]{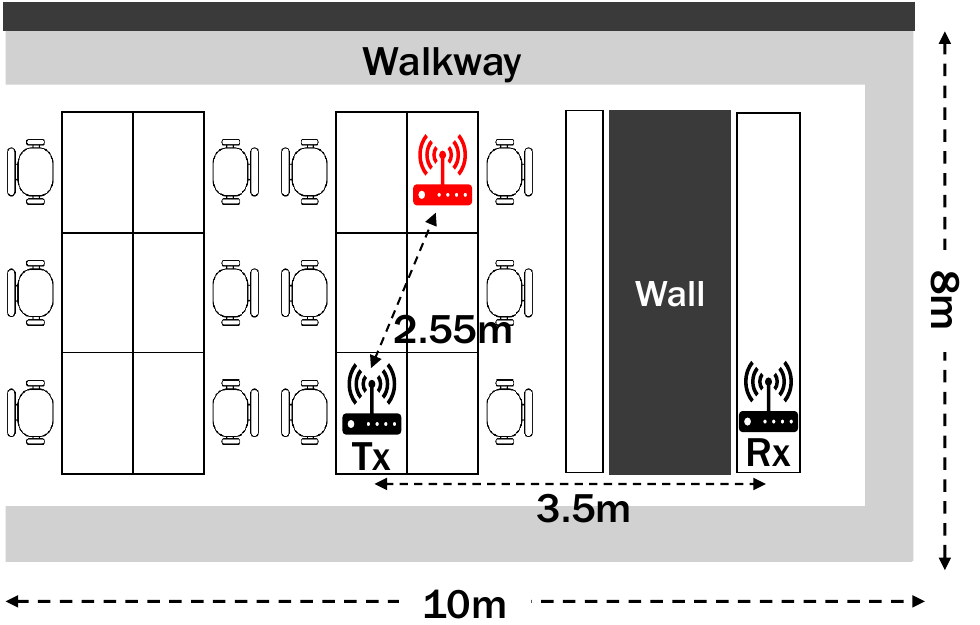} \\
        \footnotesize{(a) LoS Tx/Rx path} &
        \footnotesize{(b) NLoS Tx/Rx path}
    \end{tabular}
    \vspace{-0.2cm}
    \caption{Experimental settings established in \sys.}
    \label{fig:scenario}
    %\vspace{-0.2cm}
\end{figure}

\noindent\textbf{Symbol Shuffling Model.} 
%xyz: citation? 
%jw: I added the citation.
Previous works~\cite{hu2023robust, liu2023exploring} make the assumption that the adversary knows how the target wireless system allocates symbols to each subcarrier. This is infeasible in practice, because standard wireless communications often randomize the allocation to prevent consecutive repetition of the same symbols. 
%Therefore, each OFDM subcarrier carries symbols with an unknown distribution. 
Our adversary aims to make a subcarrier-invariant perturbation that is universally applicable to any symbol distribution of subcarriers. We define a function $\Gamma(\cdot)$ that randomly shuffles the symbols assigned to the subcarrier based on a seed $\zeta$, \jungwoo{as shown in Figure~\ref{fig:operations3}}. Consequently, we train the PGM to generate the attack signal that is robust to the unknown symbol distribution across OFDM subcarriers.

\noindent\textbf{Time and Frequency Rotation.}
%To create shift-invariant perturbation signals, adversaries can simultaneously inject perturbation signals when a benign sender begins to transmit multimodal signals. However, due to the time misalignment between the sender and the adversary in the physical attack scenario, it is difficult to match the perturbation signal accurately with the victim signal when it arrives at the receiver side.
Due to time and frequency misalignment, a random phase rotation occurs in each OFDM subcarrier. In order to enforce our perturbation to learn shift-invariant properties, we employ a phase rotation function $e^{-j2 \pi f_{k} \Delta t + j \phi}$ from the previous approaches~\cite{sadeghi2019physical, bahramali2021robust, liu2023exploring}, where $\Delta t$ and $\phi$ are time difference and phase offset between the benign signal and the adversarial perturbation, respectively.

%We model the phase shift that occurs in the $k$-th subcarrier of each OFDM symbol as $e^{-j2 \pi f_{k} \Delta t}$. Since the phase rotation is independent of the frequency of the subcarrier, all subcarriers have the same offset, $e^{j \phi}$. Using above perperties, we arbitrarily generate phase shifts through $e^{-j2 \pi f_{k} \Delta t + j \phi}$ during the offline training process. 

\begin{figure*}[t]
\centering
    \resizebox{0.7\textwidth}{!}{
    \begin{tabular}{@{}c@{}}
    \includegraphics[width=\linewidth]{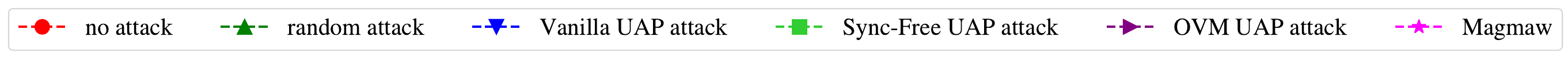} \\  
    \end{tabular}}

    \begin{tabular}{@{}cccc@{}}
    \includegraphics[width=0.17\linewidth]{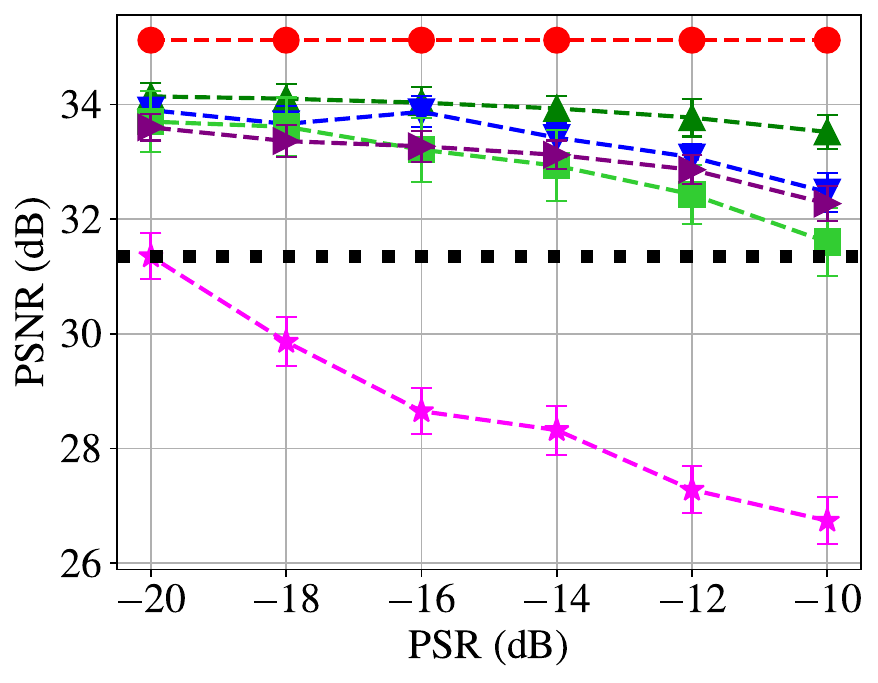} &
    \includegraphics[width=0.17\linewidth]{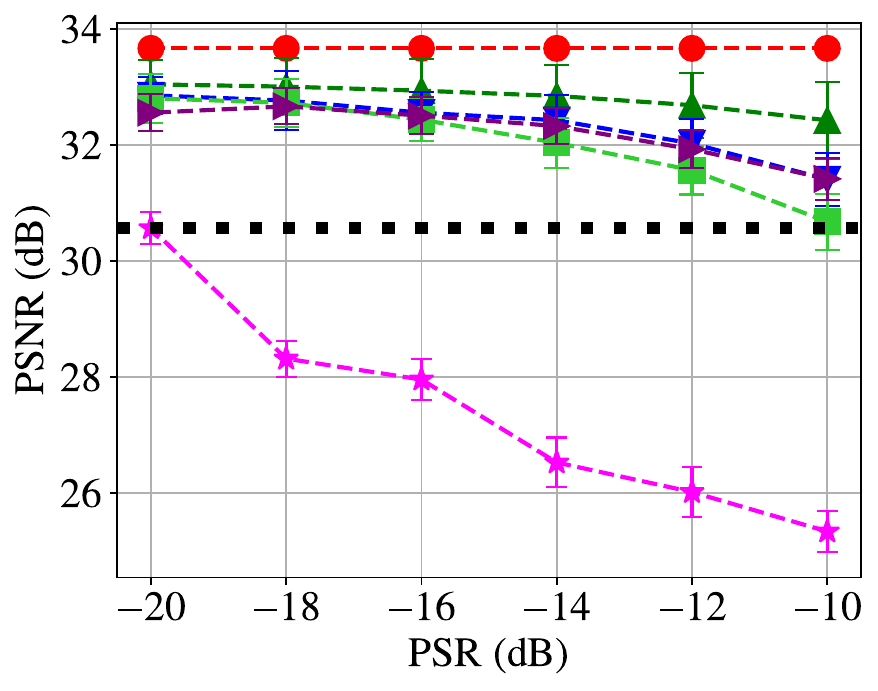} &
    \includegraphics[width=0.17\linewidth]{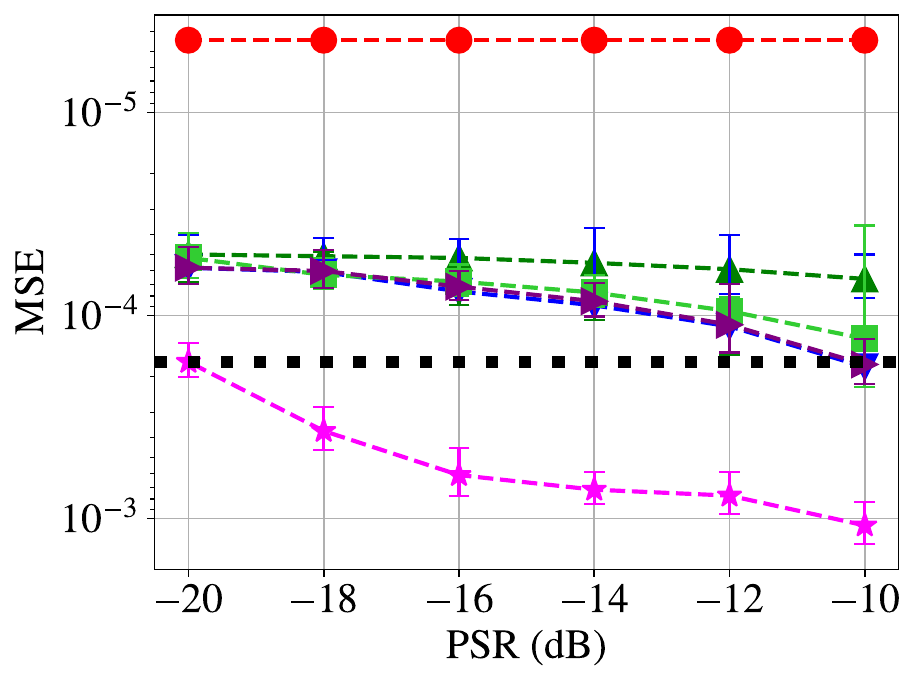} &
    \includegraphics[width=0.17\linewidth]{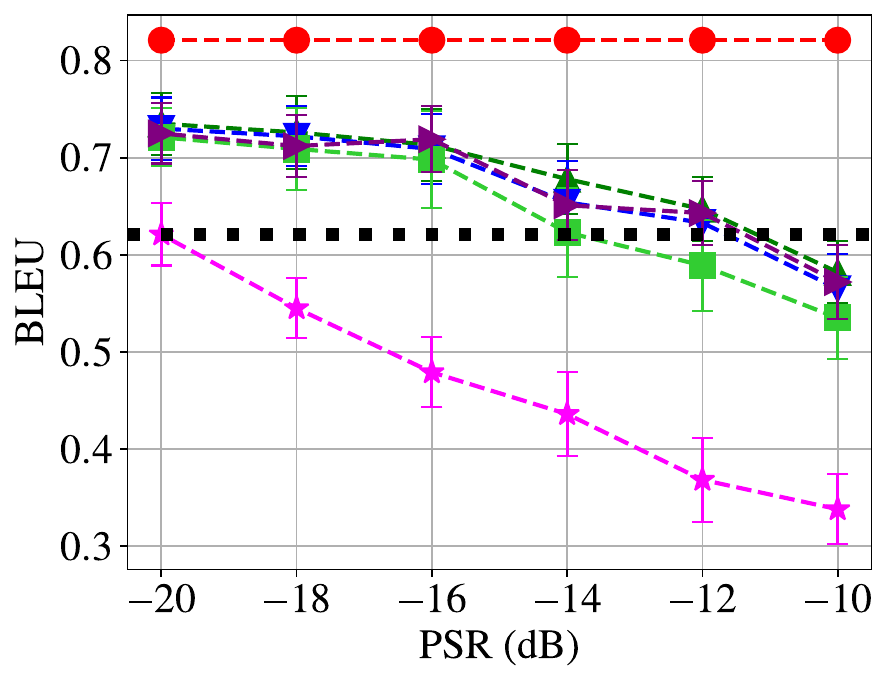} \\
    \footnotesize{Image Transmission~\cite{yang2022ofdm}}&
    \footnotesize{Video Transmission~\cite{wang2022wireless}}&
    \footnotesize{Speech Transmission~\cite{weng2021semantic}} &
    \footnotesize{Text Transmission~\cite{xie2021deep}}\\
    \end{tabular}
  \vspace{-0.2cm}
  \caption{\sys on ML-based wireless communication systems (i.e., modality-specific JSCC models).}
  \label{fig:Exp_attack1}
  \vspace{-0.4cm}
\end{figure*}

\begin{figure}[t]
\centering
    \begin{tabular}{@{}cc@{}}
        \includegraphics[width=0.34\linewidth]{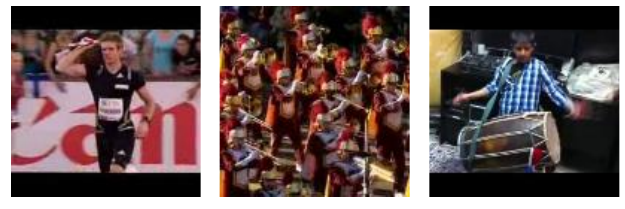} &
        \includegraphics[width=0.34\linewidth]{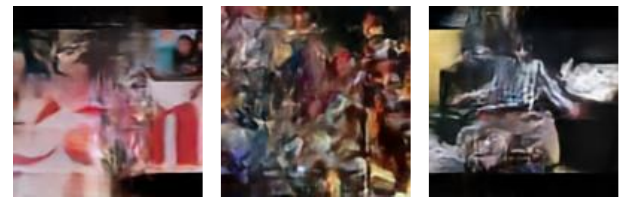} \\
        \footnotesize{(a) Transmitted Images} &
        \footnotesize{(b) Received Images} \\
        \includegraphics[width=0.34\linewidth]{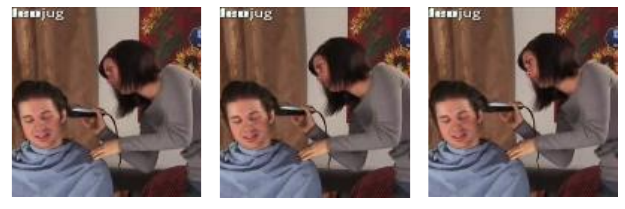} &
        \includegraphics[width=0.34\linewidth]{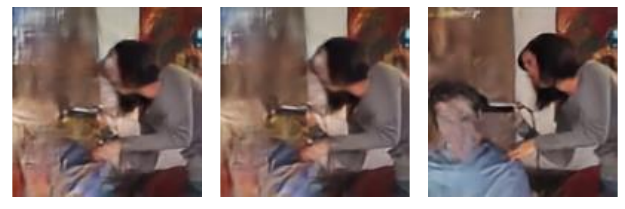} \\
        \footnotesize{(c) Transmitted Video} &
        \footnotesize{(d) Received Video} \\
        \includegraphics[width=0.34\linewidth]{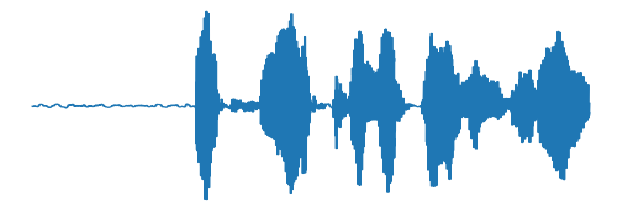} &
        \includegraphics[width=0.34\linewidth]{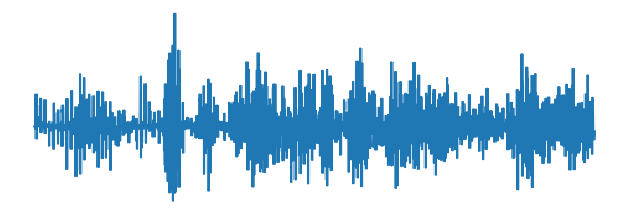} \\
        \footnotesize{(e) Transmitted Speech} &
        \footnotesize{(f) Received Speech} \\
        \includegraphics[width=0.34\linewidth]{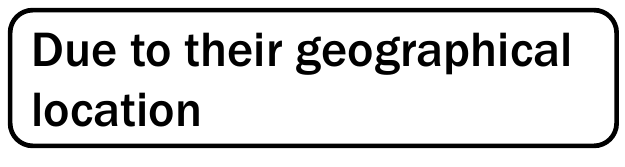} &
        \includegraphics[width=0.34\linewidth]{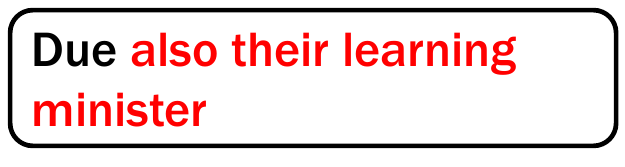} \\
        \footnotesize{(g) Transmitted Text} &
        \footnotesize{(h) Received Text} \\
    \end{tabular}
  \vspace{-0.1cm}
  \caption{Visualization of attack effects on multimodal JSCC.}
  \label{fig:visual1}
  \vspace{-0.2cm}
\end{figure}

\noindent\textbf{Power Normalization.}
$\mathcal{M}(\cdot)$ is a power normalization function that adjusts the perturbation signal according to $\epsilon$, which is the upper bound on the attacker's signal power. We follow the existing power remapping function~\cite{bahramali2021robust} to preserve the power constraint of the perturbations as follows:
\begin{equation} \label{eq:general9}
\mathcal{M}(\boldsymbol{\gamma}^{u}_{t}, \epsilon) = 
\begin{cases}
\sqrt{\epsilon} \frac{\boldsymbol{\gamma}^{u}_{t}}{\norm{\boldsymbol{\gamma}^{u}_{t}}_{2}}, & \norm{\boldsymbol{\gamma}^{u}_{t}}^{2}_{2} > \epsilon, \\
\boldsymbol{\gamma}^{u}_{t}, & \norm{\boldsymbol{\gamma}^{u}_{t}}^{2}_{2} \leq \epsilon.
\end{cases}
\end{equation}
where \jungwoo{PSR is the ratio of the power of the attack signal to the power of the victim signal. $\epsilon$ is defined as $\norm{\hat{y}_{t}}^{2}_{2} \cdot 10^{\text{PSR/10}}$, where $\hat{y}_{t}$ is the time-domain signal in Equation~\eqref{eq:channel}}. $\boldsymbol{\gamma}^{u}_{t}$ is the output of the symbol extension and symbol shuffling models.

\noindent\textbf{Transformation Function.}
Consequently, we obtain the converted perturbation signal transmitted from the $k$-th subcarrier of the $i$-th OFDM symbol through the 
%xyz: Use "transformation" consistently
%jw: Thank you for letting me know.
transformation function $P_{\mu, \zeta, \epsilon, \phi, \Delta t}(\cdot)$ as follows:
\begin{equation} \label{eq:general8}
\begin{gathered}
P_{\mu, \zeta, \epsilon, \phi, \Delta t}(\boldsymbol{\delta}_{t})[i,k] = \mathcal{M}(\boldsymbol{\gamma}^{u}_{t}, \epsilon)[i,k] e^{j \phi} e^{-j2 \pi f_{k} \Delta t}, \\ 
\text{where} \ \boldsymbol{\gamma}^{u}_{t} = \Gamma(K(\boldsymbol{\delta}^{u}_{t}, \mu), \zeta).
\end{gathered}
\end{equation}
Here, the transformation function is controlled by various parameters ${\mu, \zeta, \epsilon, \phi, \Delta t}$. 

\vspace{-0.2cm}
\subsection{Hardware Implementation}
Figure~\ref{fig:scenario} shows real-world attack scenarios in which the attacker (red device) sends a perturbation signal to the receiver. To thoroughly study radio signal propagation, we classify the physical environment into Line Of Sight (LoS) or NLoS (Non Line of Sight) between the transmitter and receiver. We obtain the experimental results from both Tx-Rx scenarios and then indicate the distribution of the results. We further showcase the difference in efficiency for each scenario in \cref{sec:ablation}.

\noindent\textbf{Target Wireless System.} We first implement the ML-based wireless communication system depicted in Figure~\ref{fig:main} through USRP B210, a software-defined radio widely used in designing wireless communication systems. We drive the USRP B210 using GNURadio software package~\cite{blossom2004gnu} that provides a graphical programming interface for configuring transceivers and allows us to model the customized blocks. The transmitter and receiver consist of a USRP B210 and a Linux laptop, respectively, and they communicate through a single antenna, where the carrier frequency is set to 2.4 GHz. The number of cyclic prefixes and subcarriers $L_{fft}$ is 16 and 64, respectively. Of the 64 subcarriers, 48 are used to carry symbols for ML-based JSCC, 4 of which are used for pilot symbols.

\noindent\textbf{Attack System.} We build an adversarial transmitter using a USRP N310 device with a single antenna and a Linux desktop. 
%Note that the attacker is not synchronized with the target wireless system. Moreover, our adversarial device has no knowledge of the victim wireless channel $\textbf{H}_{\textbf{t}}$ between the transmitter and receiver. 
We randomly move the antenna to collect 2000 random realizations of the channel $\{\textbf{H}^{l}_{\textbf{a}}\}^{2000}_{l=1}$ between the adversarial transmitter and receiver. Following the previous work~\cite{bahramali2021robust}, we set the range of PSR to [-20,-10] dB. To perform the UAP attack, we adopt surrogate models with different architectures and parameters from the target wireless communication system and the downstream classifier. We train the PGM offline according to the Algorithm~\ref{alg:pseudocode}. The hyperparameters $(\beta^{\text{VC}}_{cls}, \beta^{\text{AVE}}_{cls}, \beta_{ds}, \beta_{dv})$ are all set to 1.

%After training is complete, we collect several samples of perturbation signals $P_{\tau}(\boldsymbol{\delta}^{u}_{t})$ to facilitate online attacks. In our setup, the attacker feeds random triggers into the PGM to generate a large set of UAPs, and randomly selects UAP adversarial examples from among them. The adversarial transmitter system then loads the stored perturbation signal and transmits it through the OFDM transmitter.

\begin{comment}
\begin{figure}[t]
    \centering
    \begin{tabular}{@{}c@{}}
        \includegraphics[width=\columnwidth]{Fig/fig_main_3.png} \\
    \end{tabular}

    \caption{(a) High-level view of our attack scenario on end-to-end ML-based wireless communication system. (b) Example of our attack.}
    \label{fig:main_3}
\vspace{-0.7cm}
\end{figure}

\begin{figure}[t]
    \centering
    \begin{tabular}{@{}c@{}}
        \includegraphics[width=\columnwidth]{Fig/fig_main_4.png} \\
    \end{tabular}

    \caption{(a) High-level view of our attack scenario on end-to-end ML-based wireless communication system. (b) Example of our attack.}
    \label{fig:main_4}
\vspace{-0.7cm}
\end{figure}
\end{comment}

\vspace{-0.1cm}
\section{Attack Evaluation}
\label{sec:attack_eval}
%In this section, we first present the experimental setup. Then we evaluate the black-box \sys against ML-based wireless systems. We further compare the results with conventional wireless attacks. Finally, we show the effectiveness of \sys against downstream services by combining the wireless system and classification modules into unified framework.

\subsection{Experimental Setup}
\label{sec:exp_setup}

\noindent\textbf{ML Models.} We consider four state-of-the-art JSCC models that deliver multimodal data over the wireless channel and re-implement them based on several open-source resources~\cite{yang2022ofdm, xie2021deep}. In Appendix Table~\ref{tab:substitute_models}, we show the surrogate JSCC models\footnote{\jungwoo{We also evaluate how less similar surrogate models affect the attack performance in the supplementary document~\cite{supplement}.}}. \jungwoo{We use constellation mapping schemes $C\in\{\text{QPSK, 16-QAM, 64-QAM}\}$ adopted in wireless standards and coding rates $\lambda\in\{\frac{1}{6}, \frac{1}{12}\}$ chosen from existing literature. The coding rate is computed as channel usage per source~\cite{yang2022ofdm}.}
%
%We use the coding rates $\lambda\in\{\frac{1}{6}, \frac{1}{12}\}$ chosen from existing papers. 
Note that each JSCC model has different model weights based on the variations of $C$ and $\lambda$.
%\jungwoo{Coding rate is computed as channel use per source~\cite{yang2022ofdm}.}

%Therefore, we train each JSCC model with different $C\in \{\text{QPSK, 16-QAM, 64-QAM}\}$ and $\lambda \in \{\frac{1}{6}, \frac{1}{12}\}$. 
%We train the model with an SNR uniformly distributed in [0,20] dB, to represent a wide range of channel conditions.

\begin{figure}[t]
    \centering
    \begin{tabular}{@{}cc@{}}
        \includegraphics[width=0.44\linewidth]{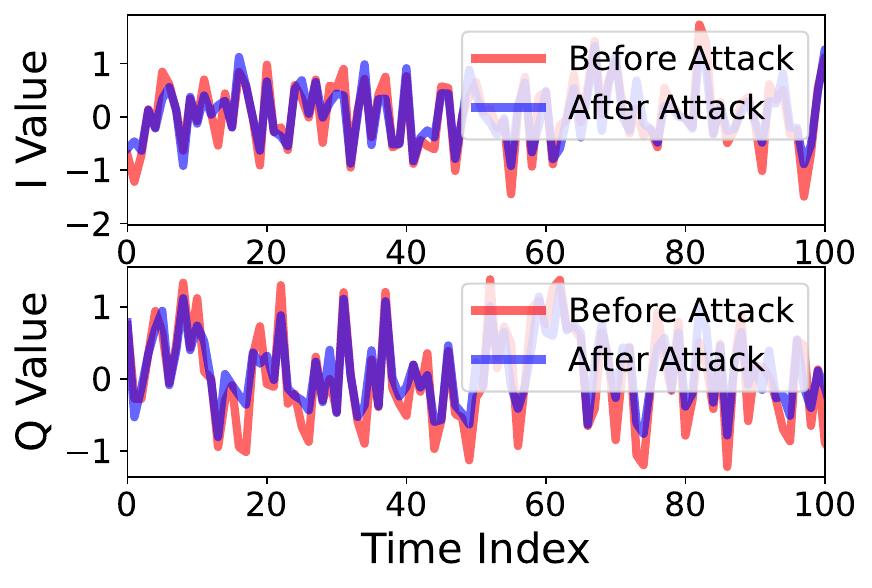} &
        \includegraphics[width=0.44\linewidth]{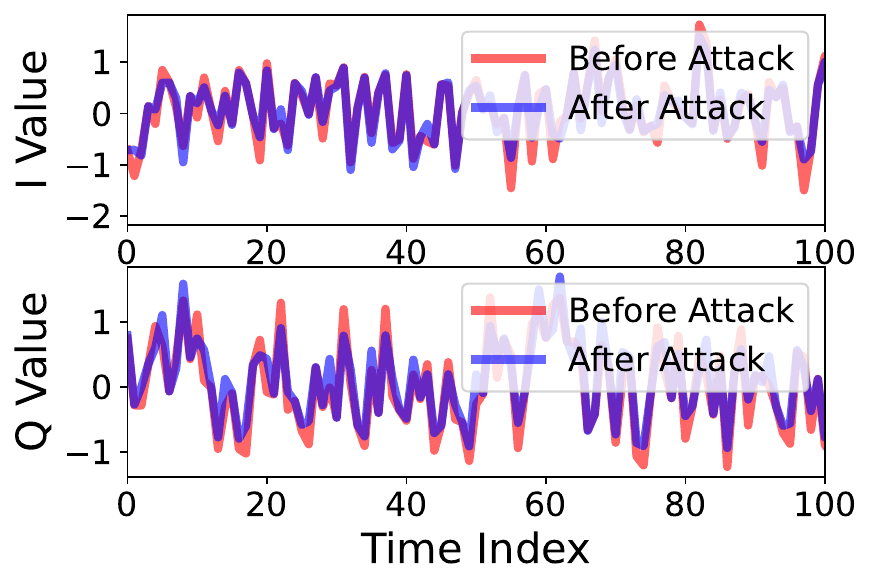} \\
        \footnotesize{(a) PSR is -10dB} &
        \footnotesize{(b) PSR is -16dB}
    \end{tabular}
    \vspace{-0.1cm}
    \caption{\jungwoo{Effect of perturbation when modulation is 16-QAM.}}
    \label{fig:pert_effect}
    %\vspace{-0.2cm}
\end{figure}

\begin{figure}[t]
    \centering
    \begin{tabular}{@{}ccc@{}}
        \includegraphics[width=0.28\linewidth]{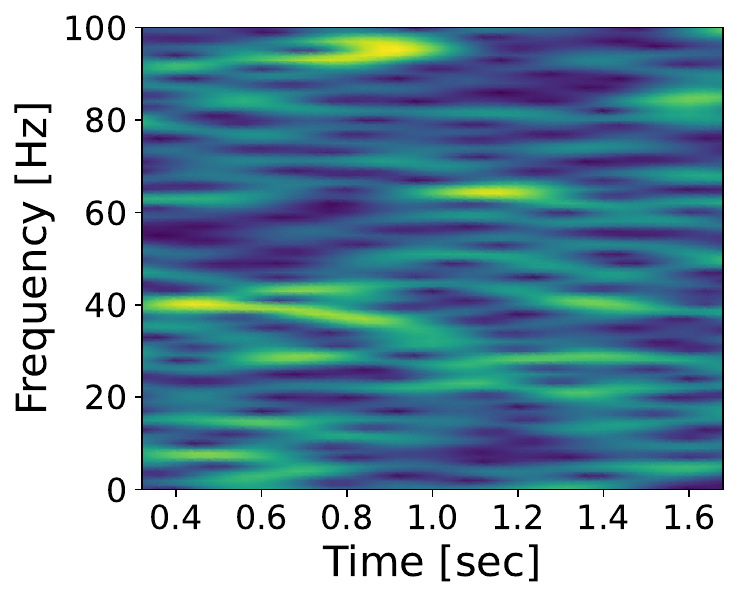} &
        \includegraphics[width=0.28\linewidth]{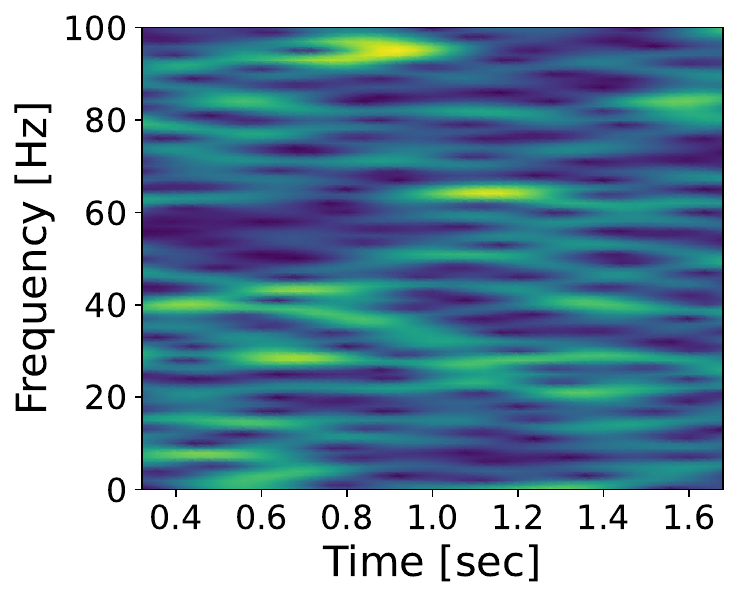} &
        \includegraphics[width=0.28\linewidth]{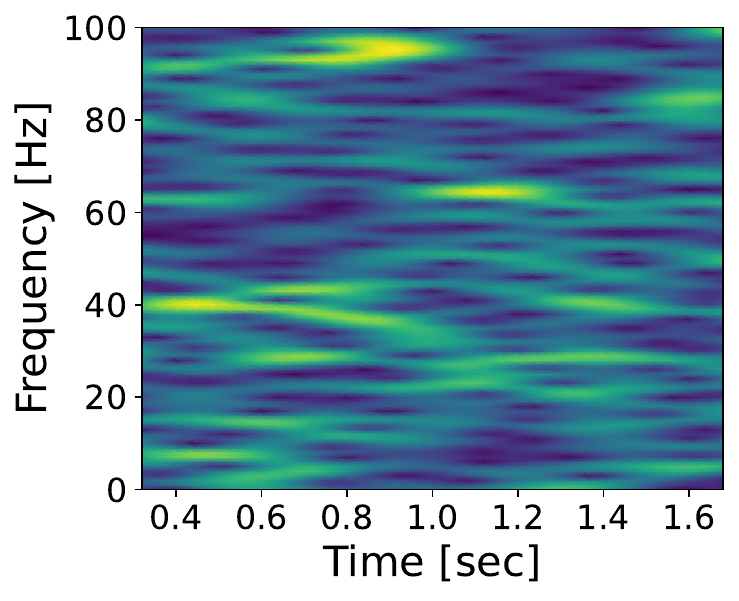} \\
        \footnotesize{(a) Before Attack} &
        \footnotesize{(b) PSR = -10dB} &
        \footnotesize{(c) PSR = -20dB}
    \end{tabular}
    \vspace{-0.1cm}
    \caption{\jungwoo{CSI heatmap when the sampling rate is 200Hz.}}
    \label{fig:pert_effect2}
    %\vspace{-0.2cm}
\end{figure}

\noindent\textbf{Downstream Tasks.} We also consider scenarios where the receiver applies the demodulated data to ML-based downstream services, such as VC and AVE. For the VC task, we benchmark three state-of-the-art models, namely, I3D~\cite{carreira2017quo}, SlowFast~\cite{feichtenhofer2019slowfast} and TPN~\cite{yang2020temporal}. As a benchmark model for a multimodal task, we choose the AVE proposed by \cite{tian2021can}. Appendix Table~\ref{tab:substitute_models2} depicts surrogate models to craft transferable attacks.

\begin{figure*}[ht]
\centering
    \begin{tabular}{@{}cccc@{}}
    \includegraphics[width=0.17\linewidth]{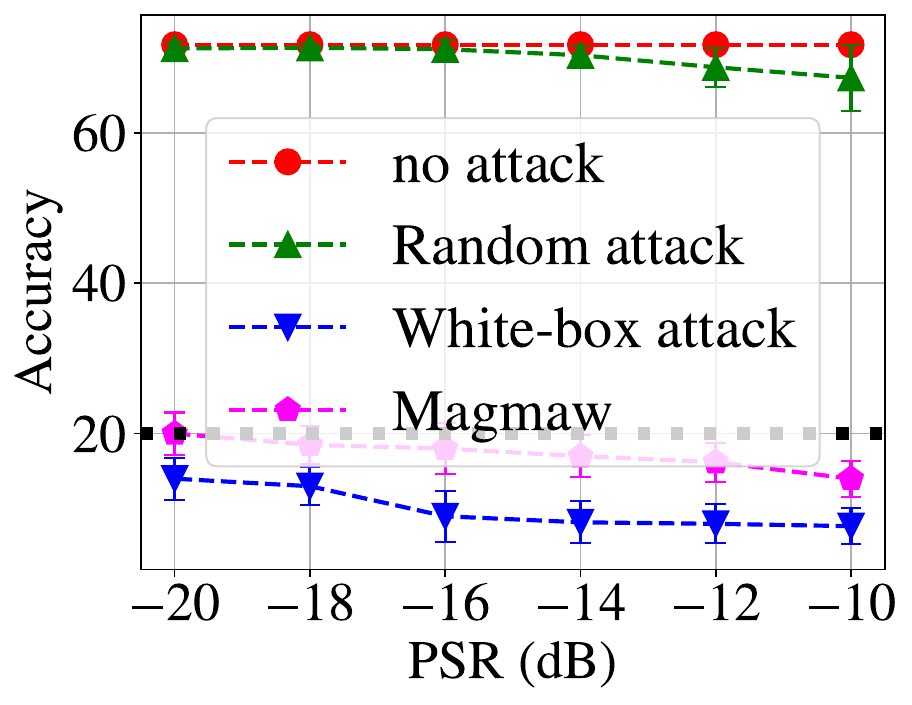} &
    \includegraphics[width=0.17\linewidth]{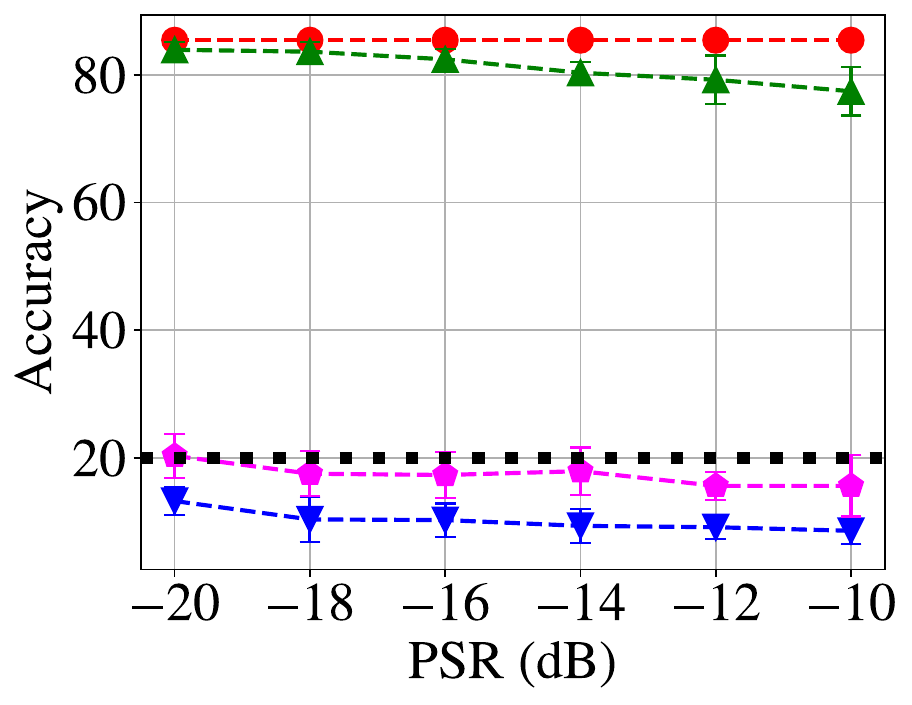} &
    \includegraphics[width=0.17\linewidth]{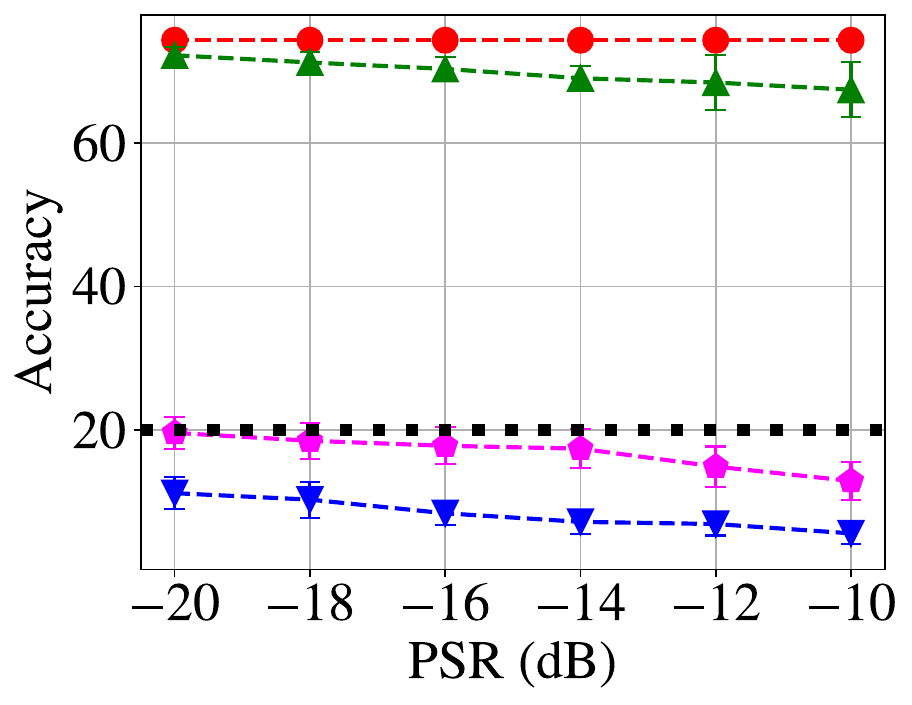} &
    \includegraphics[width=0.17\linewidth]{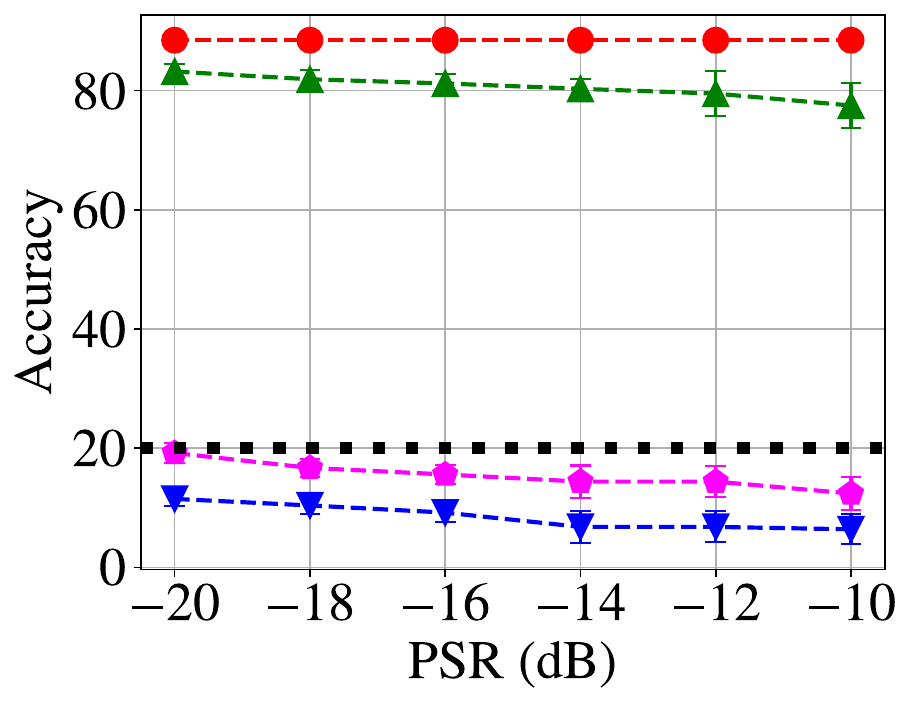} \\
    \footnotesize{I3D~\cite{carreira2017quo}}&
    \footnotesize{SlowFast~\cite{feichtenhofer2019slowfast}}&
    \footnotesize{TPN~\cite{yang2020temporal}} &
    \footnotesize{AVE~\cite{tian2021can}}\\
    \end{tabular}
  \vspace{-0.2cm}
  \caption{\sys on ML-based downstream classification tasks (i.e., VC and AVE). 
  %Magmaw's average precision, recall, and F1 score are 0.1607, 0.1837, and 0.1714, respectively.}
  }
  \label{fig:Exp_attack2}
  \vspace{-0.4cm}
\end{figure*}

\noindent\textbf{Dataset.} We choose popular multimodal datasets to train and evaluate JSCC models. For training the image and video JSCC models, we adopt the Vimeo90K dataset~\cite{xue2019video}, which is widely used in evaluating image and video processing tasks. %It consists of 89,800 video clips with 7 consecutive frames with a resolution of $448 \times 256$.
To facilitate efficient training, the video sequences are cropped to a resolution of $256 \times 256$. We then evaluate the image and video JSCC models using the UCF-101 dataset~\cite{soomro2012ucf101}.
For the speech JSCC model, we use the speech dataset from Edinburgh DataShare~\cite{valentini2017noisy}, which contains more than 10,000 training data and 800 test data with a sampling rate of 16 KHz. %Speech files follow $\textit{.wav}$ format. Each $\textit{.wav}$ file have different number of of speech samples.
We truncate the speech sample sequence to have 128 frames with a frame length of 128 after framing. For the text JSCC model, we select the proceedings of the European Parliament, which includes about 2 million sentences and 53 million words. We pre-process the dataset to have sentence lengths between 4 and 30 words. We then split it into training and test sets. We also select widely used datasets as benchmarks to evaluate VC and AVE downstream tasks. We adopt the UCF-101 human activity dataset~\cite{soomro2012ucf101} to verify \sys on the VC model. %UCF-101 includes 13320 videos from 101 human action categories. 
For evaluating the AVE model, we adopt the audio-visual event dataset~\cite{tian2018audio} which contains 4,143 video clips with 28 events.

\noindent\textbf{Evaluation Metrics.} We use evaluation metrics that effectively reflect the semantic information of each modality. In the image and video domains, we select the PSNR as the representative picture quality measurement. In the speech domain, the MSE reflects the quality of the received speech. For the text domain, the BLEU score~\cite{belghazi2018mutual} is widely used to compare the difference between the original sentence and the reconstructed one. We measure the experimental results from the two Tx-Rx scenarios (see Figure~\ref{fig:scenario}), and then plot the distribution of the results in the figure. We use the black dotted line as the quality threshold for each experimental result, indicating that the result below it is not properly restored, which can pose a serious threat to back-end users.
%to measure the performance in machine translation task~\cite{belghazi2018mutual}. Thus, we adopt BLEU score to compare the difference between the original sentence and reconstructed sentence. 

%Perturbation-to-Signal Ratio (PSR) is a metric utilized to quantify the relationship between a signal and perturbation within a signal processing system. A lower PSR means that the adversarial perturbation is more difficult to detect.

\noindent\textbf{Baseline Attacks.} We compare \sys with four types of baseline attacks: (1) Random Attack, (2) Vanilla UAP Attack, (3) Sync-Free UAP Attack, and (4) \jungwoo{One-hot Vector Modality-based (OVM) UAP Attack~\cite{bahramali2021robust}}. We design the random attack to transmit randomly sampled Gaussian noise into the air. It resembles classic jamming, as Gaussian jamming is widely used \cite{gao2015worst}. The vanilla UAP is an entry-level attack where multi-modality, protocol, and synchronization are not considered in crafting perturbations. The sync-free UAP attacker knows the perturbation undergoes time and phase shifts and tries to exploit such knowledge to devise shift-invariant attacks. \jungwoo{Following previous work~\cite{bahramali2021robust}, OVM UAP is trained with a dataset consisting of one-hot vector messages.} For downstream tasks, we compare \sys to random and white-box attacks.

%For other attacks~\cite{bahramali2021robust, hu2023robust, liu2023exploring}, we re-implement perturbation signals based on details provided in the papers. Since these studies did not consider the attack on downstream services, we evaluate our attack feasibility in downstream tasks by comparing  it to white-box and random attacks. %Our experimental results show that black-box attacks can achieve almost the same attack performance as white-box attacks.

\vspace{-0.2cm}
\subsection{Attacks against Multimodal JSCC} 
%In this section, we provide the experimental results to demonstrate the effectiveness of our attacks. 
%We study the black-box attack scenario, i.e., when the adversary has no knowledge of the DNN architecture or parameters used for ML-based wireless communication and downstream classifiers.

\noindent\textbf{Analysis of \sys.} 
Figure~\ref{fig:Exp_attack1} presents the reconstruction performance of the ML-based wireless transmission systems under adversarial attacks. We sweep PSR from -20dB to -10dB with steps of 2dB. We compare the performance of \sys to that of the baseline attacks. As shown in Figure~\ref{fig:Exp_attack1}, \sys dramatically deteriorates the performance metrics in the range of all PSRs. Note that ``no attack'' shows the original performance of the benign model. When applying the adversarial attacks on the image JSCC model, the PSNR drops by up to 8.04dB. For the video JSCC model, PSNR is lowered by 8.29dB on average by \sys. We see that the video model is more vulnerable to our adversarial signals than the image JSCC model. The main reason is that the video JSCC model encodes the current frame based on the previously decoded frame, thus propagating the reconstruction distortion to the next frame. For the speech model, we find that \sys degrades MSE loss by 3.91 times more than the baseline. We also observe that the BLEU score of the text JSCC model drops to a minimum of 0.338 points under \sys.

\noindent\textbf{Comparison with Baselines.} As depicted in Figure~\ref{fig:Exp_attack1}, \sys outperforms the baselines by a large margin. Against the image JSCC model, \sys lowers PSNR by up to 5.68dB more than the vanilla UAP attack and up to 4.85dB more than the sync-free UAP attack. 
\jungwoo{We see that OVM UAP attacks have similar results to random attacks.}
Without considering the multi-modality, wireless protocols, and vulnerabilities of the model, the evaluated baselines cannot critically hurt the JSCC.

\noindent\textbf{Attack Visualization.} 
As shown in Figure~\ref{fig:visual1}, we visualize the attack effect on the multimodal data reconstruction at the receiver. As seen, the JSCC decoder fails to retain semantic information. Specifically, the restored images and videos have noise-like artifacts, which dramatically reduce the users' quality of experience (QoE). Furthermore, the user cannot hear the speaker's voice in a speech sequence due to noticeable noise. The text JSCC decoder generates sentences with incorrect grammar and context, so the user cannot understand the sender's message. 
\jungwoo{In Figure~\ref{fig:pert_effect}, we present the differences in complex-valued symbols before and after the attack. We observe that Magmaw's low PSR results in minimal changes to the original signal. Additionally, as shown in Figure~\ref{fig:pert_effect2}, the variation in channel state information (CSI) due to perturbation is extremely low.}

\noindent\textbf{Analysis of Modulation.}
In Appendix Figure~\ref{fig:appendix1}, we further demonstrate the attack results of \sys for different constellation mapping methods. We confirm that \sys severely degrades the performance of JSCC models regardless of constellation type. As 64-QAM has slightly higher recovery performance than other modulations (16-QAM, QPSK) in all modalities, we confirm that the higher order of the modulation helps to increase the robustness.
%We see that the JSCC model with 64-QAM performs better than the JSCC models employing other constellation mapping methods. 
%This indicates that a higher order of modulation can create a more coherent mapping between source and channel inputs. However, after our attack was applied, the performance of all JSCC models deteriorated similarly regardless of constellation type.
%\noindent\textbf{Impact of Different Surrogate Models.} 
%\jungwoo{To Do}

\vspace{-0.2cm}
\subsection{Attacks against Downstream Tasks} 
\noindent\textbf{Analysis of \sys.} 
We evaluate the accuracy of each classifier when \sys is directed to a downstream classifier. Then, we provide a comparison with other baseline attacks. Figure~\ref{fig:Exp_attack2} shows the attack results for the video classifiers I3D~\cite{carreira2017quo}, SlowFast~\cite{feichtenhofer2019slowfast}, and TPN~\cite{yang2020temporal} and the audio-visual event classifier AVE~\cite{tian2021can}. We compare the performance of \sys to white-box and random attack scenarios. We see that the changes made in random attacks are not optimized to subvert the model. In the white-box attack scenario, the attacker has complete knowledge of the classification model. Figure~\ref{fig:Exp_attack2} presents the accuracy of each baseline for different PSRs. As shown, transmitting randomly sampled perturbations performs very poorly compared to \sys. As our attack consistently achieves comparable attack performance compared to the white-box attacks, we confirm that our UAP signals are successfully transferable to unseen downstream models. Specifically, \sys achieves an average attack success rate of 81.6$\%$, which is only 8.7$\%$ lower on average than white-box attacks.

\begin{figure}[t]
\centering
\setlength{\tabcolsep}{2pt}
\renewcommand{\arraystretch}{.5}
        \begin{tabular}{@{}cc@{}}
        \includegraphics[width=0.41\linewidth]{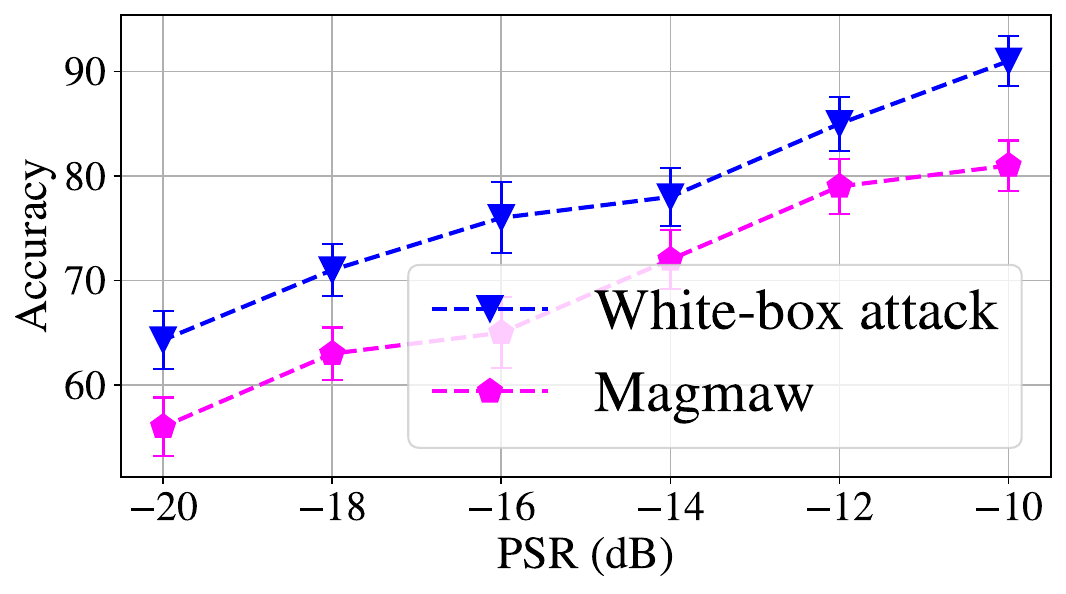} &
        \includegraphics[width=0.41\linewidth]{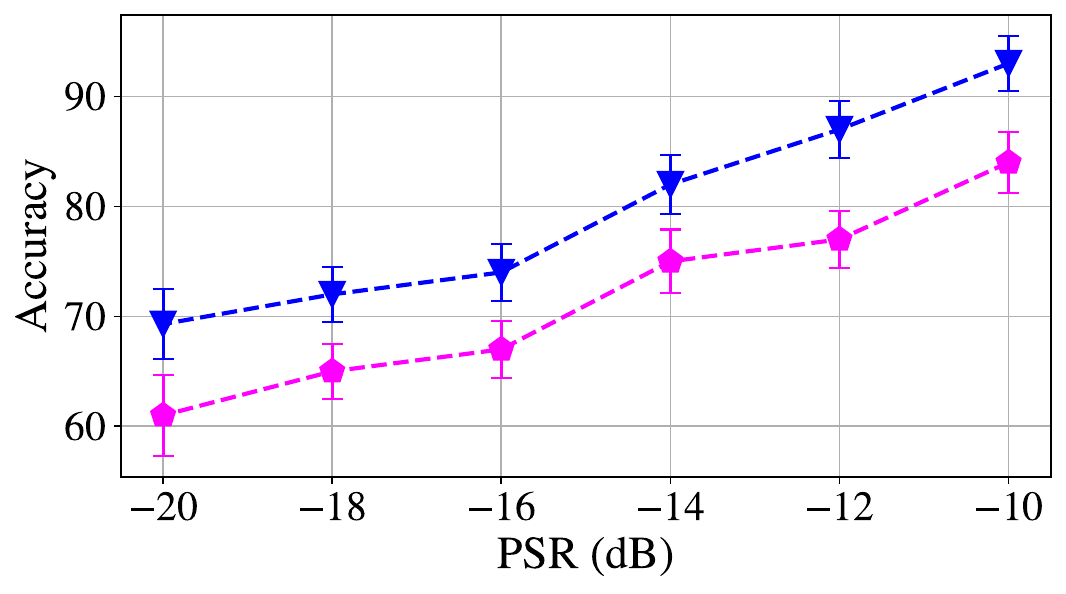} \\
        \footnotesize{(a) I3D~\cite{carreira2017quo}} &
        \footnotesize{(b) SlowFast~\cite{feichtenhofer2019slowfast}} \\
        \includegraphics[width=0.41\linewidth]{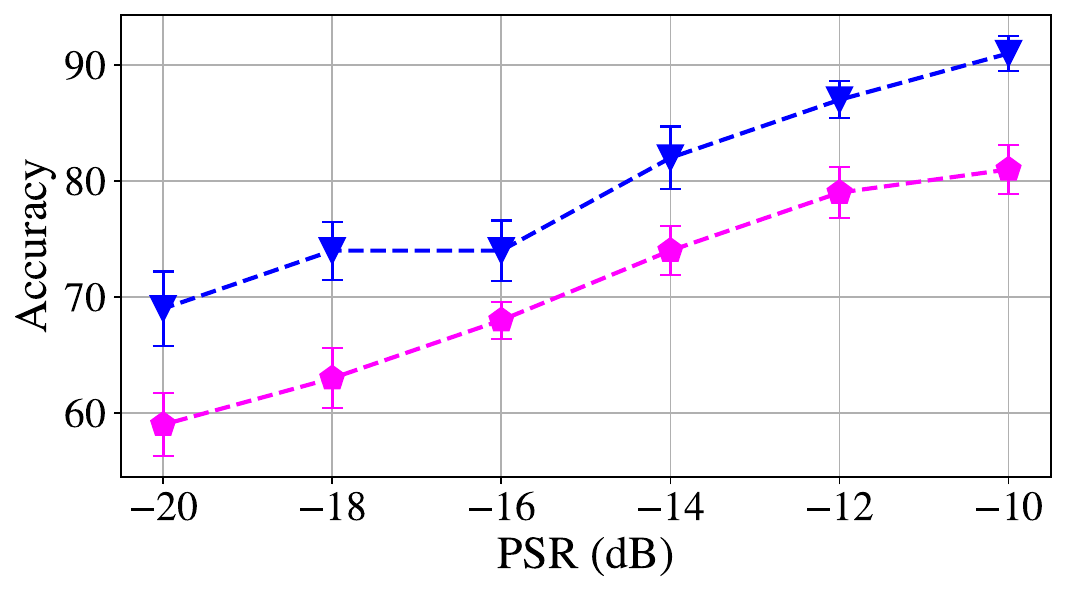} &
        \includegraphics[width=0.41\linewidth]{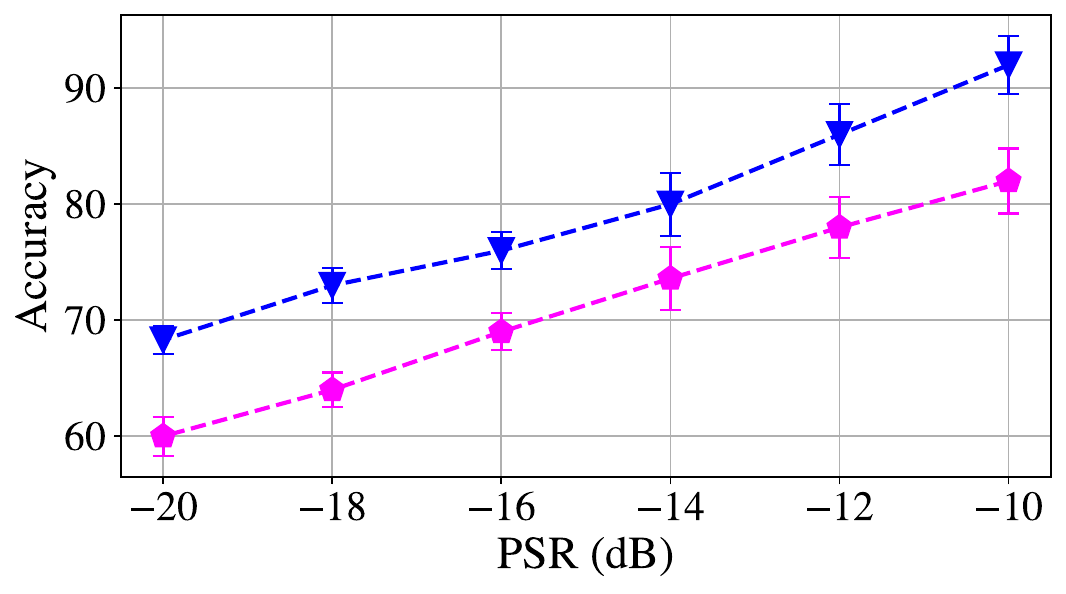} \\
        \footnotesize{(c) TPN~\cite{yang2020temporal}} &
        \footnotesize{(d) AVE~\cite{tian2021can}} \\
        \end{tabular}
  \vspace{-0.1cm}
  \caption{\jungwoo{Results for targeted UAPs on downstream tasks.}
  }
  \label{fig:targeted}
%\vspace{-0.2cm}
\end{figure}

%In Figure~\ref{fig:Exp_attack2} (b), we measure the accuracy of each classifier for different SNRs with the PSR fixed at -10dB. Our experimental results demonstrate that our black-box attacker can subvert downstream services in different SNRs. In particular, we obtain an average attack success rate of 85.2$\%$ on the AVE model, which is higher than the video classifiers. Because AVE models receive multiple input modalities, adversarial signals can interfere with larger-dimensional input signals than single-modal classifiers. 

\noindent\textbf{Analysis of Modulation.}
To analyze the influence of different constellation mapping techniques on the downstream tasks, we illustrate the attack results on the downstream classifiers when different constellation mapping methods are applied to ML-based wireless communication systems in Appendix Figure ~\ref{fig:appendix1}. Although 64-QAM can increase accuracy slightly more than other modulations, we observe that our protocol-agnostic attack defeats all modulation techniques.

\noindent\textbf{Analysis of Targeted Attacks.} 
We investigate targeted UAPs aimed at flipping the prediction of inputs to a target class. To accomplish this, we define the loss function as below:
\begin{equation} \label{eq:targeted}
\mathcal{L}^{N}_{cls} = F^{N}_{c^{*}}(\bar{X}_{N}) - \max\limits_{c \neq c^{*}}F^{N}_{c}(\bar{X}_{N}), \\
\end{equation}
where $c^{*}$ is a target class. We train the PGM by replacing $\mathcal{L}^{N}_{cls}$ in Equation~\eqref{eq:general7}. Targeted attacks gain success if and only if $\mathcal{L}^{N}_{cls}>0$.
As shown in Figure~\ref{fig:targeted}, the targeted UAPs achieve up to 82$\%$ accuracy in AVE when PSR is -10dB. 
%Thus, we change the loss function in Appendix~\ref{sec:appen_targeted}. As shown in Figure~\ref{fig:targeted}, attack performance is dependent on PSR, with up to 82$\%$ accuracy in AVE when PSR is -10dB.
Compared to untargeted UAPs, the fooling ratio is relatively low because it is more challenging to trick the predictions of all samples into a specific class~\cite{zhang2021survey}.
%See detailed results in Figure~\ref{fig:targeted}.

%This is because non-targeted transferability performs reasonably well, while targeted transferability does not. 
%\noindent\textbf{Impact of Different Surrogate Models.} 
%\jungwoo{To Do}

\vspace{-0.2cm}
\subsection{Ablation Study}
\label{sec:ablation}

\begin{figure}[t]
\centering
\setlength{\tabcolsep}{2pt}
\renewcommand{\arraystretch}{.5}
        \begin{tabular}{@{}cc@{}}
        \includegraphics[width=0.42\linewidth]{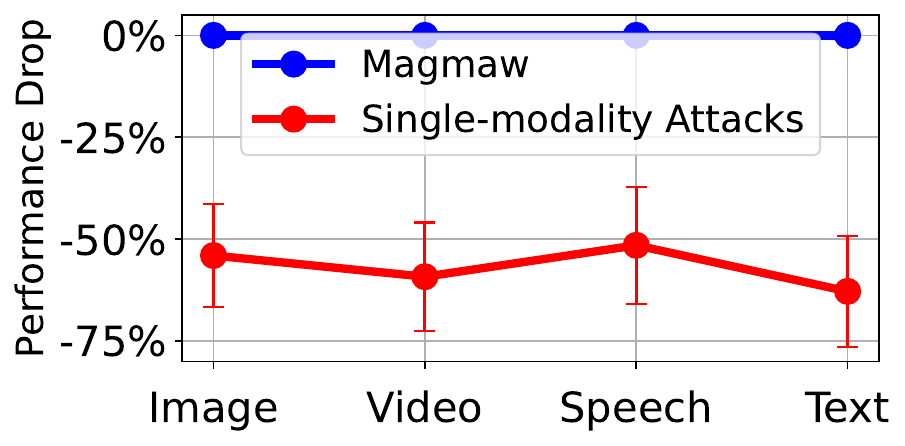} &
        \includegraphics[width=0.42\linewidth]{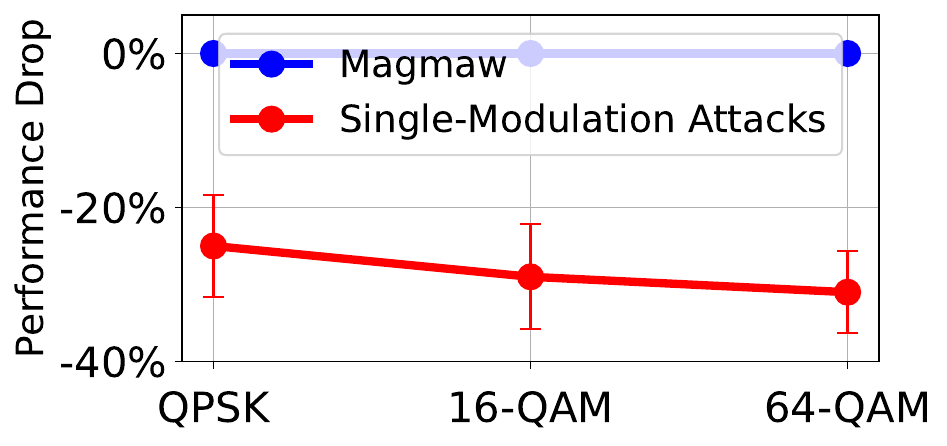} \\
        \footnotesize{(a) Modality} &
        \footnotesize{(b) Modulation Scheme} \\
        \end{tabular}
  \vspace{-0.1cm}
  \caption{Visualization of reduced attack performance when the attacker doesn't consider modality or modulation.
  }\label{fig:comp}
%\vspace{-0.5cm}
\end{figure}

\noindent\textbf{Impact of Multi-Modality.} 
To understand the importance, we study the transferability of adversarial perturbations between different modalities. For each modality, we learn a modality-specific perturbation signal and then conduct an experiment in which we inject the learned perturbation into the radio signals of other modalities. As shown in Figure~\ref{fig:comp} (a), we see that the lack of learning generalized adversarial features limits both the cross-modal and cross-model transferability.

\noindent\textbf{Effect of Modulation.} To verify the effectiveness of protocol-agnostic attacks, we conduct an ablation study on attacking JSCC without considering the constellation mapping method. As shown in Figure~\ref{fig:comp} (b), eliminating knowledge of the physical layer protocol has a significant impact on the effectiveness of the attack. We enable the transferability of adversarial examples by creating diverse modulated signals.

\begin{figure}[t]
\centering
\setlength{\tabcolsep}{2pt}
\renewcommand{\arraystretch}{.5}
        \begin{tabular}{@{}cc@{}}
        \includegraphics[width=0.42\linewidth]{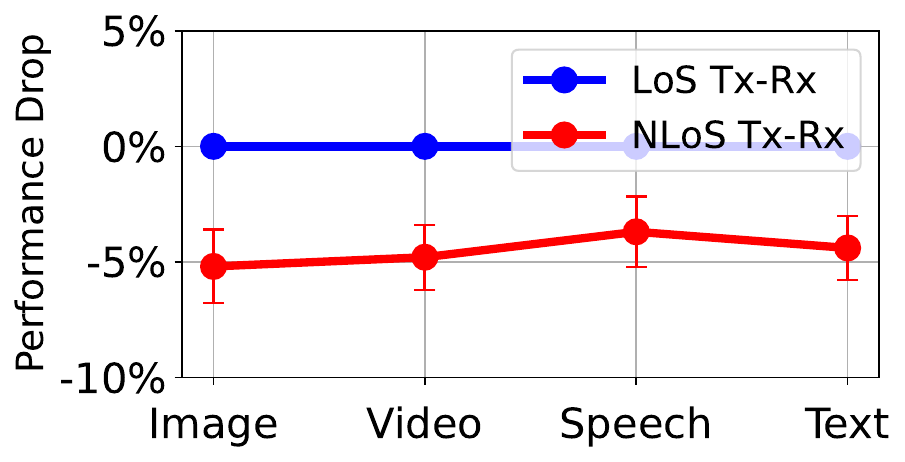} &
        \includegraphics[width=0.42\linewidth]{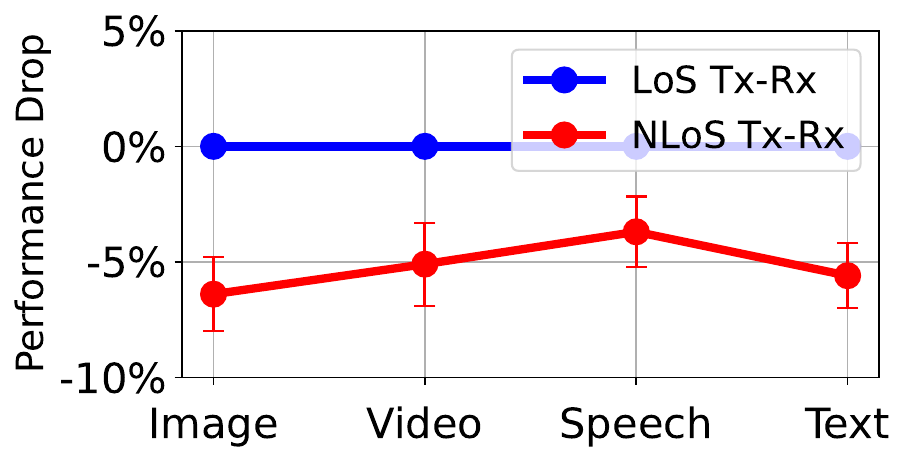} \\
        \footnotesize{(a) Before Attack} &
        \footnotesize{(b) After Attack} \\
        \end{tabular}
  \vspace{-0.1cm}
  \caption{Impact of Tx-Rx placement. We measure the performance degradation of JSCC on NLoS paths compared to the performance of JSCC on LoS paths.
  }\label{fig:comp2}
%\vspace{-0.2cm}
\end{figure}

\noindent\textbf{Impact of Tx-Rx Placement.} Each Tx-Rx scenario has different amounts of multipath because the power via the LoS path is stronger than power via the reflection path. To investigate the influence of multipath, we first compare the performance of JSCC in the two scenarios when there is no attack. As shown in Figure~\ref{fig:comp2} (a), we see that the NLoS path makes the interference issue in wireless communication, reducing the performance of JSCC by 5$\%$. We then inject our perturbations into the channel to analyze the effect of the NLoS path. As shown in Figure~\ref{fig:comp2} (b), we confirm that \sys is effective regardless of the location of Tx-Rx. A slight decrease in attack performance when the Tx/Rx path is NLoS is due to the degradation of the original performance of JSCC.

 \begin{figure*}[ht]
\centering
    \begin{tabular}{@{}cccc@{}}
        \includegraphics[width=0.17\linewidth]{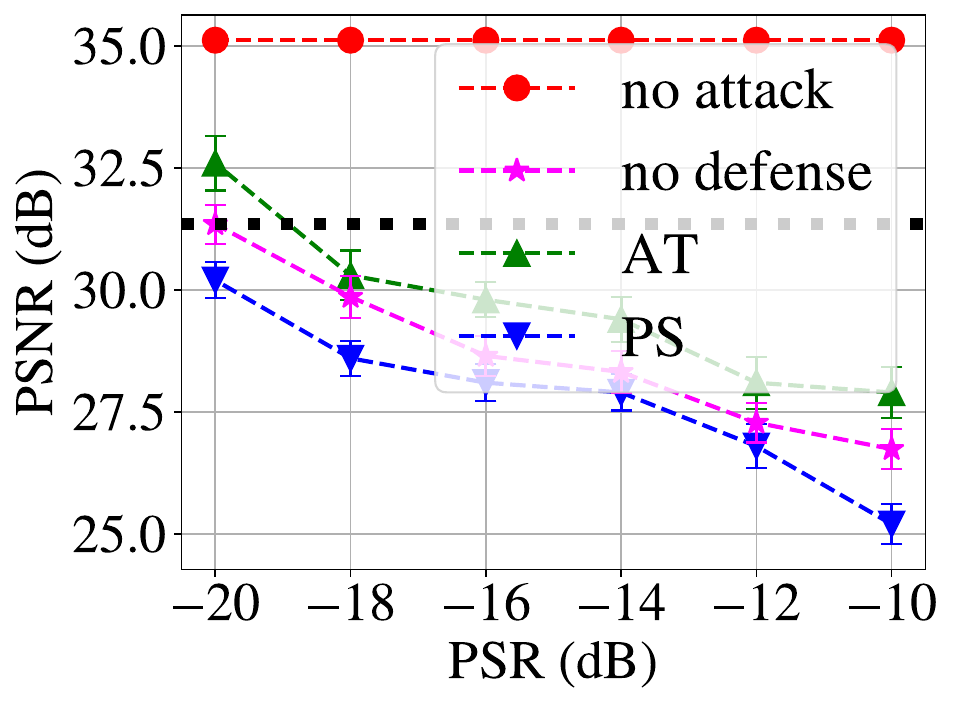} &
        \includegraphics[width=0.17\linewidth]{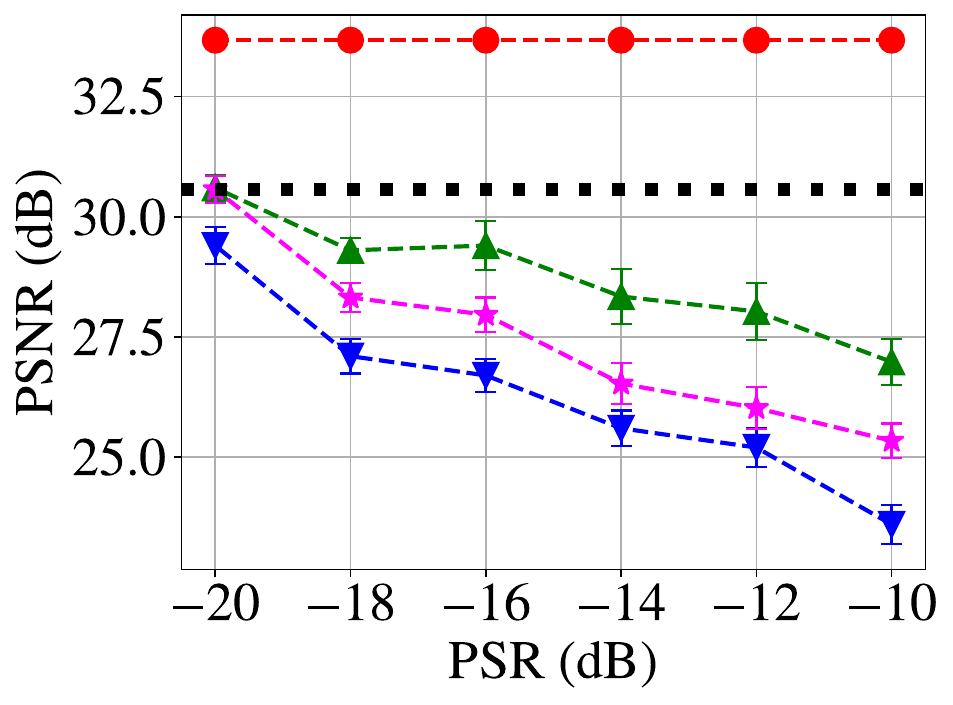} &
        \includegraphics[width=0.17\linewidth]{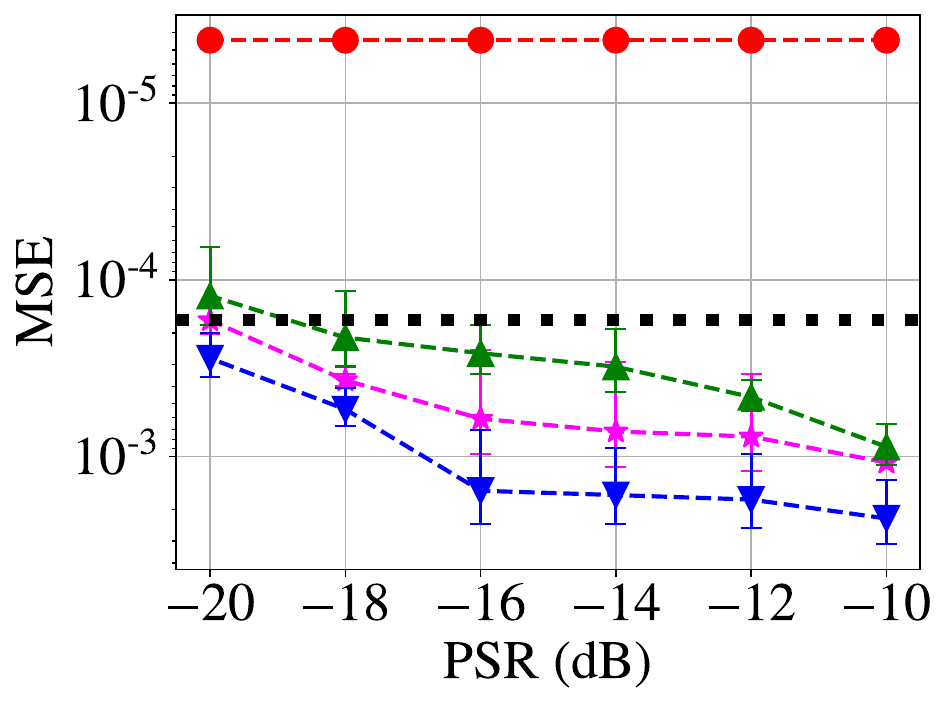} &
        \includegraphics[width=0.17\linewidth]{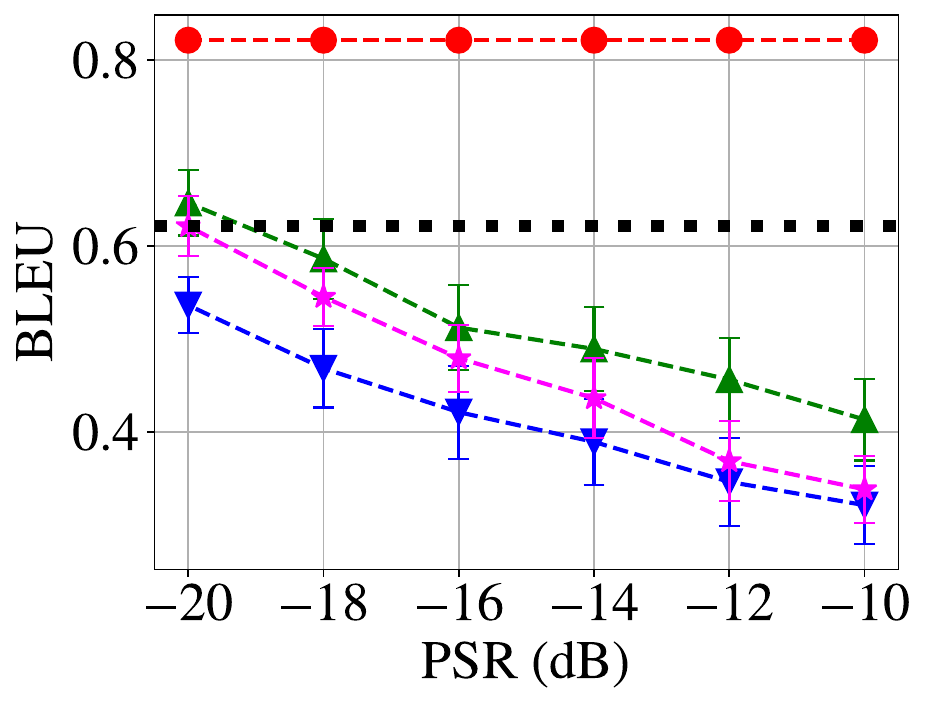} \\
        \footnotesize{Image Transmission~\cite{yang2022ofdm}}&
        \footnotesize{Video Transmission~\cite{wang2022wireless}}&
        \footnotesize{Speech Transmission~\cite{weng2021semantic}} &
        \footnotesize{Text Transmission~\cite{xie2021deep}} \\
    \end{tabular}
        \begin{tabular}{@{}c@{}}
        \footnotesize{(a) ML-based Wireless Communication Systems} \\
        \end{tabular}
%  \vspace{-0.2cm}
%  \caption{Evaluation of defenses applied to JSCC. AT and PS denote adversarial training and perturbation subtraction.}
%  \label{fig:Exp_defense1}
%  \vspace{-0.6cm}
%\end{figure*}
%\begin{figure*}[h]
%\centering
    \begin{tabular}{@{}cccc@{}}
        \includegraphics[width=0.17\linewidth]{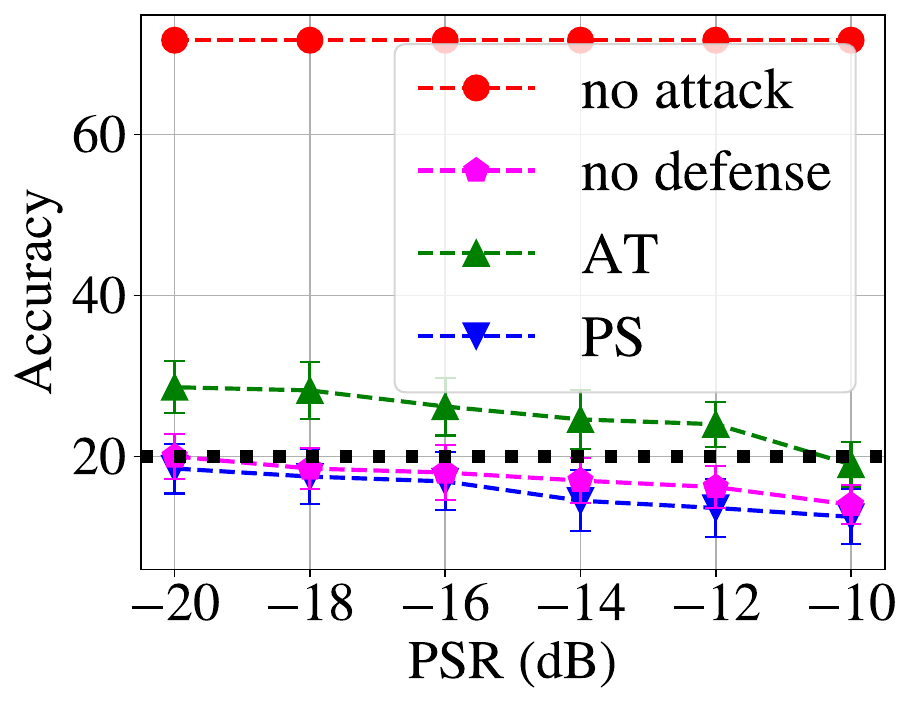} &
        \includegraphics[width=0.17\linewidth]{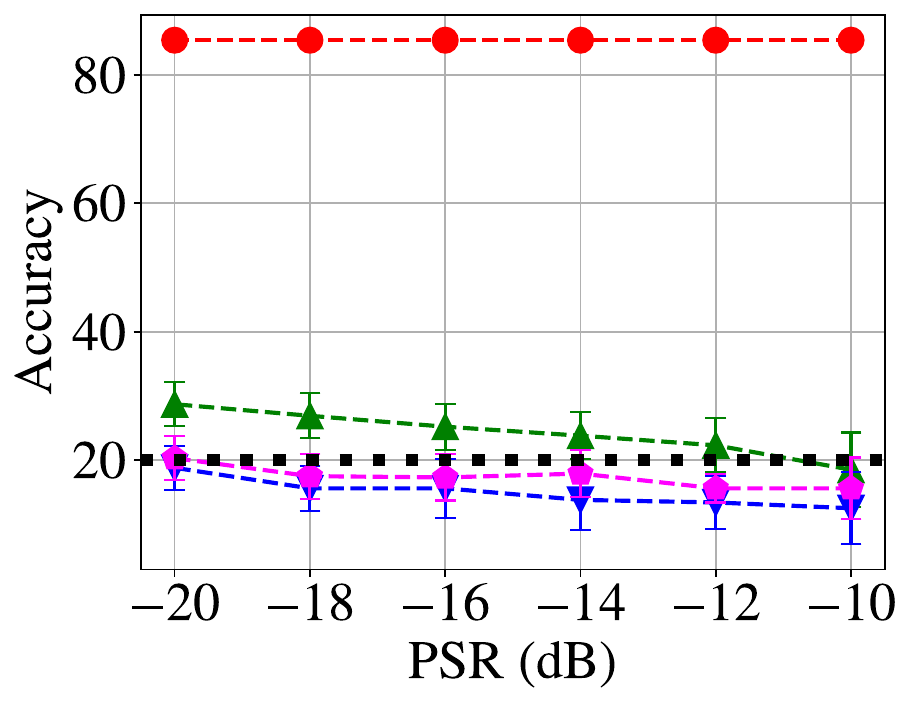} &
        \includegraphics[width=0.17\linewidth]{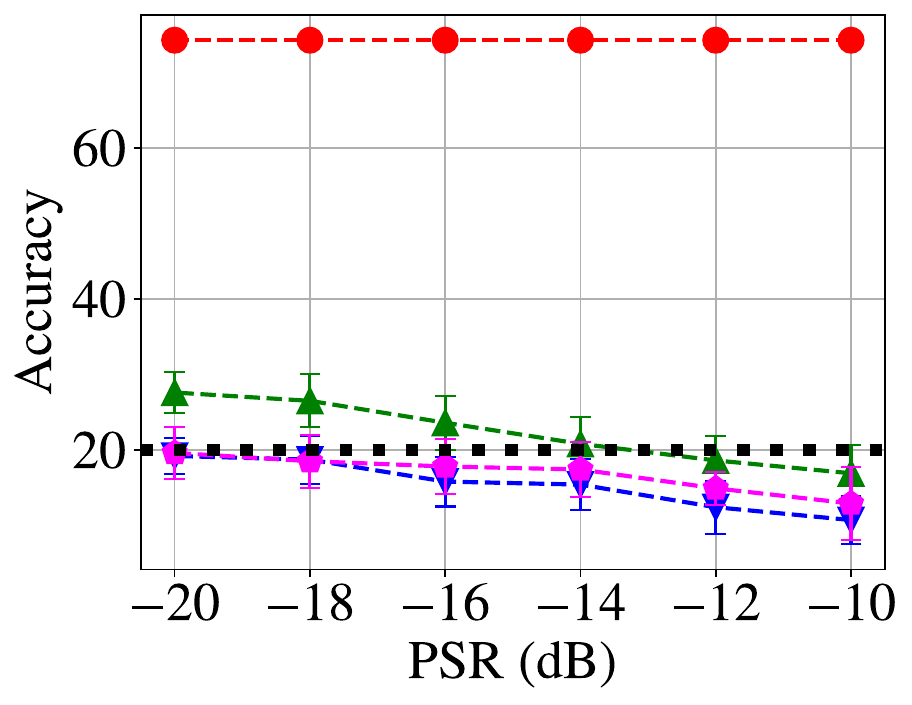} &
        \includegraphics[width=0.17\linewidth]{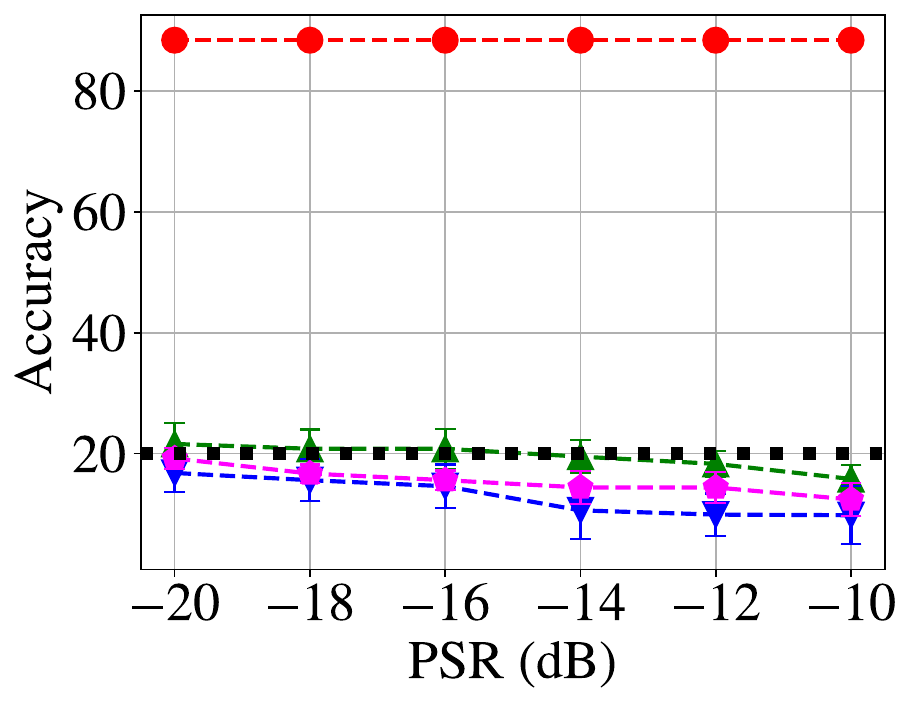} \\
        \footnotesize{I3D~\cite{carreira2017quo}}&
        \footnotesize{SlowFast~\cite{feichtenhofer2019slowfast}}&
        \footnotesize{TPN~\cite{yang2020temporal}} &
        \footnotesize{AVE~\cite{tian2021can}}\\
    \end{tabular}
        \begin{tabular}{@{}c@{}}
        \footnotesize{(b) ML-based Downstream Classification Tasks} \\
        \end{tabular}
  \vspace{-0.2cm}
  \caption{Evaluation of defenses. AT and PS denote adversarial training and perturbation subtraction.}
  \label{fig:Exp_defense}
  \vspace{-0.5cm}
\end{figure*}

\vspace{-0.1cm}
\section{Resiliency to Defense}
\label{sec:defense_eval}
%In this section, we evaluate the performance of the conventional defense mechanisms~\cite{bahramali2021robust, liu2023exploring}, namely, adversarial training and perturbation subtraction against \sys{}. 

\jungwoo{The defense performance depends on what information the defender knows about the attack formulation. From \cref{sec:adversarial_training} to \cref{sec:pdetection}, we present multiple expert defenders who know the PGM's model architecture, the channel distribution between the attacker and the receiver, and attack mechanisms illustrated in Algorithm~\ref{alg:pseudocode}. In \cref{sec:oracle_defender}, we test \sys against an oracle defender who knows every detail about \sys.}

% including transformation function.
%We consider the worst case scenario for \sys{} by assuming that there exists a strong defender who is aware of the PGM's model architecture and the attack algorithms.

%Therefore, we make several assumptions about what the defender knows about our adversarial attacks. We make slight modifications to make these defense algorithms applicable to our assumption.

\vspace{-0.2cm}
\subsection{Adversarial Training}
\label{sec:adversarial_training}
The defender aims to obtain a robust ML-based JSCC model for each modality to protect the physical layer from the \sys. Since we assume that the defender knows the model architecture of the PGM, adversarial training extends the training dataset to include all adversarial examples and then trains a JSCC model on the augmented dataset. Algorithm~\ref{alg:pseudocode2} shows detailed steps of our adversarial training. We refer to the target JSCC models as $\mathcal{J}_{Q,C,\lambda}$, and denote the PGM as $\mathcal{G}$, which is identical to the attacker's model architecture but with different model parameters.
The defender trains an ML-based JSCC model by selecting a 
%xyz: What does this mean?
%jw: I change the explanation as below.
batch from the training dataset $\mathbb{D}^{Q}$
and generating the adversarial signals controlled by several parameters of the transformation function $P_{\tau}$. We then expand the training dataset to include all adversarial examples and train the model on the augmented training dataset.

\setlength{\textfloatsep}{2pt}
\begin{algorithm}[h]
\footnotesize
\caption{Adversarial Training against \sys}\label{alg:pseudocode2}
\begin{algorithmic}
\State \textbf{Input:} Dataset $\mathbb{D}^{Q}$, ML-based JSCC model $\mathcal{J}_{Q,C,\lambda}$, PGM $\mathcal{G}$, 
\State \textbf{Output:} Robust JSCC model $\mathcal{J}_{Q,C,\lambda}$
\State $Q \gets \text{Modality}$, $C \gets \text{Modulation}$, $\lambda \gets \text{Coding rate}$
%\State $C \gets \text{Constellation mapping scheme}$
%\State $\lambda \gets \text{Coding rate}$
%\State Initialize underlying ML-based JSCC model $\mathcal{J}_{Q,C,\lambda}$

\For {\text{epoch} $l < \mathrm{MaxIter}$}
    \State $\textbf{H}_{\textbf{t}}$ is randomly sampled from channel model
    \State $\textbf{H}_{\textbf{a}}$ is sampled uniformly from training set
    \State $\mathbb{B}^{adv} \gets [ \; ]$
    \For {\text{each batch} $\textbf{B}^{Q} \in \mathbb{D}^{Q}$}
        \State Train the JSCC model $\mathcal{J}_{Q,C,\lambda}$ on $\textbf{B}^{Q}$
        \State $ z_{t} \sim \text{Uniform}(0,1)$
        \State $\tau_{l} \gets \text{randomly sampled }\{\mu, \zeta, \epsilon, \phi, \Delta t\}$
        %\State $\bar{Y}^{Q}_{t}[i,k] \gets \hat{Y}^{Q}_{t}[i,k] + P_{\tau_{l}}(\mathcal{G}(z_{t}))[i,k]$
        %\State $\bar{x}^{Q}_{t} \gets$ Equation~\ref{eq:general2}
        \State Store $P_{\tau_{l}}(\mathcal{G}(z_{t}))$ in $\mathbb{B}^{adv}$ for each data in $\textbf{B}^{Q}$
        %\State $\mathbb{B}_{adv}$.append($ P_{\tau_{l}}(\mathcal{G}(z_{t}))$)
    \EndFor
    \State $\mathbb{D}^{Q}$.append($\mathbb{D}^{Q} + \mathbb{B}^{adv}$)
\EndFor
\State \textbf{Return:} Robust JSCC model $\mathcal{J}_{Q,C,\lambda}$
\end{algorithmic}
\end{algorithm}

\noindent\textbf{ML-based Wireless System.} We validate \sys against the ML-based wireless communication systems, whose resiliency has been improved by adversarial training. As shown in Figure~\ref{fig:Exp_defense} (a), incorporating adversarial examples inside the model training process results in a lower ability to restore source data even if the underlying victim model is not attacked. Moreover, we observe that adversarial training cannot protect ML-based wireless communication from \sys. The reason is that the JSCC model has to be trained on a huge set of perturbations that the defender generates with PGM. Yet it is not feasible for the defender to train JSCC models that are resilient to all possible perturbations. Another reason is that the defender uses a PGM with different parameters from the attacker's model, so the distribution of adversarial signals generated by the two models is different.

\noindent\textbf{Downstream Tasks.} Figure~\ref{fig:Exp_defense} (b) shows the accuracy of the downstream models trained by adversarial training. Adversarial training significantly reduces the accuracy of benign models, hindering their applicability. We observe that \sys still achieves a high attack success rate even though the benign model undergoes adversarial training. This is because training a model that is universally robust to different types of perturbed signals, while being able to correctly classify input data, is a fundamentally challenging problem.

\vspace{-0.2cm}
\subsection{Perturbation Signal Subtraction}
\label{sec:perturbation_subtraction}
This defense scheme can be performed at the physical layer before the signal is passed through the OFDM receiver. Defenders aim to mitigate the effects of perturbations and reconstruct the originally transmitted signal. %Thus, the defender takes action on incoming signals that are attacked based on the knowledge of the adversary. 
As we assume that the defender has knowledge of \sys's model architecture, the receiver generates a perturbation signal via the defender's PGM and then subtracts it from the received wireless signal.

\noindent\textbf{ML-based Wireless System.} The defense results are summarized in Figure~\ref{fig:Exp_defense} (a). We observe that the source data restored by each JSCC model is more degraded than before the defense. This is because the cancellation of the adversarial signal fails and further amplifies the power of the perturbation. Even if the defender knows the structure of the PGM, the defender cannot generate exactly the same perturbation signal if the model parameters of the PGM are different.

\noindent\textbf{Downstream Tasks.} As shown in Figure~\ref{fig:Exp_defense} (b), applying perturbation signal subtraction reduces the accuracy of the downstream services by an average of 3.6$\%$. We see that the defender cannot increase the accuracy of the downstream classifier by simply subtracting an estimate of the perturbation. The accuracy of the classifier tends to depend heavily on the quality of the input source. 
%xyz: Unclear what this means
%Therefore, if the physical layer of the wireless communication system cannot be improved, the downstream service will not be able to perform the intellectual task correctly.

\begin{table}[b]
\centering
\vspace{+0.2cm}
\caption{Detection AUC of perturbation detection. The first row shows the result before fine-tuning, and the second row shows the result after fine-tuning.} \label{tab:detection2}
\resizebox{\columnwidth}{!}{
%\vspace{-0.6cm}
\begin{tabular}{ccccc}
\hline
& Image Signal
& Video Signal
& Speech Signal
& Text Signal
\\ \hline
\multirow{2}{*}{\begin{tabular}[c]{@{}c@{}}Detection\\  AUC\end{tabular}}
& 53.2$\%$       
& 52.5$\%$     
& 52.8$\%$      
& 53.4$\%$  \\ \cline{2-5}
& 55.6$\%$      
& 54.4$\%$ 
& 56.5$\%$         
& 57.1$\%$ \\ \hline
\end{tabular}}
\end{table}

\vspace{-0.2cm}
\subsection{Adversarial Perturbation Detection}
\label{sec:pdetection}
We define an input-level detection that aims to correctly find adversarially manipulated signals at the receiver side. The underlying hypothesis for this defense follows previous studies~\cite{xu2021detecting, yang2023jigsaw} that show that UAPs may leave signatures observable by ML-based anomaly detection algorithms. Based on this, we design a perturbation detector~\cite{simonyan2014very} that acts as a discriminator to distinguish the clean signal $\hat{Y}^{Q}_{t}$ from the perturbed signal $\bar{Y}^{Q}_{t}$. Leveraging the trace of UAPs, we design the binary classifier as follows. First, we train the detector offline using the training dataset constructed from the defender's PGM. In the online process, we label the received signals as adversarial attacks when the efficiency of JSCC deteriorates and include them in the training data. Finally, we fine-tune a well-trained model with newly collected data.

\begin{figure*}[h]
\centering
    \resizebox{0.6\textwidth}{!}{
    \begin{tabular}{@{}c@{}}
    \includegraphics[width=\linewidth]{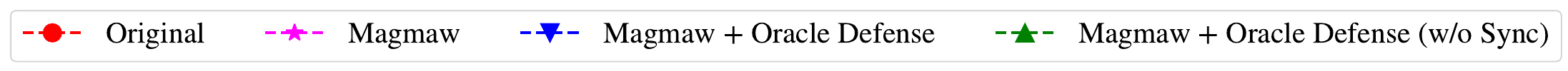} \\  
    \end{tabular}}
    \begin{tabular}{@{}cccc@{}}
    \includegraphics[width=0.17\linewidth]{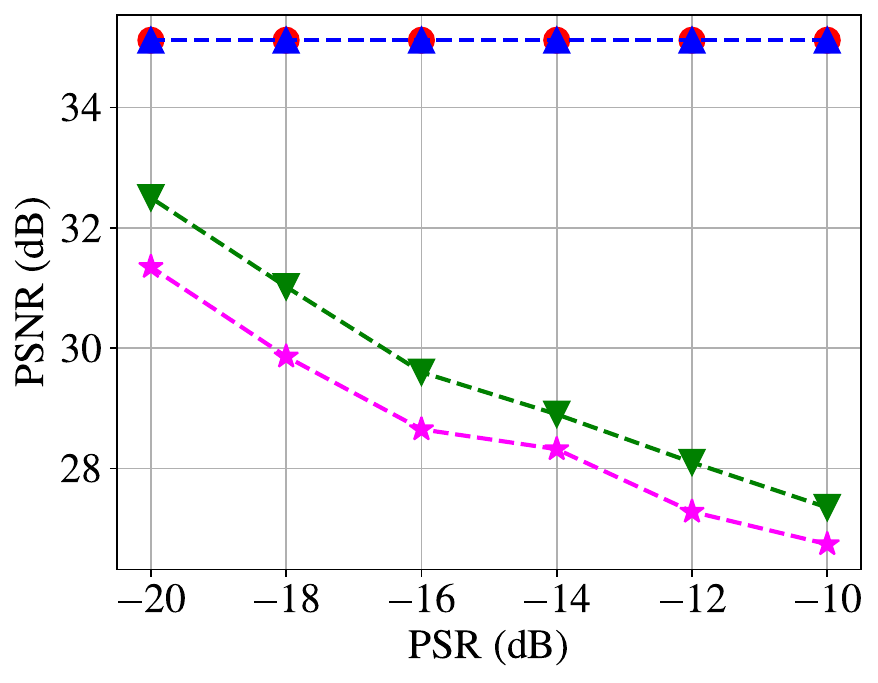} &
    \includegraphics[width=0.17\linewidth]{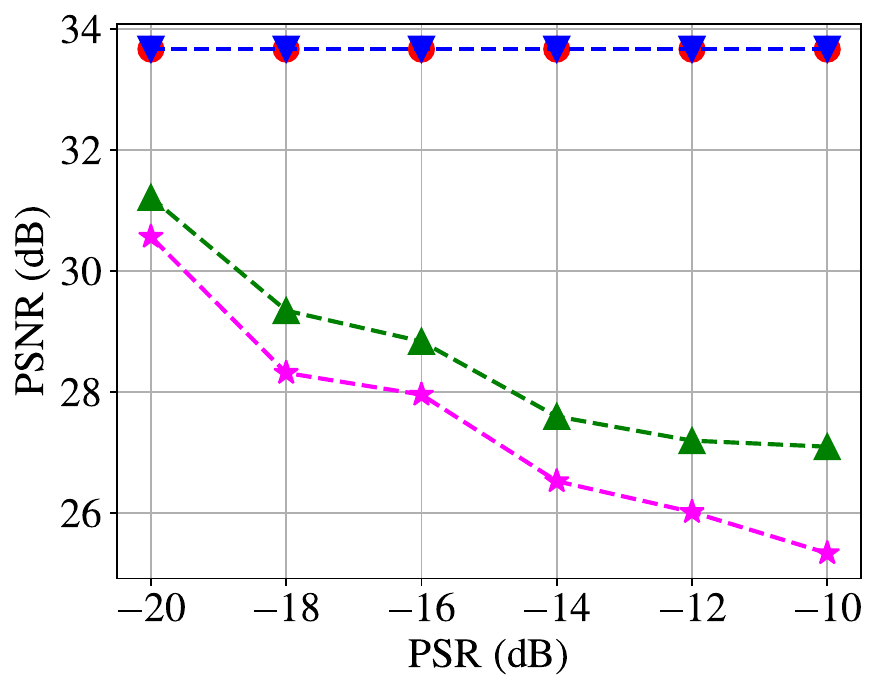} &
    \includegraphics[width=0.17\linewidth]{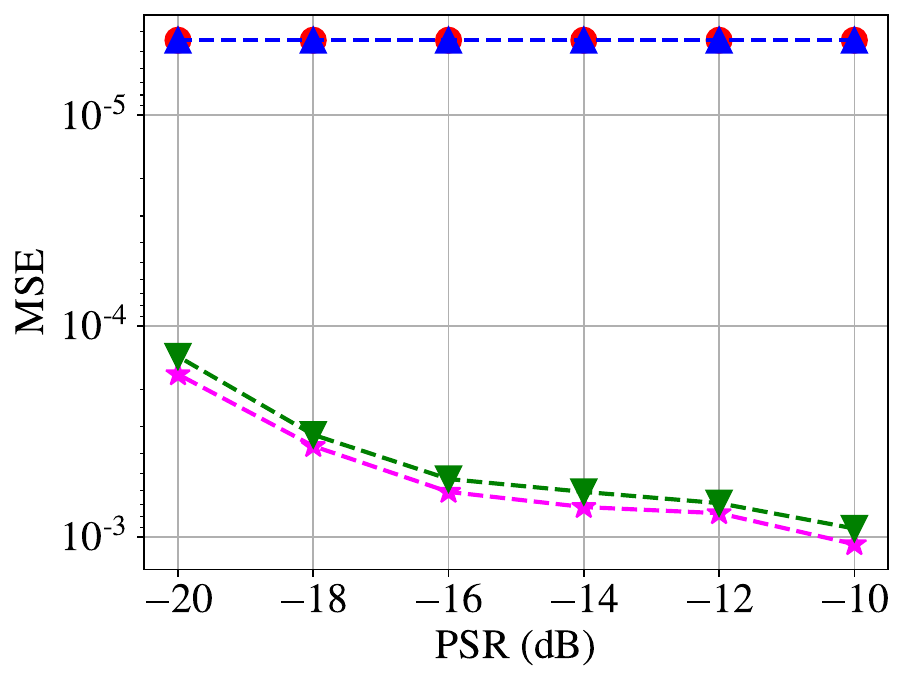} &
    \includegraphics[width=0.17\linewidth]{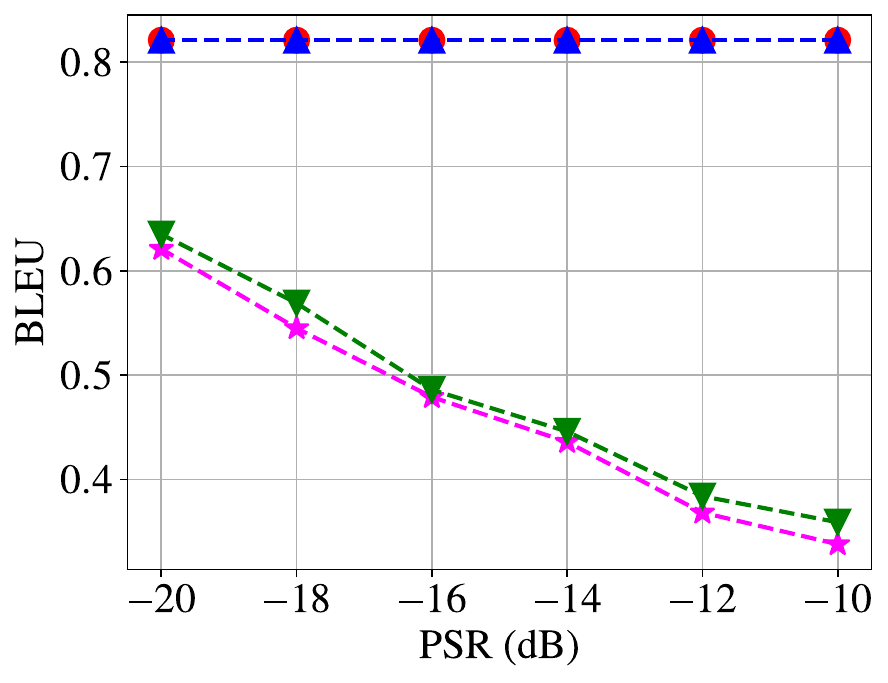} \\
    \footnotesize{Image Transmission~\cite{yang2022ofdm}}&
    \footnotesize{Video Transmission~\cite{wang2022wireless}}&
    \footnotesize{Speech Transmission~\cite{weng2021semantic}} &
    \footnotesize{Text Transmission~\cite{xie2021deep}}\\
    \end{tabular}
        \begin{tabular}{@{}c@{}}
        \footnotesize{(a) ML-based Wireless Communication Systems} \\
        \end{tabular}
    \begin{tabular}{@{}cccc@{}}
    \includegraphics[width=0.17\linewidth]{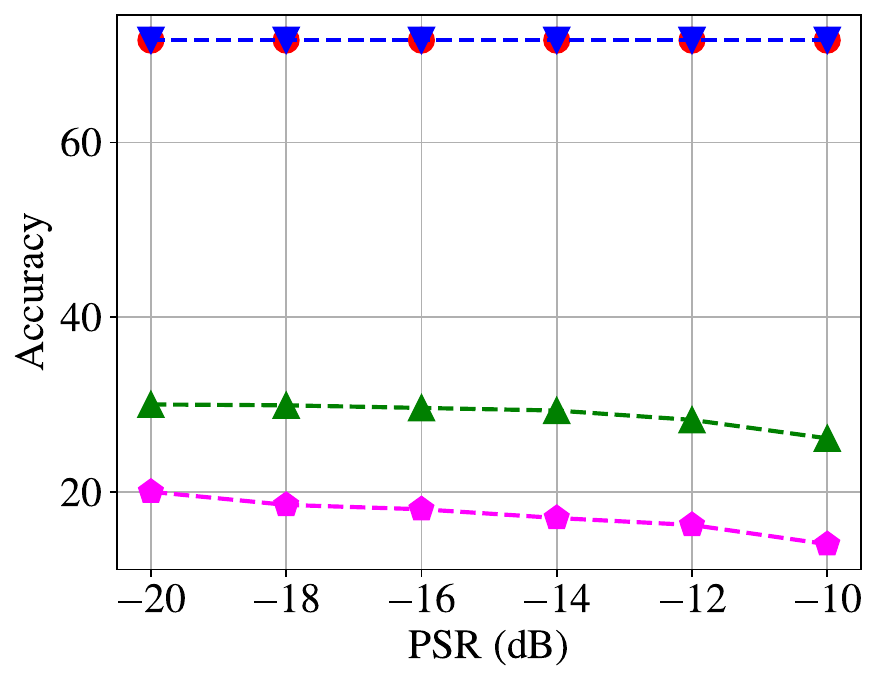} &
    \includegraphics[width=0.17\linewidth]{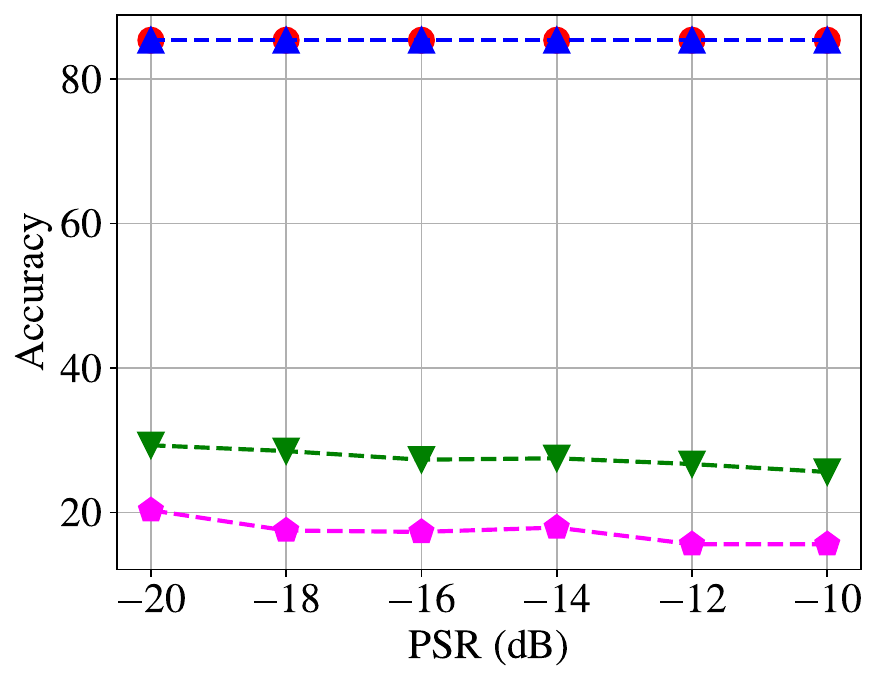} &
    \includegraphics[width=0.17\linewidth]{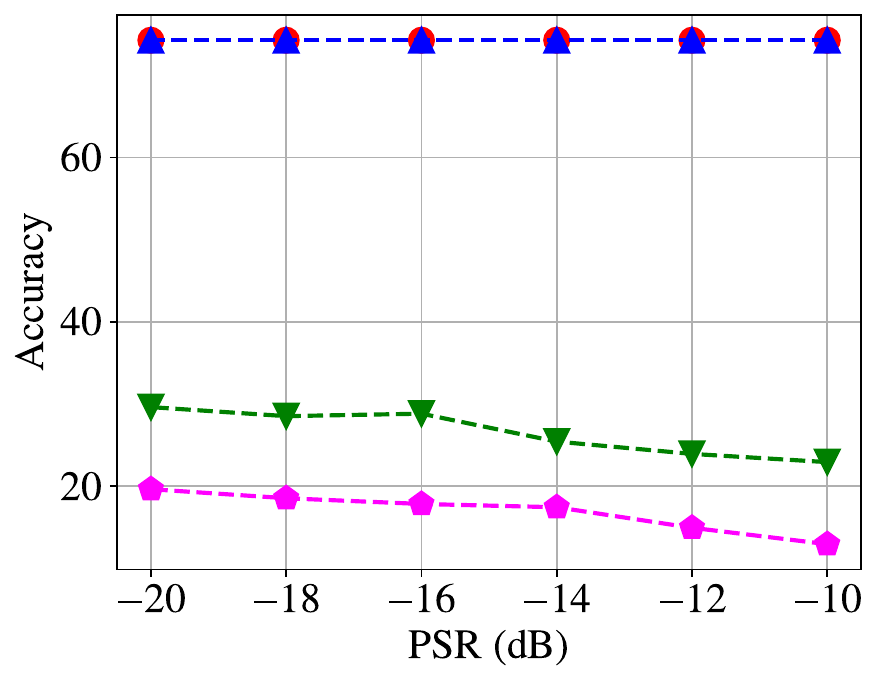} &
    \includegraphics[width=0.17\linewidth]{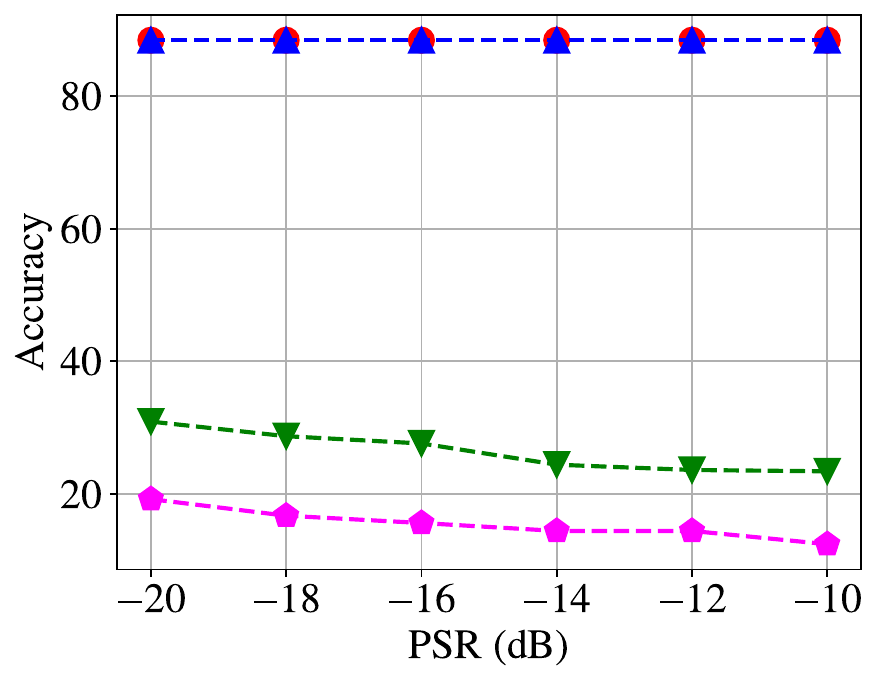} \\
    \footnotesize{I3D~\cite{carreira2017quo}}&
    \footnotesize{SlowFast~\cite{feichtenhofer2019slowfast}}&
    \footnotesize{TPN~\cite{yang2020temporal}} &
    \footnotesize{AVE~\cite{tian2021can}}\\
    \end{tabular}
        \begin{tabular}{@{}c@{}}
        \footnotesize{(b) ML-based Downstream Classification Tasks} \\
        \end{tabular}
  \vspace{-0.2cm}
  \caption{\jungwoo{Evaluation of \sys against oracle defenders in wireless systems and downstream tasks. The first and second rows are the results of JSCC and downstream tasks, respectively.}}
  \label{fig:appendix_oracle}
  \vspace{-0.5cm}
\end{figure*}

\begin{figure*}[h]
\centering
        \begin{tabular}{@{}cccc@{}}
        \includegraphics[width=0.17\linewidth]{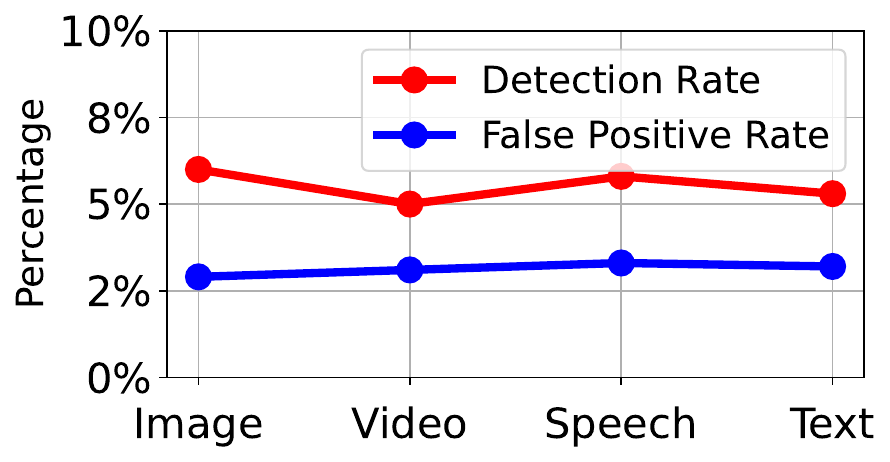} &
        \includegraphics[width=0.17\linewidth]{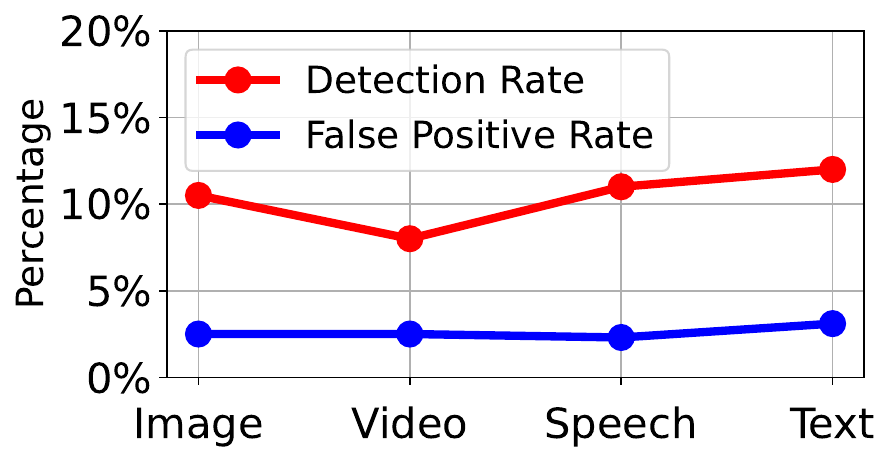} &
        \includegraphics[width=0.17\linewidth]{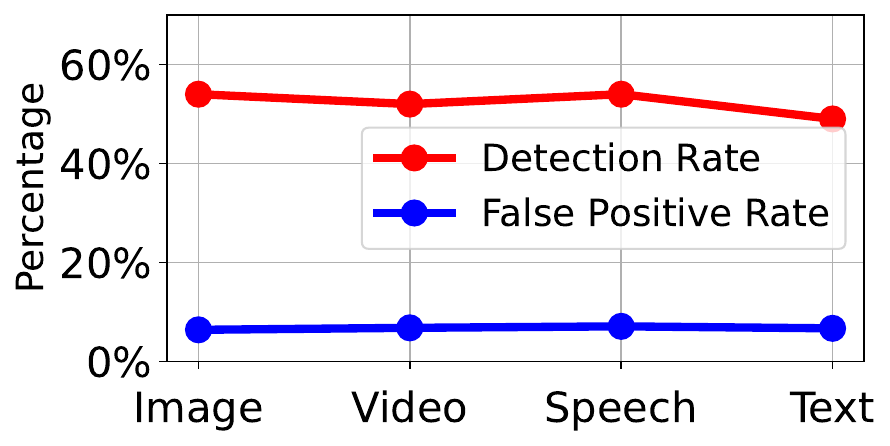} &
        \includegraphics[width=0.17\linewidth]{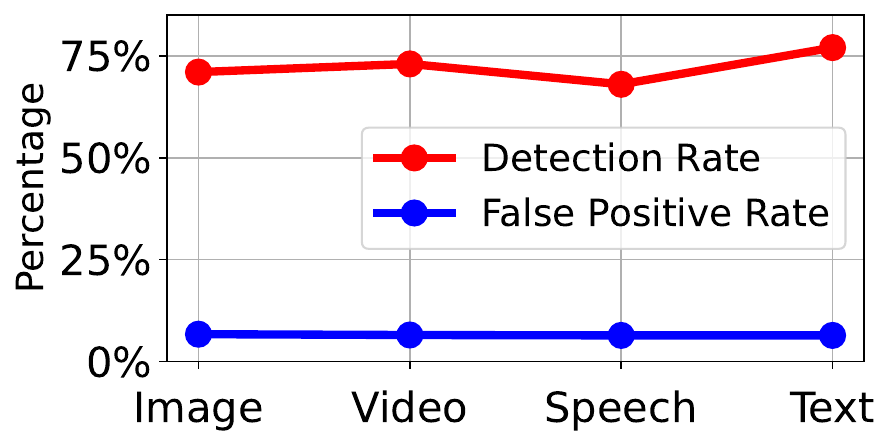} \\
        \footnotesize{Before Fine-Tuning} &
        \footnotesize{After Fine-Tuning} &
        \footnotesize{Before Fine-Tuning} &
        \footnotesize{After Fine-Tuning} \\
        \end{tabular}
        \begin{tabular}{@{}cc@{}}
        \footnotesize{(a) When \sys involves the discriminator} &
        \footnotesize{(b) When \sys doesn't involve the discriminator} \\
        \end{tabular}
  \vspace{-0.2cm}
  \caption{Detection and false positive rates of perturbation detector.
  }\label{fig:detection2}
  \vspace{-0.4cm}
\end{figure*}

Appendix Figure~\ref{fig:detection2} (a) summarizes the detection accuracy and false positive rate of our perturbation detector. It shows that \sys can bypass detection, even though the fine-tuning improves the accuracy of the detector. This is because \sys is trained to generate perturbed signals, which are indistinguishable from the clean signal, as shown in Equation~\eqref{eq:general_div} and Equation~\eqref{eq:general_div2}. For example, the fine-tuned detector only obtains up to 12$\%$ accuracy to detect perturbed radio signals in the text transmission. The results in Appendix Figure~\ref{fig:detection2} (b) have shown the detection rate of the perturbation detector when \sys conducted the training without regularization loss. The detector achieves about 75$\%$ detection rate after fine-tuning. We verify that
ML-based detectors can offer strong generalization capability in distinguishing PGM-generated perturbations. In order to train undetectable and robust UAPs, we should leverage a discriminator to enforce stealthiness.

As shown in Table~\ref{tab:detection2}, we report the Area Under Curve (AUC) of Receiver Operation Characteristic Curve (ROC) of the perturbation detection. The AUC metric shows the probability that the detector will assign a higher score to a perturbed signal than to a clean signal. We verify that the AUC results are close to the random guess, which means that \sys can achieve high undetectability. Another drawback of malware classifiers is that when an attacker changes position, the channel matrix between the attacker and the receiver also changes, requiring the defender to collect new datasets to adapt to the new environment.

%Furthermore, if the attacker changes the location of the attack device, the channel matrix between the attacker and the receiver will also be changed, forcing the defender to collect new datasets.

%Although our attack might not avoid all the detection of the defender, more than 88$\%$ of the perturbation signals can cause a significant disruptions in quality-sensitive services.

\vspace{-0.2cm}
\subsection{Oracle Defender}
\label{sec:oracle_defender}
\jungwoo{It is crucial to identify the lower bound of the effectiveness of attacks~\cite{carlini2019evaluating}. We define two strong defenders as follows:
\begin{itemize}
\setlength\itemsep{0.3em}
    \item \textit{Oracle Defender} knows complete details of \sys, including PGM architecture and parameters, $\textbf{H}_{\textbf{a}}$, 
   %xyz1: Do you mean time and frequency offset?
   %jw1: Yes, I changed "phase" to "frequency".
    the time and frequency offsets between \sys and receiver, and how \sys assigns symbols to OFDM subcarriers. 
    \item \textit{Oracle Defender without Sync Assumption} is aware of all details except the time/frequency offset. This is practical because otherwise, the receiver has to coordinate with the attacker to estimate the time/frequency offset and convey the information to the defender. 
    %but is not synchronized with \sys. Due to the broadcast nature of the wireless channel, it is challenging to be aligned with \sys's signal in time or frequency.
\end{itemize}
%An oracle defender knows all details of \sys, including the PGM architecture/parameters, the channel matrix between the attacker and receiver $\textbf{H}_{\textbf{a}}$, the time and frequency offsets between the attacker and receiver, and how \sys assigns symbols to OFDM subcarriers. 
%This is an unrealistic assumption for the following reasons: 1) the defender does not have access to the learned parameters of the PGM, and 2) the defender is not synchronized with the attacker, leading to random time and phase offsets. However, 
%Although this is an impractical assumption, we aim to identify the lower bound of the effectiveness of \sys. 
These defenders reconstruct the signal by removing the attack effect from the received wireless signal by utilizing the same perturbations generated by \sys.
%The oracle defender removes the attack effect from the received wireless signal by utilizing the same perturbations generated by \sys.
%leverages the same perturbations produced by \sys to remove the effects of the disturbance from the received radio signal and reconstruct the originally transmitted signal.
}

\noindent\textbf{ML-based Wireless System.}
\jungwoo{The oracle defender can completely neutralize \sys, as shown in Figure~\ref{fig:appendix_oracle}. These results are consistent with those reported in \cite{bahramali2021robust}, which also points out that this defense is impractical.
We further measure defense performance by eliminating the assumption that the attacker and receiver are synchronized. The oracle defense without sync assumption can only reduce the efficiency of \sys by up to 21.42$\%$. This is because lack of synchronization causes inaccuracies in the results of perturbation removal. See detailed results in Figure~\ref{fig:appendix_oracle}.}

\noindent\textbf{Downstream Tasks.} \jungwoo{We also investigate the adversarial robustness of downstream tasks in the presence of oracle defenders. We verify that addressing synchronization robustness is essential to increasing the effectiveness of the oracle defender. Specifically, without sync assumption, the oracle defender only improved the robustness of the classifier by at most 16.8$\%$. Detailed results can be found in Figure~\ref{fig:appendix_oracle}.}

\vspace{-0.1cm}
\section{Case Study}
\label{sec:case}
\subsection{Attacks on Encrypted Communication}

\begin{comment}
\begin{figure}[t]
    \centering
    \begin{tabular}{@{}c@{}}
        \includegraphics[width=0.80\columnwidth]{Fig/fig_enc.pdf} \\
    \end{tabular}
    %\vspace{-0.1cm}
    \caption{Overview of \sys on secure image transmission.}
    \label{fig:enc}
\vspace{-0.2cm}
\end{figure}
\end{comment}

\begin{figure}[t]
\centering
    \begin{tabular}{@{}ccc@{}}
        \includegraphics[width=0.34\linewidth]{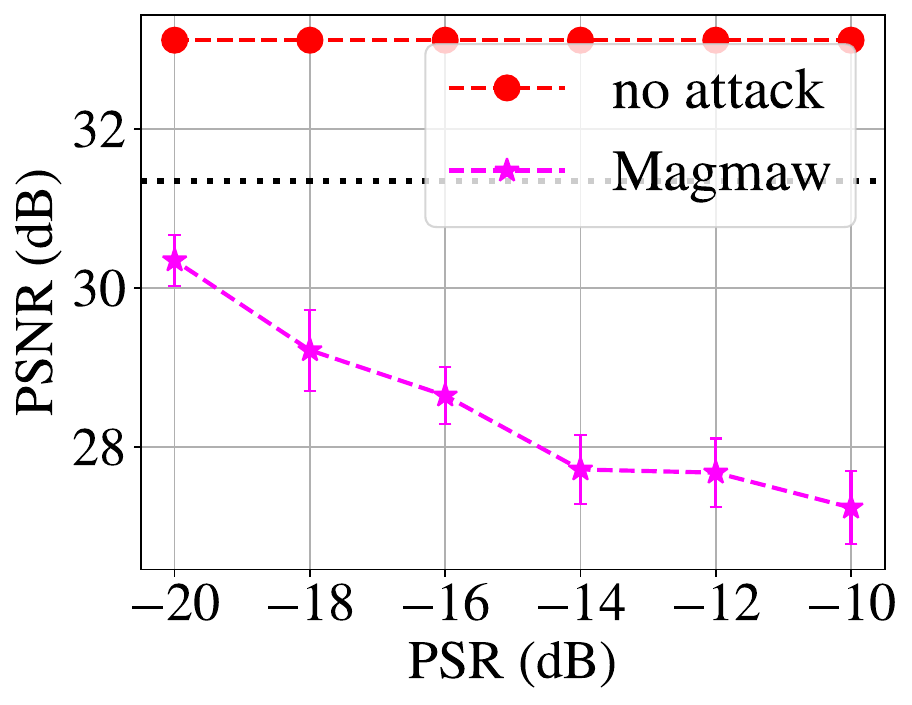} &
        \includegraphics[width=0.28\linewidth]{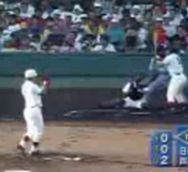} &
        \includegraphics[width=0.28\linewidth]{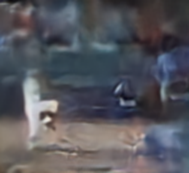} \\
        \footnotesize{(a) Results} &
        \footnotesize{(b) Transmitted Image} &
        \footnotesize{(c) Received Image} \\
    \end{tabular}
  \caption{Attack results on secure image communication. We visualize the attack results, input data, and reconstructed data at the receiver side.}
  \label{fig:case1}
  %\vspace{-0.2cm}
\end{figure}

%\subsection{Attacks on Encrypted Communication}
\label{sec:secure}
%ML-based wireless systems can suffer from privacy issues where private information can be leaked by eavesdropping channels. To solve this problem, 
Encryption schemes are commonly applied in the communication pipeline to protect users' private data~\cite{tung2023deep}. While the robustness of privacy-preserving communications with ML-based JSCC has not been investigated before, we add the encryption and decryption blocks in image JSCC to examine the impact of \sys on secure transmission, and analyze the vulnerability of encrypted signals. 

\noindent\textbf{Experiment Design.} ML-based JSCC encoder directly maps the source to the complex-valued symbols without converting it to bits. To handle this new feature, public-key encryption with LWE~\cite{regev2009lattices} rather than classical AES-based schemes~\cite{mitchell2005security} is applied in JSCC~\cite{tung2023deep}. In public-key encryption, any user can send encrypted messages to the receiver using the public key. Thus, we assume that the adversary knows the public key, but does not know the secret key. %The adversary will encrypt the perturbations using a public key, modulate the ciphertexts, and inject them into receiver's antenna.

%While the robustness of privacy-preserving communications with ML-based JSCC has not been investigated before, we add the encryption and decryption blocks in our system model to examine the impact of \sys on secure image transmission and analyze the vulnerability of encrypted wireless signal. 

%In public-key encryption, any users can send encrypted messages to the receiver using the public key. Thus, we assume that the adversary knows public key, but does not know secret key. The receiver can reconstruct the transmitted data

%We consider privacy-preserving image transmission~\cite{tung2022deep} with learning with errors (LWE)-based encryption~\cite{regev2009lattices}, where the transmitter and receiver share a public key, but only the receiver knows the secret key. 
%We set a security level of 192 according to ~\cite{lindner2011better}. 

\noindent\textbf{Attack Results.} Figure~\ref{fig:case1} shows the attack results on the secure communication system. We see that the OFDM symbols carrying the ciphertext of the image data are vulnerable to our perturbation signal. Specifically, \sys lowers the performance of secure image transmission by up to 5.88dB. This is because the decrypted output of ciphertext operations in LWE is similar to performing plaintext operations on the original plaintext data. By showing that secure communication does not provide adversarial robustness, we promote the need for new defense techniques against \sys.

%Even though encryption of channel inputs can change the arrangement of symbols, the adversary can encrypt the perturbations using a public key, modulate the ciphertexts through OFDM transmitter and inject them into the receiver's antenna. 

\vspace{-0.2cm}
\subsection{Attacks on ML Systems with Channel States as Input}
\label{sec:csi}
\jungwoo{
%There are several ML-based wireless systems that rely on a preamble (a predefined OFDM modulated training sequence) to obtain CSI.
Standard-defined preambles~\cite{biswas2004channel} are widely used in ML-driven wireless systems to obtain the CSI. 
%We discuss how \sys subverts the normal functions of these models. 
We consider two ML models that are also used as target models in RAFA~\cite{liu2023exploring}: (a) DLoc~\cite{ayyalasomayajula2020deep} performs localization task via CSI received from four fixed access points, and (b) FIRE~\cite{liu2021fire} takes the CSI of the uplink channel as input and then predicts the downlink CSI. It can address the overhead of feedback exchange in the Frequency Domain Duplex (FDD) system.}

\begin{figure}[t]
    \centering
    \begin{tabular}{@{}cc@{}}
        \includegraphics[height=0.118\textheight]{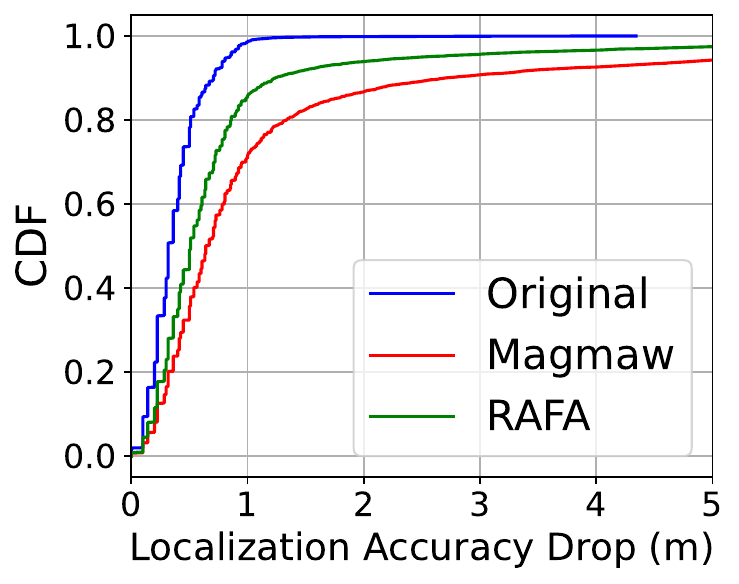} &
        \includegraphics[height=0.118\textheight]{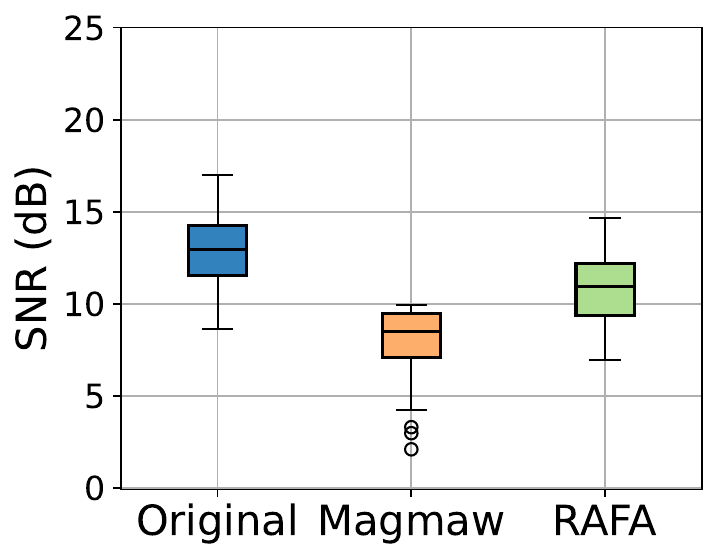} \\
        %\vspace{-0.1cm}
        \footnotesize{(a) Localization (DLoc~\cite{ayyalasomayajula2020deep}}) &
        \footnotesize{(b) Downlink Prediction (FIRE~\cite{liu2021fire}})
    \end{tabular}
    \vspace{-0.2cm}
    \caption{\jungwoo{Attack results on CSI-based models (PSR is -12dB).}}
    \label{fig:case_csi}
    %\vspace{-0.5cm}
\end{figure}

\begin{figure}[t]
    \centering
    \begin{tabular}{@{}cc@{}}
        \includegraphics[width=0.43\linewidth]{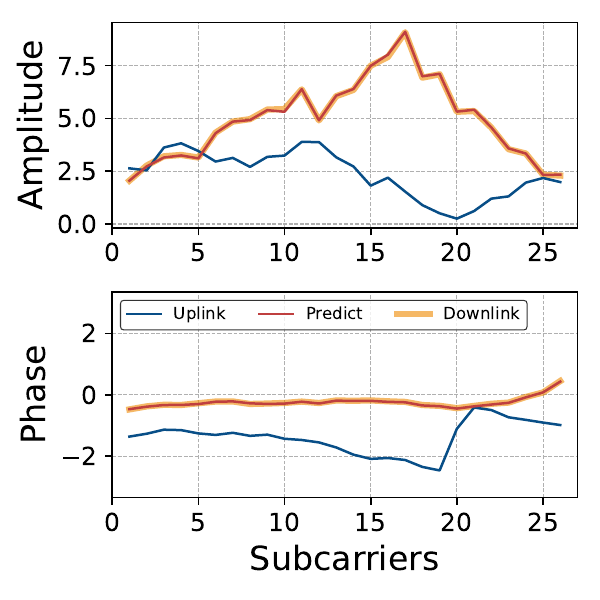} &
        \includegraphics[width=0.43\linewidth]{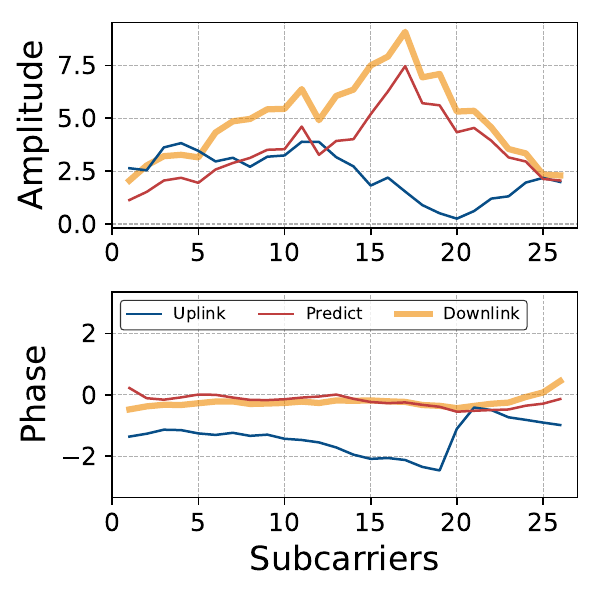} \\
        %\vspace{-0.1cm}
        \footnotesize{(a) Before Attack} &
        \footnotesize{(b) After Attack}
    \end{tabular}
    \vspace{-0.2cm}
    \caption{\jungwoo{Attack results on FIRE~\cite{liu2021fire}. We visualize the channel amplitude and phase. FIRE takes the uplink channel (blue line) as input and predicts the downlink channel (red line) that is expected to be the same as the ground truth (yellow line).}}
    \label{fig:channel_estimation}
    %\vspace{-0.2cm}
\end{figure}

%is a channel estimation model designed to address the overhead of feedback exchange in Frequency Domain Duplex (FDD) system. FIRE takes the CSI of the uplink channel as input and predicts the downlink CSI through the variational autoencoder.}

\noindent\textbf{Experiment Design.} \jungwoo{In the experiment setup, the sender allocates preambles $Y^{\mathcal{P}}_{t}$ to OFDM subcarriers (i.e., 64 subcarriers for 20MHz) and then transmits them to the receiver. Here, $\mathcal{P}$ denotes the preamble. Since \sys injects attack signals into the channel according to Equation~\eqref{eq:general4}, the received preamble is $\bar{Y}^{\mathcal{P}}_{t}$. Thus, the receiver acquires the perturbed CSI via $\mathcal{H}^{\mathcal{P}}_{t}=\bar{Y}^{\mathcal{P}}_{t}/Y^{\mathcal{P}}_{t}$ and feeds it into the target ML model. 
We re-implement the target models via details provided in their papers, as well as open source~\cite{dloc} and the dataset~\cite{shepard2016understanding}. We further improve the robustness of DLoc and FIRE through adversarial training proposed by RAFA. For a fair comparison, we utilize surrogate models used in RAFA to train \sys.}

%The receiver obtains the perturbed CSI via $\mathcal{H}^{\mathcal{P}}_{t}=\bar{Y}^{\mathcal{P}}_{t}/Y^{\mathcal{P}}_{t}$. The target model $\mathcal{A}$ then predicts the output $\mathcal{A}(\mathcal{H}^{\mathcal{P}}_{t})$ from the perturbed CSI.

\noindent\textbf{Attack Results.} \jungwoo{We compare our attack with RAFA for a comprehensive evaluation. As shown in Figure~\ref{fig:case_csi} (a), DLoc achieves 0.71m and 1.03m localization errors at the 90th and 99th percentile. However, when \sys is present, the results go up to 2.7m and 9.8m. We can observe that \sys outperforms RAFA by $1.73\times$ on average. Figure~\ref{fig:case_csi} (b) describes the accuracy of channel estimation by FIRE. The SNR measures the similarity between the estimated downlink channel and the ground truth. We confirm that \sys drops the SNR of the predicted channel by 3.8dB more than RAFA. The observed underperformance of RAFA can be due to the lack of consideration for improving the robustness of adversarial attacks during the training. In contrast, \sys achieves high robustness by leveraging a discriminator and diversity loss that increases the variability of perturbation patterns.
%
%RAFA focuses solely on the preamble and transmits short-length adversarial signals, so when time misalignment occurs, only part of the preamble is perturbed. In contrast, since \sys includes data symbols in its attack target, it transmits a longer adversarial signal. Even if temporal misalignment exists, \sys can perturb almost the entire preamble. 
%
To provide an intuition of the attack effect, we visualize an example of channel estimation in Figure~\ref{fig:channel_estimation}. 
%In Appendix Figure~\ref{fig:channel_estimation} (a), FIRE outputs red downlink estimates from blue uplink CSI, showing close results with real downlink channels in the absence of attacks. 
Figure~\ref{fig:channel_estimation} (b) shows that \sys causes subcarriers to have symbols that are significantly different from the actual symbols.}

\vspace{-0.1cm}
\section{Conclusion and Future Works}
%This paper studies practical physical-layer attacks on ML-based wireless systems. 
We present \sys, a novel attack framework to subvert semantic communication for AI-native networks.
%by superimposing a small amount of attack signals on a victim radio channel. 
%\sys further exploits the wireless systems to subvert downstream recognition systems. \sys generates modality-invariant perturbations with the input signal, and is robust to various changes in physical layer conditions (e.g. modulation, coding rate, unknown channel effects, etc.). \sys is resistant to time and phase shifts caused by the lack of synchronization with the legitimate transmitter and receiver. 
Our results show that the \sys is feasible in the real world, and can degrade the performance of both wireless communication and downstream tasks simultaneously. \sys maintains a high attack success rate by evading several defenses.
In case studies, we evaluate \sys on encrypted communication and \jungwoo{CSI modality-based models}, proving that \sys is transferable.

While \sys demonstrates great success, the perturbation designed in this work needs to be powered by software-defined radios for flexible generation of the UAPs. Promising future work is to explore new attack methods, 
e.g., intelligent reflecting surfaces~\cite{chen2023metawave}, to induce small adversarial perturbations.
%rapid phase changes of multimodal radio signals. 
Another area of future research is to establish practical defense techniques to prevent the proposed attacks.

\vspace{-0.1cm}
\section*{Acknowledgement}
We thank the NDSS reviewers for their valuable feedback. This work is partially supported by the U.S. Army/Department of Defense award number W911NF2020267 and Google Ph.D. Fellowship.

\vspace{-0.4cm}

\bibliographystyle{plain}
\bibliography{main}

\begin{thebibliography}{100}

\bibitem{supplement}
Supplementary document for ``magmaw: Modality-agnostic adversarial attacks on machine learning-based wireless communication systems``.
\newblock \url{https://drive.google.com/file/d/1Ol7Ebn9JspTNvTUmeNPB-gs5woCIs7jF/view?usp=drive_link}.

\bibitem{abdulla2014toward}
Ahmed~EAA Abdulla, Zubair~Md Fadlullah, Hiroki Nishiyama, Nei Kato, Fumie Ono, and Ryu Miura.
\newblock Toward fair maximization of energy efficiency in multiple uas-aided networks: A game-theoretic methodology.
\newblock {\em IEEE Transactions on Wireless Communications}, 2014.

\bibitem{abu2013uplink}
Najah Abu-Ali, Abd-Elhamid~M Taha, Mohamed Salah, and Hossam Hassanein.
\newblock Uplink scheduling in lte and lte-advanced: Tutorial, survey and evaluation framework.
\newblock {\em IEEE Communications surveys \& tutorials}, 16(3):1239--1265, 2013.

\bibitem{acemoglu20205g}
Alperen Acemoglu, Jan Krieglstein, Darwin~G Caldwell, Francesco Mora, Luca Guastini, Matteo Trimarchi, et~al.
\newblock 5g robotic telesurgery: Remote transoral laser microsurgeries on a cadaver.
\newblock {\em IEEE Transactions on Medical Robotics and Bionics}, 2(4):511--518, 2020.

\bibitem{albaseer2020performance}
Abdullatif Albaseer, Bekir~Sait Ciftler, and Mohamed~M Abdallah.
\newblock Performance evaluation of physical attacks against e2e autoencoder over rayleigh fading channel.
\newblock In {\em 2020 IEEE International Conference on Informatics, IoT, and Enabling Technologies}, pages 177--182, 2020.

\bibitem{alzantot2018did}
Moustafa Alzantot, Bharathan Balaji, and Mani Srivastava.
\newblock Did you hear that? adversarial examples against automatic speech recognition.
\newblock {\em arXiv preprint arXiv:1801.00554}, 2018.

\bibitem{Nokia}
Faycal~Ait Aoudia, Jakob Hoydis, Alvaro Valcarce, and Harish Viswanathan.
\newblock Toward a 6g ai-native air interface.
\newblock https://onestore.nokia .com/asset/210299, 2021.

\bibitem{apruzzese2023real}
Giovanni Apruzzese, Hyrum~S Anderson, Savino Dambra, David Freeman, Fabio Pierazzi, and Kevin Roundy.
\newblock “real attackers don't compute gradients”: bridging the gap between adversarial ml research and practice.
\newblock In {\em 2023 IEEE Conference on Secure and Trustworthy Machine Learning (SaTML)}, pages 339--364. IEEE, 2023.

\bibitem{apruzzese2022wild}
Giovanni Apruzzese, Rodion Vladimirov, Aliya Tastemirova, and Pavel Laskov.
\newblock Wild networks: Exposure of 5g network infrastructures to adversarial examples.
\newblock {\em IEEE Transactions on Network and Service Management}, 19(4):5312--5332, 2022.

\bibitem{aras2017selective}
Emekcan Aras, Nicolas Small, Gowri~Sankar Ramachandran, St{\'e}phane Delbruel, Wouter Joosen, and Danny Hughes.
\newblock Selective jamming of lorawan using commodity hardware.
\newblock In {\em Proceedings of the 14th EAI International Conference on Mobile and Ubiquitous Systems: Computing, Networking and Services}, pages 363--372, 2017.

\bibitem{dloc}
Roshan Ayyalasomayajula.
\newblock Dloc network architecture codes.
\newblock https://github.com/ucsdwcsng/, 2024.

\bibitem{ayyalasomayajula2020deep}
Roshan Ayyalasomayajula, Aditya Arun, Chenfeng Wu, Sanatan Sharma, Abhishek~Rajkumar Sethi, Deepak Vasisht, and Dinesh Bharadia.
\newblock Deep learning based wireless localization for indoor navigation.
\newblock In {\em Proceedings of the 26th Annual International Conference on Mobile Computing and Networking}, pages 1--14, 2020.

\bibitem{bahramali2021robust}
Alireza Bahramali, Milad Nasr, Amir Houmansadr, Dennis Goeckel, and Don Towsley.
\newblock Robust adversarial attacks against dnn-based wireless communication systems.
\newblock In {\em Proceedings of the 2021 ACM SIGSAC Conference on Computer and Communications Security}, 2021.

\bibitem{banerji2013ieee}
Sourangsu Banerji and Rahul~Singha Chowdhury.
\newblock On ieee 802.11: wireless lan technology.
\newblock {\em arXiv preprint arXiv:1307.2661}, 2013.

\bibitem{belghazi2018mutual}
Mohamed~Ishmael Belghazi, Aristide Baratin, Sai Rajeshwar, Sherjil Ozair, Yoshua Bengio, Aaron Courville, and Devon Hjelm.
\newblock Mutual information neural estimation.
\newblock In {\em International conference on machine learning}, pages 531--540. PMLR, 2018.

\bibitem{biggio2018wild}
Battista Biggio and Fabio Roli.
\newblock Wild patterns: Ten years after the rise of adversarial machine learning.
\newblock In {\em Proceedings of the 2018 ACM SIGSAC Conference on Computer and Communications Security}, pages 2154--2156, 2018.

\bibitem{biswas2004channel}
A~Biswas, S~Lakshmipathi, and S~Sandhu.
\newblock Channel estimation techniques with long training sequence for ieee802. 11a.
\newblock In {\em 2004 International Conference on Signal Processing and Communications, 2004. SPCOM'04.}, pages 136--139. IEEE, 2004.

\bibitem{blossom2004gnu}
Eric Blossom.
\newblock Gnu radio: tools for exploring the radio frequency spectrum.
\newblock {\em Linux journal}, 2004(122):4, 2004.

\bibitem{bourtsoulatze2019deep}
Eirina Bourtsoulatze, David~Burth Kurka, and Deniz G{\"u}nd{\"u}z.
\newblock Deep joint source-channel coding for wireless image transmission.
\newblock {\em IEEE Transactions on Cognitive Communications and Networking}, 5(3):567--579, 2019.

\bibitem{cao2023stylefool}
Yuxin Cao, Xi~Xiao, Ruoxi Sun, Derui Wang, Minhui Xue, and Sheng Wen.
\newblock Stylefool: Fooling video classification systems via style transfer.
\newblock In {\em 2023 IEEE Symposium on Security and Privacy (SP)}, 2023.

\bibitem{carlini2019evaluating}
Nicholas Carlini, Anish Athalye, Nicolas Papernot, Wieland Brendel, Jonas Rauber, Dimitris Tsipras, Ian Goodfellow, Aleksander Madry, and Alexey Kurakin.
\newblock On evaluating adversarial robustness.
\newblock {\em arXiv preprint arXiv:1902.06705}, 2019.

\bibitem{carlini2017towards}
Nicholas Carlini and David Wagner.
\newblock Towards evaluating the robustness of neural networks.
\newblock In {\em 2017 ieee symposium on security and privacy (sp)}, pages 39--57. Ieee, 2017.

\bibitem{carreira2017quo}
Joao Carreira and Andrew Zisserman.
\newblock Quo vadis, action recognition? a new model and the kinetics dataset.
\newblock In {\em proceedings of the IEEE Conference on Computer Vision and Pattern Recognition}, 2017.

\bibitem{chang2022rovisq}
Jung-Woo Chang, Mojan Javaheripi, Seira Hidano, and Farinaz Koushanfar.
\newblock Rovisq: Reduction of video service quality via adversarial attacks on deep learning-based video compression.
\newblock In {\em NDSS}, 2023.

\bibitem{chang2023videoflip}
Jung-Woo Chang, Mojan Javaheripi, and Farinaz Koushanfar.
\newblock Videoflip: Adversarial bit flips for reducing video service quality.
\newblock In {\em ACM/IEEE Design Automation Conference (DAC)}, pages 1--6, 2023.

\bibitem{chang2023netflick}
Jung-Woo Chang, Nojan Sheybani, et~al.
\newblock Netflick: Adversarial flickering attacks on deep learning based video compression.
\newblock {\em arXiv preprint arXiv:2304.01441}, 2023.

\bibitem{chen2023metawave}
Xingyu Chen, Zhengxiong Li, Baicheng Chen, Yi~Zhu, Chris~Xiaoxuan Lu, Zhengyu Peng, Feng Lin, Wenyao Xu, Kui Ren, and Chunming Qiao.
\newblock Metawave: Attacking mmwave sensing with meta-material-enhanced tags.
\newblock In {\em NDSS Symposium}, volume 2023, 2023.

\bibitem{choi2020stargan}
Yunjey Choi, Youngjung Uh, Jaejun Yoo, and Jung-Woo Ha.
\newblock Stargan v2: Diverse image synthesis for multiple domains.
\newblock In {\em Proceedings of the IEEE/CVF conference on computer vision and pattern recognition}, pages 8188--8197, 2020.

\bibitem{chowdhury20206g}
Mostafa~Zaman Chowdhury, Md~Shahjalal, Shakil Ahmed, and Yeong~Min Jang.
\newblock 6g wireless communication systems: Applications, requirements, technologies, challenges, and research directions.
\newblock {\em IEEE Open Journal of the Communications Society}, 1:957--975, 2020.

\bibitem{erni2022adaptover}
Simon Erni, Martin Kotuliak, Patrick Leu, Marc Roeschlin, and Srdjan Capkun.
\newblock Adaptover: adaptive overshadowing attacks in cellular networks.
\newblock In {\em Proceedings of the 28th Annual International Conference on Mobile Computing And Networking}, pages 743--755, 2022.

\bibitem{feichtenhofer2019slowfast}
Christoph Feichtenhofer, Haoqi Fan, Jitendra Malik, and Kaiming He.
\newblock Slowfast networks for video recognition.
\newblock In {\em Proceedings of the IEEE/CVF international conference on computer vision}, 2019.

\bibitem{flowers2019evaluating}
Bryse Flowers, R~Michael Buehrer, and William~C Headley.
\newblock Evaluating adversarial evasion attacks in the context of wireless communications.
\newblock {\em IEEE Transactions on Information Forensics and Security}, 2019.

\bibitem{gao2015worst}
Jie Gao, Sergiy~A Vorobyov, Hai Jiang, and H~Vincent Poor.
\newblock Worst-case jamming on mimo gaussian channels.
\newblock {\em IEEE Transactions on Signal Processing}, 63(21):5821--5836, 2015.

\bibitem{goodfellow2014generative}
Ian Goodfellow, Jean Pouget-Abadie, Mehdi Mirza, Bing Xu, David Warde-Farley, Sherjil Ozair, Aaron Courville, and Yoshua Bengio.
\newblock Generative adversarial nets.
\newblock {\em Advances in NeurIPS}, 27, 2014.

\bibitem{gulrajani2020towards}
Ishaan Gulrajani, Colin Raffel, and Luke Metz.
\newblock Towards gan benchmarks which require generalization.
\newblock {\em ICLR}, 2020.

\bibitem{halperin2010802}
Daniel Halperin, Wenjun Hu, Anmol Sheth, and David Wetherall.
\newblock 802.11 with multiple antennas for dummies.
\newblock {\em ACM SIGCOMM Computer Communication Review}, 40(1):19--25, 2010.

\bibitem{he20206g}
Jianhua He, Kun Yang, and Hsiao-Hwa Chen.
\newblock 6g cellular networks and connected autonomous vehicles.
\newblock {\em IEEE Network}, 2020.

\bibitem{he2016deep}
Kaiming He, Xiangyu Zhang, Shaoqing Ren, and Jian Sun.
\newblock Deep residual learning for image recognition.
\newblock In {\em Proceedings of the IEEE conference on computer vision and pattern recognition}, 2016.

\bibitem{hoydis2022sionna}
Jakob Hoydis, Sebastian Cammerer, Fay{\c{c}}al~Ait Aoudia, Avinash Vem, Nikolaus Binder, Guillermo Marcus, and Alexander Keller.
\newblock Sionna: An open-source library for next-generation physical layer research.
\newblock {\em arXiv preprint arXiv:2203.11854}, 2022.

\bibitem{hu2023robust}
Qiyu Hu, Guangyi Zhang, Zhijin Qin, Yunlong Cai, Guanding Yu, and Geoffrey~Ye Li.
\newblock Robust semantic communications with masked vq-vae enabled codebook.
\newblock {\em IEEE Transactions on Wireless Communications}, 2023.

\bibitem{hwang2008ofdm}
Taewon Hwang, Chenyang Yang, Gang Wu, Shaoqian Li, and Geoffrey~Ye Li.
\newblock Ofdm and its wireless applications: A survey.
\newblock {\em IEEE transactions on Vehicular Technology}, 58(4):1673--1694, 2008.

\bibitem{jin2023pla}
Zizhi Jin, Xiaoyu Ji, Yushi Cheng, Bo~Yang, Chen Yan, and Wenyuan Xu.
\newblock Pla-lidar: Physical laser attacks against lidar-based 3d object detection in autonomous vehicle.
\newblock In {\em 2023 IEEE Symposium on Security and Privacy (SP)}, pages 1822--1839. IEEE, 2023.

\bibitem{kim2021adversarial}
Brian Kim, Y~Sagduyu, Tugba Erpek, and Sennur Ulukus.
\newblock Adversarial attacks on deep learning based mmwave beam prediction in 5g and beyond.
\newblock In {\em IEEE Statistical Signal Processing Workshop (SSP)}, 2021.

\bibitem{kim2021channel}
Brian Kim, Yalin~E Sagduyu, Kemal Davaslioglu, Tugba Erpek, and Sennur Ulukus.
\newblock Channel-aware adversarial attacks against deep learning-based wireless signal classifiers.
\newblock {\em IEEE Transactions on Wireless Communications}, 21(6):3868--3880, 2021.

\bibitem{kingma2014adam}
Diederik~P Kingma and Jimmy Ba.
\newblock Adam: A method for stochastic optimization.
\newblock {\em arXiv preprint arXiv:1412.6980}, 2014.

\bibitem{li2024practical}
Changming Li, Mingjing Xu, Yicong Du, Limin Liu, Cong Shi, Yan Wang, Hongbo Liu, and Yingying Chen.
\newblock Practical adversarial attack on wifi sensing through unnoticeable communication packet perturbation.
\newblock In {\em Proceedings of the 30th Annual International Conference on Mobile Computing and Networking}, pages 373--387, 2024.

\bibitem{li2021mixed}
Haoran Li and Wei Lu.
\newblock Mixed cross entropy loss for neural machine translation.
\newblock In {\em International Conference on Machine Learning}, 2021.

\bibitem{li2023boosting}
Zeju Li, Xinghan Liu, Guoshun Nan, Jinfei Zhou, Xinchen Lyu, Qimei Cui, and Xiaofeng Tao.
\newblock Boosting physical layer black-box attacks with semantic adversaries in semantic communications.
\newblock In {\em ICC 2023-IEEE International Conference on Communications}, 2023.

\bibitem{lin2021wireless}
Peng Lin, Qingyang Song, F~Richard Yu, Dan Wang, Abbas Jamalipour, and Lei Guo.
\newblock Wireless virtual reality in beyond 5g systems with the internet of intelligence.
\newblock {\em IEEE Wireless Communications}, 28(2):70--77, 2021.

\bibitem{lin2022overview}
Xingqin Lin.
\newblock An overview of 5g advanced evolution in 3gpp release 18.
\newblock {\em IEEE Communications Standards Magazine}, 6(3):77--83, 2022.

\bibitem{liu2023sst}
Chenyao Liu, Jiejie Guo, Yimeng Zhang, Wenjun Xu, and Yiming Liu.
\newblock Sst-v: A scalable semantic a scalable semantic transmission framework for video.
\newblock {\em ZTE COMMUNICATIONS}, 21(2), 2023.

\bibitem{liu2022physical}
Jianwei Liu, Yinghui He, Chaowei Xiao, Jinsong Han, Le~Cheng, and Kui Ren.
\newblock Physical-world attack towards wifi-based behavior recognition.
\newblock In {\em IEEE INFOCOM 2022-IEEE Conference on Computer Communications}, pages 400--409. IEEE, 2022.

\bibitem{liu2023time}
Jianwei Liu, Yinghui He, Chaowei Xiao, Jinsong Han, and Kui Ren.
\newblock Time to think the security of wifi-based behavior recognition systems.
\newblock {\em IEEE Transactions on Dependable and Secure Computing}, 2023.

\bibitem{liu2021fire}
Zikun Liu, Gagandeep Singh, Chenren Xu, and Deepak Vasisht.
\newblock Fire: enabling reciprocity for fdd mimo systems.
\newblock In {\em Proceedings of the 27th Annual International Conference on Mobile Computing and Networking}, pages 628--641, 2021.

\bibitem{liu2023exploring}
Zikun Liu, Changming Xu, Emerson Sie, Gagandeep Singh, and Deepak Vasisht.
\newblock Exploring practical vulnerabilities of machine learning-based wireless systems.
\newblock In {\em 20th USENIX Symposium on Networked Systems Design and Implementation (NSDI 23)}, pages 1801--1817, 2023.

\bibitem{lovisotto2021slap}
Giulio Lovisotto, Henry Turner, Ivo Sluganovic, Martin Strohmeier, and Ivan Martinovic.
\newblock $\{$SLAP$\}$: Improving physical adversarial examples with $\{$Short-Lived$\}$ adversarial perturbations.
\newblock In {\em 30th USENIX Security Symposium (USENIX Security 21)}, pages 1865--1882, 2021.

\bibitem{manoj2021adversarial}
BR~Manoj, Meysam Sadeghi, and Erik~G Larsson.
\newblock Adversarial attacks on deep learning based power allocation in a massive mimo network.
\newblock In {\em International Conference on Communications}. IEEE, 2021.

\bibitem{mitchell2005security}
CHJC Mitchell and Changhua He.
\newblock Security analysis and improvements for ieee 802.11 i.
\newblock In {\em NDSS}, pages 90--110, 2005.

\bibitem{moosavi2017universal}
Seyed-Mohsen Moosavi-Dezfooli, Alhussein Fawzi, Omar Fawzi, and Pascal Frossard.
\newblock Universal adversarial perturbations.
\newblock In {\em Proceedings of the IEEE conference on computer vision and pattern recognition}, pages 1765--1773, 2017.

\bibitem{nan2023physical}
Guoshun Nan, Zhichun Li, Jinli Zhai, Qimei Cui, Gong Chen, Xin Du, Xuefei Zhang, Xiaofeng Tao, Zhu Han, and Tony~QS Quek.
\newblock Physical-layer adversarial robustness for deep learning-based semantic communications.
\newblock {\em IEEE journal on selected areas in communications}, 2023.

\bibitem{o2017introduction}
Timothy O’shea and Jakob Hoydis.
\newblock An introduction to deep learning for the physical layer.
\newblock {\em IEEE Transactions on Cognitive Communications and Networking}, 3(4):563--575, 2017.

\bibitem{pirayesh2022jamming}
Hossein Pirayesh and Huacheng Zeng.
\newblock Jamming attacks and anti-jamming strategies in wireless networks: A comprehensive survey.
\newblock {\em IEEE communications surveys \& tutorials}, 24(2):767--809, 2022.

\bibitem{qin2021semantic}
Zhijin Qin, Xiaoming Tao, Jianhua Lu, Wen Tong, and Geoffrey~Ye Li.
\newblock Semantic communications: Principles and challenges.
\newblock {\em arXiv preprint arXiv:2201.01389}, 2021.

\bibitem{Qualcommwhite}
Qualcomm.
\newblock Qualcomm whitepaper vision market-drivers and research directions on the path to 6g.
\newblock https://www.qualcomm.com/content/dam/ qcomm-martech/dm-assets/documents/Qualcomm-Whitepaper-Vision-market-drivers-and-research-directions-on-the-path-to-6G.pdf, 2022.

\bibitem{quiring2020adversarial}
Erwin Quiring, David Klein, Daniel Arp, Martin Johns, and Konrad Rieck.
\newblock Adversarial preprocessing: Understanding and preventing $\{$Image-Scaling$\}$ attacks in machine learning.
\newblock In {\em 29th USENIX Security Symposium (USENIX Security 20)}, pages 1363--1380, 2020.

\bibitem{regev2009lattices}
Oded Regev.
\newblock On lattices, learning with errors, random linear codes, and cryptography.
\newblock {\em Journal of the ACM (JACM)}, 56(6):1--40, 2009.

\bibitem{saad2019vision}
Walid Saad, Mehdi Bennis, and Mingzhe Chen.
\newblock A vision of 6g wireless systems: Applications, trends, technologies, and open research problems.
\newblock {\em IEEE network}, 34(3):134--142, 2019.

\bibitem{sadeghi2018adversarial}
Meysam Sadeghi and Erik~G Larsson.
\newblock Adversarial attacks on deep-learning based radio signal classification.
\newblock {\em IEEE Wireless Communications Letters}, 8(1):213--216, 2018.

\bibitem{sadeghi2019physical}
Meysam Sadeghi and Erik~G Larsson.
\newblock Physical adversarial attacks against end-to-end autoencoder communication systems.
\newblock {\em IEEE Communications Letters}, 23(5):847--850, 2019.

\bibitem{sambhwani2022transitioning}
Sharad Sambhwani, Zdravko Boos, Sidharth Dalmia, Arman Fazeli, Bertram Gunzelmann, Anatoliy Ioffe, Murali Narasimha, Francesco Negro, Laxminarayana Pillutla, and John Zhou.
\newblock Transitioning to 6g part 1: Radio technologies.
\newblock {\em IEEE Wireless Communications}, 2022.

\bibitem{sato2021dirty}
Takami Sato, Junjie Shen, Ningfei Wang, Yunhan Jia, Xue Lin, and Qi~Alfred Chen.
\newblock Dirty road can attack: Security of deep learning based automated lane centering under $\{$Physical-World$\}$ attack.
\newblock In {\em 30th USENIX Security Symposium}, pages 3309--3326, 2021.

\bibitem{shepard2016understanding}
Clayton Shepard, Jian Ding, Ryan~E Guerra, and Lin Zhong.
\newblock Understanding real many-antenna mu-mimo channels.
\newblock In {\em Asilomar Conference on Signals, Systems and Computers}. IEEE, 2016.

\bibitem{shi2018spectrum}
Yi~Shi, Tugba Erpek, Yalin~E Sagduyu, and Jason~H Li.
\newblock Spectrum data poisoning with adversarial deep learning.
\newblock In {\em IEEE Military Communications Conference (MILCOM)}, pages 407--412, 2018.

\bibitem{simonyan2014very}
Karen Simonyan and Andrew Zisserman.
\newblock Very deep convolutional networks for large-scale image recognition.
\newblock {\em arXiv preprint arXiv:1409.1556}, 2014.

\bibitem{soomro2012ucf101}
Khurram Soomro, Amir~Roshan Zamir, and Mubarak Shah.
\newblock Ucf101: A dataset of 101 human actions classes from videos in the wild.
\newblock {\em arXiv preprint arXiv:1212.0402}, 2012.

\bibitem{strasser2008jamming}
Mario Strasser, Christina Popper, Srdjan Capkun, and Mario Cagalj.
\newblock Jamming-resistant key establishment using uncoordinated frequency hopping.
\newblock In {\em 2008 IEEE Symposium on Security and Privacy (sp 2008)}, pages 64--78. IEEE, 2008.

\bibitem{szegedy2013intriguing}
Christian Szegedy, Wojciech Zaremba, Ilya Sutskever, Joan Bruna, Dumitru Erhan, Ian Goodfellow, and Rob Fergus.
\newblock Intriguing properties of neural networks.
\newblock {\em arXiv preprint arXiv:1312.6199}, 2013.

\bibitem{tian2018audio}
Yapeng Tian, Jing Shi, Bochen Li, Zhiyao Duan, and Chenliang Xu.
\newblock Audio-visual event localization in unconstrained videos.
\newblock In {\em Proceedings of the European conference on computer vision (ECCV)}, pages 247--263, 2018.

\bibitem{tian2021can}
Yapeng Tian and Chenliang Xu.
\newblock Can audio-visual integration strengthen robustness under multimodal attacks?
\newblock In {\em Proceedings of the IEEE/CVF conference on computer vision and pattern recognition}, pages 5601--5611, 2021.

\bibitem{tung2023deep}
Tze-Yang Tung and Deniz G{\"u}nd{\"u}z.
\newblock Deep joint source-channel and encryption coding: Secure semantic communications.
\newblock In {\em International Conference on Communications}. IEEE, 2023.

\bibitem{valentini2017noisy}
Cassia Valentini-Botinhao et~al.
\newblock Noisy speech database for training speech enhancement algorithms and tts models.
\newblock {\em University of Edinburgh. School of Informatics. Centre for Speech Technology Research (CSTR)}, 2017.

\bibitem{vaswani2017attention}
Ashish Vaswani, Noam Shazeer, Niki Parmar, Jakob Uszkoreit, Llion Jones, Aidan~N Gomez, {\L}ukasz Kaiser, and Illia Polosukhin.
\newblock Attention is all you need.
\newblock {\em Advances in neural information processing systems}, 30, 2017.

\bibitem{wang2022wireless}
Sixian Wang, Jincheng Dai, Zijian Liang, Kai Niu, Zhongwei Si, Chao Dong, Xiaoqi Qin, and Ping Zhang.
\newblock Wireless deep video semantic transmission.
\newblock {\em IEEE Journal on Selected Areas in Communications}, 41(1):214--229, 2022.

\bibitem{wang2022transformer}
Yang Wang, Zhen Gao, Dezhi Zheng, Sheng Chen, Deniz Gunduz, and H~Vincent Poor.
\newblock Transformer-empowered 6g intelligent networks: From massive mimo processing to semantic communication.
\newblock {\em IEEE Wireless Communications}, 2022.

\bibitem{6g_field}
Alan Weissberger.
\newblock Chinese engineers field test a “6g” network with semantic communications on 4g infrastructure.
\newblock \url{https://techblog.comsoc.org/2024/07/15/chinese-engineers-field-test-a-6g-network-with-semantic-communications-on-4g-infrastructure/}.

\bibitem{welleck2019neural}
Sean Welleck, Ilia Kulikov, Stephen Roller, Emily Dinan, Kyunghyun Cho, and Jason Weston.
\newblock Neural text generation with unlikelihood training.
\newblock {\em arXiv preprint arXiv:1908.04319}, 2019.

\bibitem{wen20245g}
Haohuang Wen, Phillip Porras, Vinod Yegneswaran, Ashish Gehani, and Zhiqiang Lin.
\newblock 5g-spector: An o-ran compliant layer-3 cellular attack detection service.
\newblock In {\em Proceedings of the 31st Annual Network and Distributed System Security Symposium (NDSS'24), San Diego, California, USA. The Internet Society. Google Scholar Google Scholar Cross Ref Cross Ref}, 2024.

\bibitem{Huawei}
Tong Wen, Ma~Jianglei, Zhu Peiying, and Chen Yan.
\newblock Ai: The bridge to 6g.
\newblock https://www.huawei.com/en/huaweitech/publication/202401/ai-bridge-to-6g, 2024.

\bibitem{weng2021semantic}
Zhenzi Weng and Zhijin Qin.
\newblock Semantic communication systems for speech transmission.
\newblock {\em IEEE Journal on Selected Areas in Communications}, 39(8):2434--2444, 2021.

\bibitem{ieee2021wireless}
IEEE~802.11 working group et~al.
\newblock Wireless lan medium access control (mac) and physical layer (phy) specifications-amendment 2: Enhanced throughput for operation in license-exempt bands above 45 ghz.
\newblock {\em IEEE Standard}, 802, 2021.

\bibitem{xie2021deep}
Huiqiang Xie, Zhijin Qin, Geoffrey~Ye Li, and Biing-Hwang Juang.
\newblock Deep learning enabled semantic communication systems.
\newblock {\em IEEE Transactions on Signal Processing}, 69:2663--2675, 2021.

\bibitem{xie2015pistream}
Xiufeng Xie, Xinyu Zhang, Swarun Kumar, and Li~Erran Li.
\newblock pistream: Physical layer informed adaptive video streaming over lte.
\newblock In {\em Proceedings of the 21st Annual International Conference on Mobile Computing and Networking}, pages 413--425, 2015.

\bibitem{xu2005feasibility}
Wenyuan Xu, Wade Trappe, Yanyong Zhang, and Timothy Wood.
\newblock The feasibility of launching and detecting jamming attacks in wireless networks.
\newblock In {\em Proceedings of the 6th ACM international symposium on Mobile ad hoc networking and computing}, pages 46--57, 2005.

\bibitem{xu2021detecting}
Xiaojun Xu, Qi~Wang, Huichen Li, Nikita Borisov, Carl~A Gunter, and Bo~Li.
\newblock Detecting ai trojans using meta neural analysis.
\newblock In {\em IEEE Symposium on Security and Privacy (SP)}, pages 103--120. IEEE, 2021.

\bibitem{xue2019video}
Tianfan Xue, Baian Chen, Jiajun Wu, Donglai Wei, and William~T Freeman.
\newblock Video enhancement with task-oriented flow.
\newblock {\em International Journal of Computer Vision}, 127:1106--1125, 2019.

\bibitem{yang2020temporal}
Ceyuan Yang, Yinghao Xu, Jianping Shi, Bo~Dai, and Bolei Zhou.
\newblock Temporal pyramid network for action recognition.
\newblock In {\em Proceedings of the IEEE/CVF conference on computer vision and pattern recognition}, pages 591--600, 2020.

\bibitem{yang2019hiding}
Hojoon Yang, Sangwook Bae, Mincheol Son, Hongil Kim, Song~Min Kim, and Yongdae Kim.
\newblock Hiding in plain signal: Physical signal overshadowing attack on $\{$LTE$\}$.
\newblock In {\em 28th USENIX Security Symposium (USENIX Security 19)}, pages 55--72, 2019.

\bibitem{yang2023jigsaw}
Limin Yang, Zhi Chen, Jacopo Cortellazzi, Feargus Pendlebury, Kevin Tu, Fabio Pierazzi, Lorenzo Cavallaro, and Gang Wang.
\newblock Jigsaw puzzle: Selective backdoor attack to subvert malware classifiers.
\newblock In {\em IEEE Symposium on Security and Privacy (SP)}, 2023.

\bibitem{yang2022ofdm}
Mingyu Yang, Chenghong Bian, and Hun-Seok Kim.
\newblock Ofdm-guided deep joint source channel coding for wireless multipath fading channels.
\newblock {\em IEEE Transactions on Cognitive Communications and Networking}, 8(2):584--599, 2022.

\bibitem{you2024next}
Changsheng You, Yunlong Cai, Yuanwei Liu, Marco Di~Renzo, Tolga~M Duman, Aylin Yener, and A~Lee Swindlehurst.
\newblock Next generation advanced transceiver technologies for 6g.
\newblock {\em arXiv preprint arXiv:2403.16458}, 2024.

\bibitem{zhang2021survey}
Chaoning Zhang, Philipp Benz, Chenguo Lin, Adil Karjauv, Jing Wu, and In~So Kweon.
\newblock A survey on universal adversarial attack.
\newblock {\em arXiv preprint arXiv:2103.01498}, 2021.

\bibitem{zhang2022near}
Haiyang Zhang, Nir Shlezinger, Francesco Guidi, Davide Dardari, Mohammadreza~F Imani, and Yonina~C Eldar.
\newblock Near-field wireless power transfer for 6g internet of everything mobile networks: Opportunities and challenges.
\newblock {\em IEEE Communications Magazine}, 60(3):12--18, 2022.

\bibitem{zhang2015wireless}
Tan Zhang, Ashish Patro, Ning Leng, and Suman Banerjee.
\newblock A wireless spectrum analyzer in your pocket.
\newblock In {\em Proceedings of the 16th International Workshop on Mobile Computing Systems and Applications}, pages 69--74, 2015.

\end{thebibliography}

\begin{table*}[t]
\centering
\scriptsize
\caption{Surrogate JSCCs used for \sys. We use the $n_1 \Rightarrow n_2$ notation where $n_1$ is the number of layers/kernels for the corresponding module in the template model and $n_2$ is the altered number of layers/kernels in the new victim model.}
\label{tab:substitute_models}
\resizebox{0.85\textwidth}{!}{
\midsepremove
\begin{tabular}{llcccccccc}
\toprule
& & M1 
& M2 
& M3 
& M4 
& M5 
& M6 
& M7
& M8 \\
\midrule
\multicolumn{2}{c}{Template Model}
& Image JSCC~\cite{yang2022ofdm}
& Image JSCC~\cite{yang2022ofdm}
& Video JSCC~\cite{wang2022wireless}
& Video JSCC~\cite{wang2022wireless} 
& Speech JSCC~\cite{weng2021semantic} 
& Speech JSCC~\cite{weng2021semantic}
& Text JSCC~\cite{xie2021deep}
& Text JSCC~\cite{xie2021deep} \\
\midrule
\multirow{3}{*}{\begin{tabular}[c]{@{}l@{}}JSCC \\ Encoder\end{tabular}}
& \# Layers & 8 $\Rightarrow$ 6 & 8 $\Rightarrow$ 10 & 6 $\Rightarrow$ 5 & 6 $\Rightarrow$ 8 & 19 $\Rightarrow$ 16 & 19 $\Rightarrow$ 22 & 5 $\Rightarrow$ 6 & 5 $\Rightarrow$ 7 \\ 
\cline{2-10}
& \multirow{2}{*}{\# Kernels}  & 64 $\Rightarrow$ 56 & 64 $\Rightarrow$ 72& 128 $\Rightarrow$ 120 & 128 $\Rightarrow$ 136& \multirow{2}{*}{64 $\Rightarrow$ 56} & \multirow{2}{*}{64 $\Rightarrow$ 72} & \multirow{2}{*}{256 $\Rightarrow$ 248} & \multirow{2}{*}{256 $\Rightarrow$ 264}\\ 
& & 128 $\Rightarrow$ 120 & 128 $\Rightarrow$ 136 & 192 $\Rightarrow$ 184 & 192 $\Rightarrow$ 200 & & & & \\ 
\midrule
\multirow{3}{*}{\begin{tabular}[c]{@{}l@{}}JSCC \\ Decoder\end{tabular}}
& \# Layers & 8 $\Rightarrow$ 6 & 8 $\Rightarrow$ 10 & 6 $\Rightarrow$ 5 & 6 $\Rightarrow$ 8 & 19 $\Rightarrow$ 16 & 19 $\Rightarrow$ 22 & 6 $\Rightarrow$ 7 & 6 $\Rightarrow$ 8\\ 
\cline{2-10}
& \multirow{2}{*}{\# Kernels}  & 64 $\Rightarrow$ 56 & 64 $\Rightarrow$ 72& 128 $\Rightarrow$ 120 & 128 $\Rightarrow$ 136 & \multirow{2}{*}{64 $\Rightarrow$ 56} & \multirow{2}{*}{64 $\Rightarrow$ 72} & \multirow{2}{*}{256 $\Rightarrow$ 248} & \multirow{2}{*}{256 $\Rightarrow$ 264} \\ 
& & 128 $\Rightarrow$ 120 & 128 $\Rightarrow$ 136 & 192 $\Rightarrow$ 184 & 192 $\Rightarrow$ 200 & & & & \\ 
\midrule
\multirow{2}{*}{\begin{tabular}[c]{@{}l@{}}Video \\ Analysis\end{tabular}}
& \# Layers & \textemdash & \textemdash & 10 $\Rightarrow$ 7 & 10 $\Rightarrow$ 13 & \textemdash & \textemdash & \textemdash & \textemdash \\ 
& \# Kernels & \textemdash & \textemdash & 96 $\Rightarrow$ 88 & 96 $\Rightarrow$ 104 & \textemdash & \textemdash & \textemdash & \textemdash \\
\midrule
\multirow{2}{*}{\begin{tabular}[c]{@{}l@{}}Video \\ Synthesis\end{tabular}}
& \# Layers & \textemdash & \textemdash & 13 $\Rightarrow$ 10 & 13 $\Rightarrow$ 16& \textemdash & \textemdash & \textemdash & \textemdash \\ 
& \# Kernels & \textemdash & \textemdash & 96 $\Rightarrow$ 88 & 96 $\Rightarrow$ 104 & \textemdash & \textemdash & \textemdash & \textemdash \\
\bottomrule
\end{tabular}}
\vspace{-0.2cm}
\end{table*}

\begin{table*}[h]
\centering
\scriptsize
\caption{Surrogate downstream models for \sys. We use the $n_1 \Rightarrow n_2$ notation where $n_1$ is the number of layers/kernels for the corresponding module in the template model and $n_2$ is the altered number of layers/kernels in the new victim model.}
%\tiny
\label{tab:substitute_models2}
\resizebox{0.7\textwidth}{!}{
\midsepremove
\begin{tabular}{llcccccccc}
\toprule
& & M1 
& M2 
& M3 
& M4 
& M5 
& M6 
& M7
& M8 \\
\midrule
\multicolumn{2}{c}{Template Model}
& I3D~\cite{carreira2017quo}
& I3D~\cite{carreira2017quo}
& SlowFast~\cite{feichtenhofer2019slowfast}
& SlowFast~\cite{feichtenhofer2019slowfast} 
& TPN~\cite{yang2020temporal} 
& TPN~\cite{yang2020temporal}
& AVE~\cite{tian2021can}
& AVE~\cite{tian2021can} \\
\midrule
\multirow{2}{*}{Classifier}
& \# Layers & 57 $\Rightarrow$ 51 & 57 $\Rightarrow$ 63 & 30 $\Rightarrow$ 26 & 30 $\Rightarrow$ 34 & 17 $\Rightarrow$ 15 & 17 $\Rightarrow$ 19 & 27 $\Rightarrow$ 24 & 27 $\Rightarrow$ 30 \\ 
\cline{2-10}
& Kernels  & 64 $\Rightarrow$ 56 & 64 $\Rightarrow$ 72& 128 $\Rightarrow$ 120 & 128 $\Rightarrow$ 136& 64 $\Rightarrow$ 56 & 64 $\Rightarrow$ 72 & 64 $\Rightarrow$ 56 & 64 $\Rightarrow$ 72\\ 
\bottomrule
\end{tabular}}
\vspace{-0.2cm}
\end{table*}

\begin{figure*}[h]
\centering
    \resizebox{0.7\textwidth}{!}{
    \begin{tabular}{@{}c@{}}
    \includegraphics[width=\linewidth]{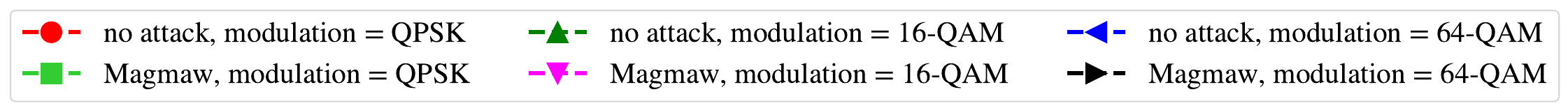} \\  
    \end{tabular}}
    \begin{tabular}{@{}cccc@{}}
    \includegraphics[width=0.17\linewidth]{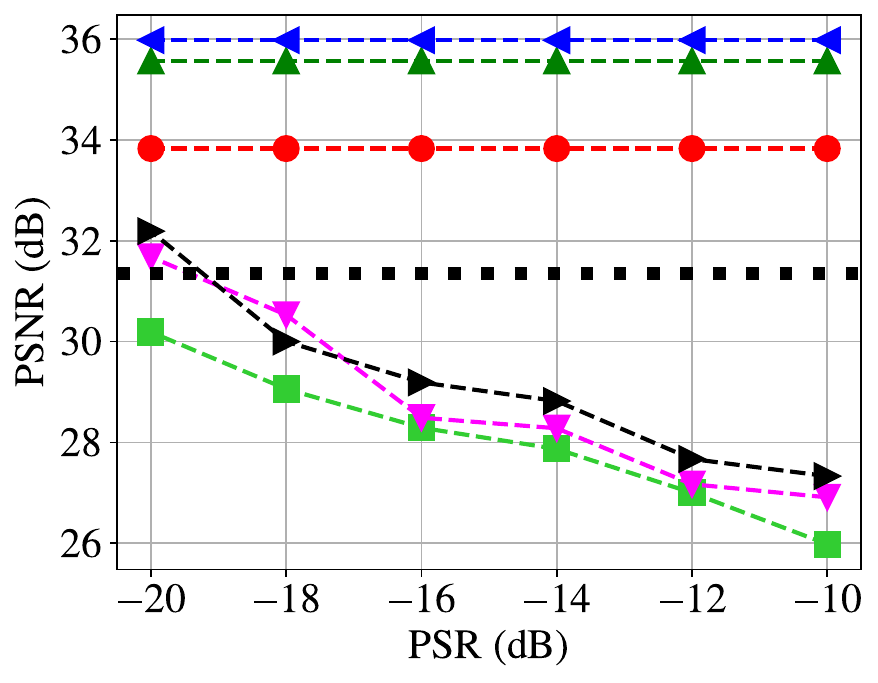}   &
    \includegraphics[width=0.17\linewidth]{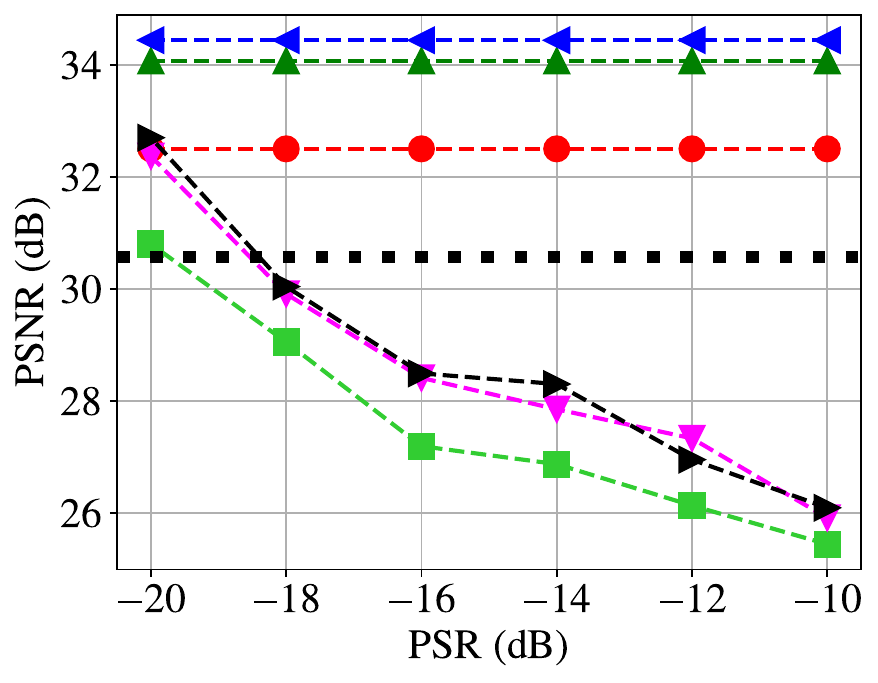}    &
    \includegraphics[width=0.17\linewidth]{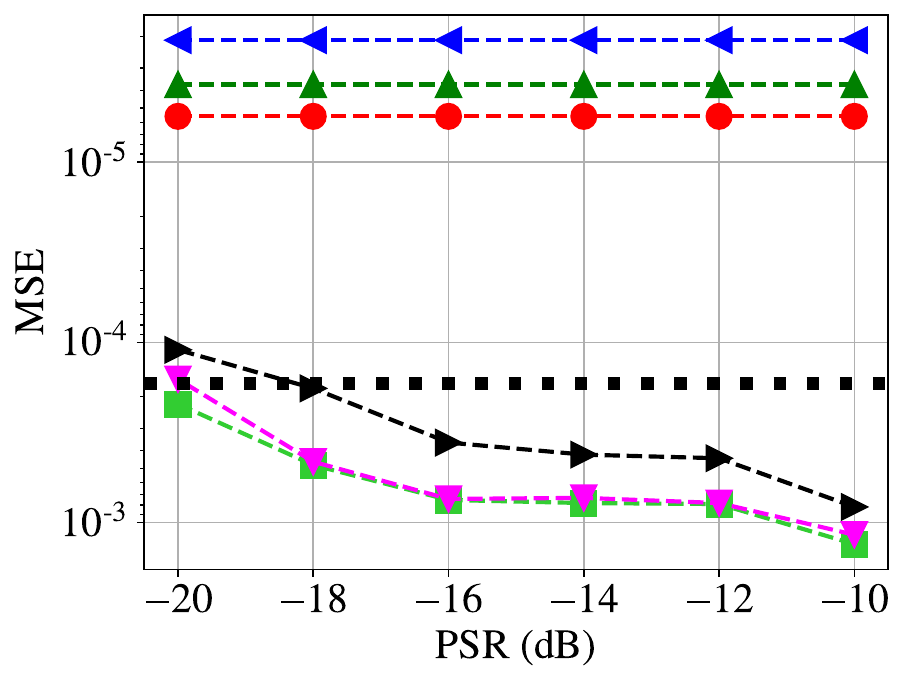} &
    \includegraphics[width=0.17\linewidth]{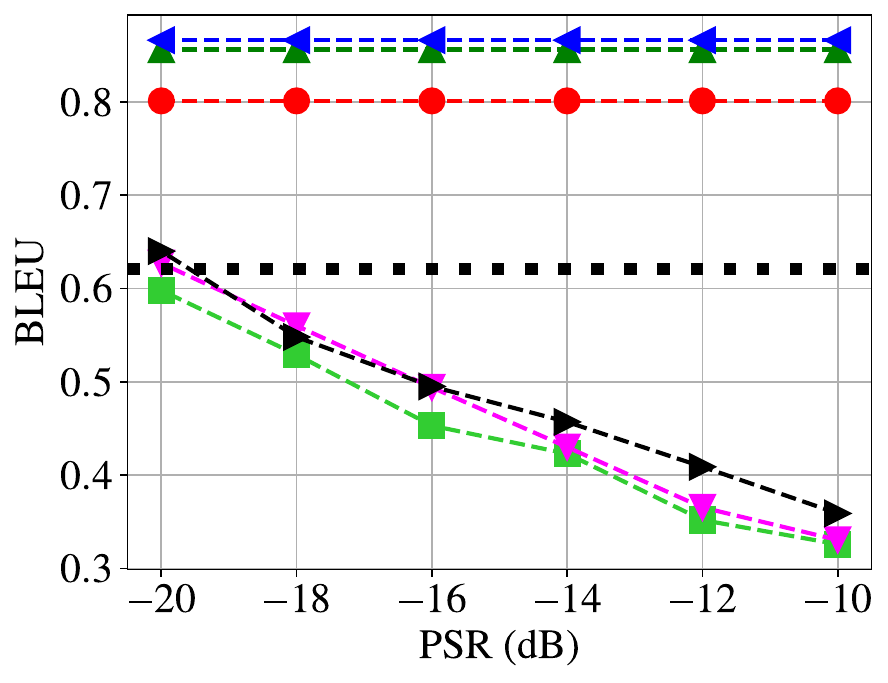} \\
    \footnotesize{Image Transmission~\cite{yang2022ofdm}}&
    \footnotesize{Video Transmission~\cite{wang2022wireless}}&
    \footnotesize{Speech Transmission~\cite{weng2021semantic}} &
    \footnotesize{Text Transmission~\cite{xie2021deep}}\\
    \end{tabular}
    \begin{tabular}{@{}cccc@{}}
    \includegraphics[width=0.17\linewidth]{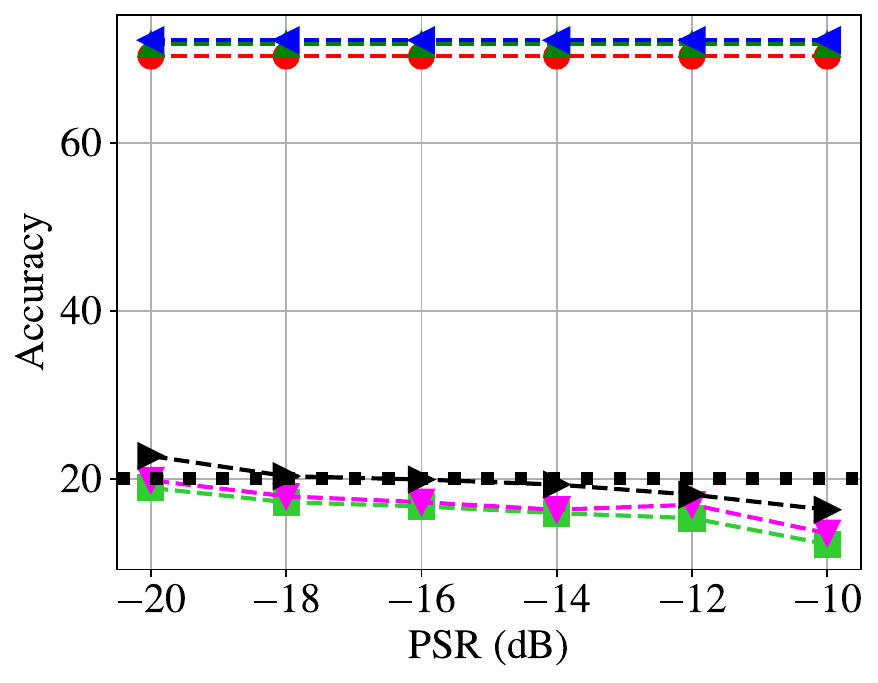} &
    \includegraphics[width=0.17\linewidth]{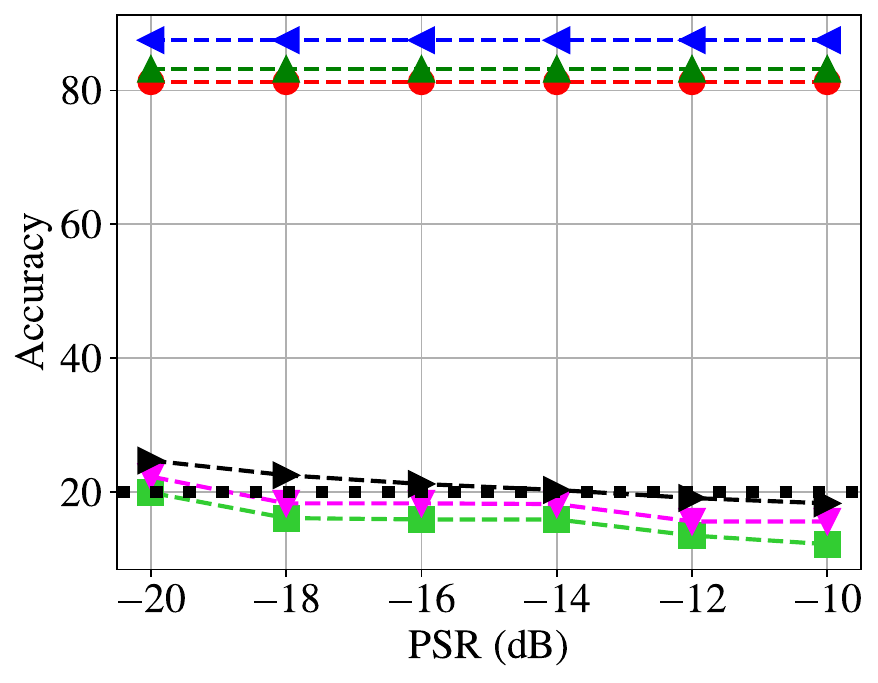} &
    \includegraphics[width=0.17\linewidth]{Fig/Exp/Exp_modulation_psr_slowfast.pdf} &
    \includegraphics[width=0.17\linewidth]{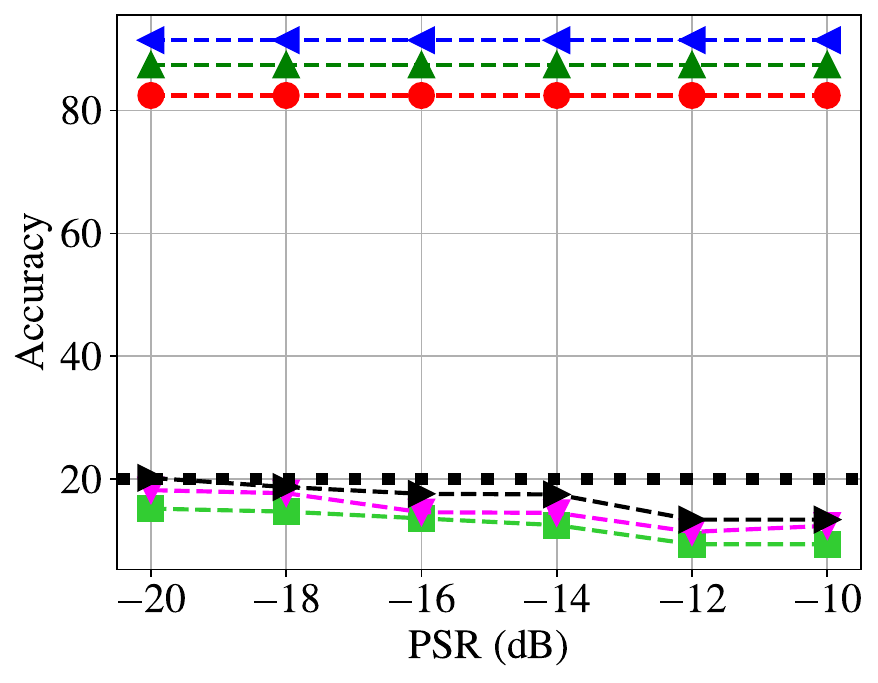} \\
    \footnotesize{I3D~\cite{carreira2017quo}}&
    \footnotesize{SlowFast~\cite{feichtenhofer2019slowfast}}&
    \footnotesize{TPN~\cite{yang2020temporal}} &
    \footnotesize{AVE~\cite{tian2021can}}\\
    \end{tabular}
  \vspace{-0.2cm}
  \caption{\sys on wireless systems with different types of constellation mapping schemes (i.e., QPSK, 16-QAM, 64-QAM). The first and second rows are the results of JSCC and downstream tasks, respectively.}
  \label{fig:appendix1}
  \vspace{-0.3cm}
\end{figure*}

\begin{comment}
\begin{figure*}[h]
\centering
    \resizebox{0.7\textwidth}{!}{
    \begin{tabular}{@{}c@{}}
    \includegraphics[width=\linewidth]{Fig/Exp/Exp_legend_modulation.pdf} \\  
    \end{tabular}}
    \begin{tabular}{@{}cccc@{}}
    \includegraphics[width=0.19\linewidth]{Fig/Exp/Exp_modulation_psr_i3d.pdf} &
    \includegraphics[width=0.19\linewidth]{Fig/Exp/Exp_modulation_psr_slowfast.pdf} &
    \includegraphics[width=0.19\linewidth]{Fig/Exp/Exp_modulation_psr_slowfast.pdf} &
    \includegraphics[width=0.19\linewidth]{Fig/Exp/Exp_modulation_psr_ave.pdf} \\
    \footnotesize{I3D~\cite{carreira2017quo}}&
    \footnotesize{SlowFast~\cite{feichtenhofer2019slowfast}}&
    \footnotesize{TPN~\cite{yang2020temporal}} &
    \footnotesize{AVE~\cite{tian2021can}}\\
    \end{tabular}
  \caption{\sys on ML-based downstream tasks with different types of constellation mapping schemes.}
  \label{fig:appendix2}
\end{figure*}
\end{comment}

%\begin{comment}
%\end{comment}

\begin{appendices}
%\vspace{-0.3cm}
\section{Real-World Experimental Settings}
\label{sec:appendix_real_world}
\begin{figure}[h]
    \centering
    \vspace{-0.3cm}
    \begin{tabular}{@{}c@{}}
        \includegraphics[width=0.95\columnwidth]{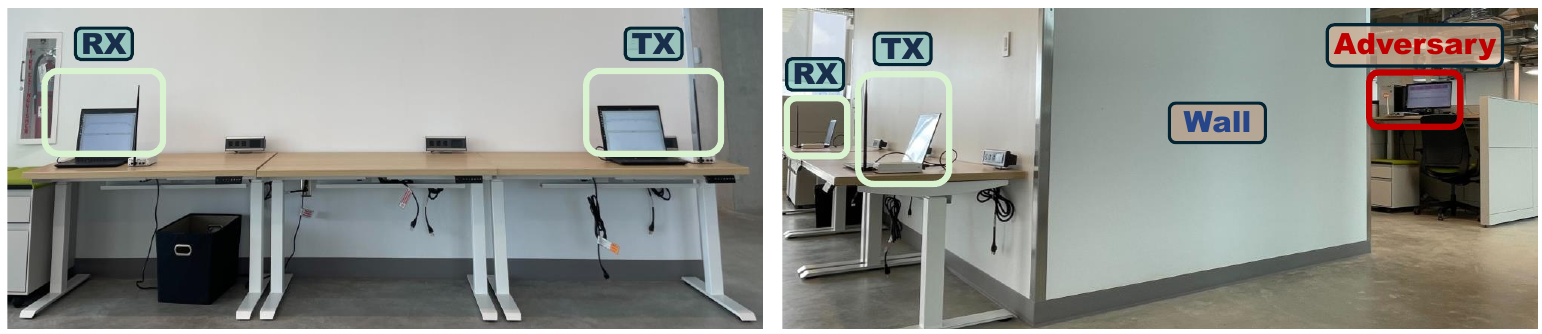} \\
    \end{tabular}
    \caption{Experiment Settings. A scenario where an adversary sends an adversarial signal from behind a wall in LoS Tx-Rx.}
    \label{fig:real_world}
    \vspace{-0.3cm}
\end{figure}

%\vspace{-0.2cm}
We choose a representative indoor environment, as depicted in Figure~\ref{fig:real_world}. In this setting, an unidentified adversary transmits an adversarial signal from behind a wall in Line-of-Sight (LoS) between the transmitter (Tx) and receiver (Rx). \jungwoo{We also consider the scenario where multiple users share the spectrum. When multiple devices try to transmit data simultaneously, the Wi-Fi protocol allows only one device to transmit to prevent interference between transmitters~\cite{halperin2010802}. There are two main anti-collision mechanisms: (1) carrier sensing and (2) collision avoidance. Before sending the data, a wireless device first listens to the shared medium to determine whether another device is sending signals. The transmitter detects the signal power of the target channel on the shared medium. If the signal power is greater than a threshold, the transmitter stops transmitting packets and waits for a certain amount of time (usually random). The transmitter repeats the above anti-collision process until it determines that the shared medium is clear. 
\sys disrupts packets whenever a transmitter sends data by continuously sending adversarial signals.
%This allows \sys to perturb the packets every time the transmitters send data.
}

\end{appendices}

% IEEEtran.cls defaults to using nonbold math in the Abstract.
% This preserves the distinction between vectors and scalars. However,
% if the conference you are submitting to favors bold math in the abstract,
% then you can use LaTeX's standard command \boldmath at the very start
% of the abstract to achieve this. Many IEEE journals/conferences frown on
% math in the abstract anyway.

% no keywords

% For peer review papers, you can put extra information on the cover
% page as needed:
% \ifCLASSOPTIONpeerreview
% \begin{center} \bfseries EDICS Category: 3-BBND \end{center}
% \fi
%
% For peerreview papers, this IEEEtran command inserts a page break and
% creates the second title. It will be ignored for other modes.
%\IEEEpeerreviewmaketitle

\end{document}